\documentclass[aps,pra, twocolumn,superscriptaddress,longbibliography]{revtex4-2}
\usepackage{graphicx}
\usepackage{float}
\usepackage[dvipsnames]{xcolor}
\usepackage{derivative}
\usepackage{xcolor}
\usepackage[margin=0.68in]{geometry}
\usepackage{siunitx}
\usepackage{soul}
\usepackage{minitoc}
\usepackage{color}
\usepackage{amssymb}
\usepackage{amsmath, scalerel, breqn}
\usepackage{colortbl}
\usepackage{bbm}
\usepackage{physics}
\usepackage{bbold}
\usepackage[normalem]{ulem}
\usepackage{comment}
\usepackage{enumitem}

\definecolor{dark-red}{rgb}{0.84,0.15,0.15}
\definecolor{dark-blue}{rgb}{0.15,0.15,0.4}
\definecolor{medium-blue}{rgb}{0,0,0.5}
\definecolor{copper}{rgb}{0.72, 0.45, 0.2}

\usepackage{hyperref}
\hypersetup{
    colorlinks, linkcolor={dark-red},
    citecolor={dark-blue}, urlcolor={medium-blue}
}

\usepackage[most]{tcolorbox}
\usetikzlibrary{patterns}
\pgfdeclarepatternformonly{editstrikeoutpattern}{\pgfqpoint{-1pt}{-1pt}}{\pgfqpoint{11pt}{11pt}}{\pgfqpoint{10pt}{10pt}}%
{
  \pgfsetlinewidth{0.4pt}
  \pgfpathmoveto{\pgfqpoint{0pt}{0pt}}
  \pgfpathlineto{\pgfqpoint{10.1pt}{10.1pt}}
  \pgfusepath{stroke}
}
\newtcolorbox{tcbstrikeout}{breakable,
 enhanced jigsaw,
 opacityback=0,
 parbox=false,
 boxrule=0mm,
 top=0mm,bottom=0pt,left=0pt,right=0pt,
 boxsep=0pt,
 frame hidden,
 finish={\fill[pattern=editstrikeoutpattern] (frame.north west) rectangle (frame.south east);}
}

\usepackage[capitalize]{cleveref}

\definecolor{green}{rgb}{0.08,0.7,0.05}

\newcommand{\CNOT}[2]{\mathsf{C}_{#1}\mathsf{NOT}_{#2}}

\newcommand{\ulgt}{\hat{U}_\mathrm{LGT}}
\makeatletter
\def\@fnsymbol#1{\ensuremath{\ifcase#1\or *\or \dagger\or \dagger\or
		\mathsection\or \mathparagraph\or \|\or **\or \dagger\dagger
		\or \ddagger\ddagger \else\@ctrerr\fi}}
\makeatother

\begin{document}
	\doparttoc 
	\faketableofcontents 
	\part{}
	
	\title{Observation of disorder-free localization using a (2+1)D lattice gauge theory on a quantum processor}
	\author{Google Quantum AI and Collaborators\hyperlink{authorlist}{$\,^\dagger$}}
	%\author{Google Quantum AI and Collaborators$\,^\dagger$}
	% \date{\today}

	\begin{abstract}
		Disorder-induced phenomena in quantum many-body systems pose significant challenges for analytical methods and numerical simulations at relevant time and system scales. To reduce the cost of disorder-sampling, we investigate quantum circuits initialized in states tunable to superpositions over all disorder configurations. In a translationally-invariant lattice gauge theory (LGT), these states can be interpreted as a superposition over gauge sectors. We observe localization in this LGT  in the absence of disorder in one and two dimensions: perturbations fail to diffuse despite fully disorder-free evolution and initial states. However, R\'enyi entropy measurements reveal that superposition-prepared states fundamentally differ from those obtained by direct disorder sampling. Leveraging superposition, we propose an algorithm with a polynomial speedup in sampling disorder configurations, a longstanding challenge in many-body localization studies.
	\end{abstract}

	\maketitle 
	
	Understanding electric and thermal conduction in solid state systems is at the core of condensed matter physics and has significance for modern technology\,\cite{Mott1968,AMermin_text,Dobrosavljevic}. In the absence of disorder and at late times, such systems relax to thermal states that are well described by equilibrium statistical mechanics under the so-called ergodic hypothesis. Mechanisms for violating ergodicity have attracted interest related to the creation and probing of nonequilibrium matter, and have been proposed for protecting coherence in quantum systems\,\cite{bahri2015localization,Kim_PRL_2016,SchulzPRL2019,van2019bloch,ribeiro2020many,scherg2021observing,morong2021observation,Guy_PRB_2022,adler2024observation}. Introducing spatial disorder in a local potential is a well-known mechanism for breaking ergodicity and can result in the localization of excitations. In non-interacting systems, disorder can change the structure of eigenstates from being extended to exponentially localized\,\cite{Anderson1958}, and has been studied in many classical and quantum systems\,\cite{Wiersma1997nature,Schwartz2007,Aspect2008nature,kondov2011three,50years, Karamlou2022}. For many years, the conventional wisdom was that interacting systems do not localize, regardless of the disorder magnitude. However, theoretical and experimental studies of the past two decades have shown that localization may persist on very large timescales even in the presence of interactions between particles\,\cite{basko2006metal,Serbyn_PRL_2013,Huse_PRB_2014}. Increasing disorder strength in these systems could result in a non-equilibrium many-body localized (MBL) phase\,\cite{basko2006metal,nandkishore2015many,abanin2017recent,schreiber2015observation}.
	
\begin{figure*}[t!]
		\centering
		\includegraphics[width=0.95\textwidth]{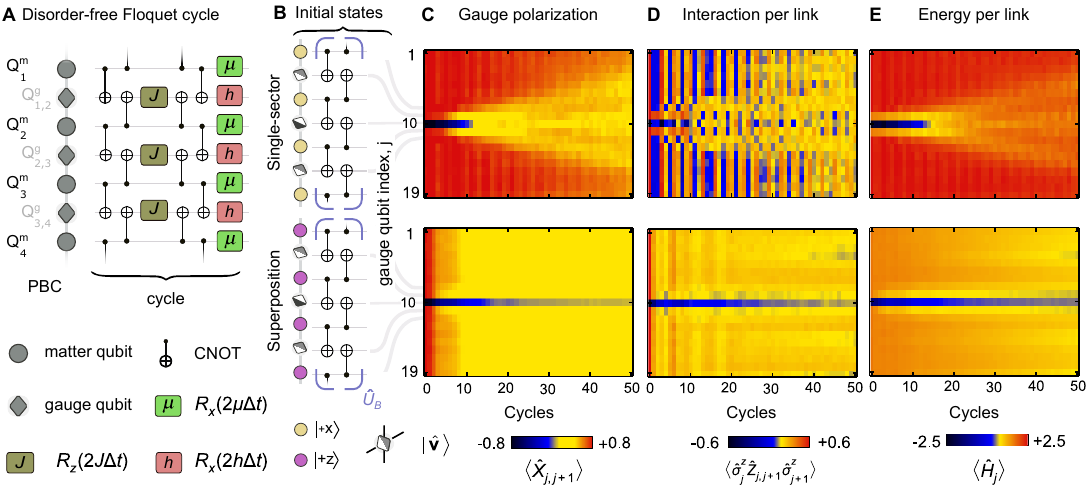}
		\caption{\small \textbf{Transitionally invariant unitary and dynamics of a local perturbation in 1d}.  (\textbf{A}) Schematic of the unitary gate sequence showing a typical Floquet cycle that implements the second-order Trotter unitary corresponding to the lattice gauge theory Hamiltonian $\hat{H}_{\textrm{LGT}}$ in \cref{eq:lgt-hamiltonian}, with $\mathbb{Z}_2$ matter sites (circles) and $\mathbb{Z}_2$ gauge sites (diamonds) in 1d with periodic boundary condition\,(PBC). (\textbf{B}) Schematic of the circuits to prepare translationally invariant initial states of 38 qubits ($N_m=N_g=19$) with an energy perturbation at the center link ($j=10$). The center gauge qubit is initialized in a direction opposite to the rest on the $XZ$-plane of the Bloch sphere. Gauge qubits are shown as half-filled diamonds with the direction of the filled half indicating the direction of the qubit Bloch vector (\(\hat{\mathbf{v}} \propto J \hat{\mathbf{z}} + h \hat{\mathbf{x}}\) or \(- \hat{\mathbf{v}}\)).  Likewise, matter qubits are prepared in an eigenstate of either Pauli-$X$ (yellow circles) or -$Z$ (purple circles). Then two layers of CNOT gates\,($\hat U_B$) are applied to these product states resulting in a single-sector or superposition initial states respectively. Dynamics of the initial perturbation were examined through the gauge polarization (\textbf{C}), matter-gauge-matter interaction (\textbf{D}), and the energy per link (\textbf{E}). Gauge qubit index $j$ is used to label the gauge qubit $Q^g_{j,j+1}$ on link $(j,j+1)$. The Floquet parameters used were $J=1, \, \Delta t =0.25, \, h=1.3$, and  $\mu=1.5$.} 
		\label{fig:1d_energy_dynamics} 
	\end{figure*}

	Given the association of quenched disorder with localization, it is
	interesting to ask whether it is possible to localize excitations without
	breaking translational
	invariance\,\cite{Grover_2014,Schiulaz_PRB_2015,yao2016quasi,Hickey_2016,Mondaini_PRB_2017,Mazza_PRB_2019,Bernien2017Scars,Serbyn2021quantum,Chandran2023quantum}. Here we use -- seemingly paradoxically -- translationally invariant evolutions of nearly translationally invariant states to study disordered systems. We study quantum evolution corresponding to the Trotterized dynamics of a lattice gauge theory (LGT) Hamiltonian in (1+1)D and (2+1)D (see Eq.~\eqref{eq:lgt-hamiltonian}). Using a superconducting quantum processor, we observe that energy excitations remain localized for certain nearly translationally invariant initial conditions in both one and two dimensions (\cref{fig:1d_energy_dynamics} and  \cref{fig:2d_energy_dynamics}, respectively).

	This observation of disorder-free localization (DFL) in the dynamics generated by \cref{eq:lgt-hamiltonian} is not entirely surprising. This Hamiltonian has an extensive number of local symmetries, whose eigenvalues correspond to an effective background potential. We will show this explicitly by a local unitary transformation under which the symmetry generators appear as couplings. The localizing translationally invariant states correspond to superpositions over all symmetry sectors (i.e. background potentials). In most sectors, the dynamics take place in a disordered background, providing an explanation for the localization of energy. Moreover, from this perspective, our experiment is a demonstration of a general scheme for the study of disordered systems on superconducting processors. 
	
	Generally, studying a disordered system requires averaging over disordered realizations. Consider a Hamiltonian that consists of ordered and disordered parts: $\hat{H} = \hat{H}_{\mathrm{ord}} + \hat{H}_{\mathrm{dis}}$ with $\hat{H}_{\mathrm{dis}} = \sum_{j} g_{j} \hat{D}_j$, where $\hat{D}_j$ is local and $g_j$ is binary disorder $g_{j} = \pm 1$. We are interested in the expectation value of the quantum observable at time $t$ for a quenched disorder realization $\mathbf{g}$, $\langle \hat{O}(t)\rangle_{\mathbf{g}}$, averaged over many $\mathbf{g} = \{g_j\}$ i.e., 
	\begin{equation}
		\overline{\langle\,{\hat{O}(t)}\,\rangle} = \sum_{\mathbf{g}} w_{\mathbf{g}}\, \langle\,{\hat{O} (t) \,\rangle}_{\mathbf{g}},
		\label{eqn:disorder_averaging}
	\end{equation}
	where $w_{\mathbf{g}}$ is the weight corresponding to each disordered realization and is independent of the Hamiltonian. To predict the behavior in the thermodynamic limit via computational studies in a finite system of size $N$, averaging over many disorder realizations is required. While polynomial (in $N$) number of disorder realizations are sufficient for self-averaging systems such as good metals, an exponential number $(e^{\# N})$ is required for systems dominated by rare events. 
	
\begin{figure}[b!]
		\centering
		\includegraphics[width=0.37\textwidth]{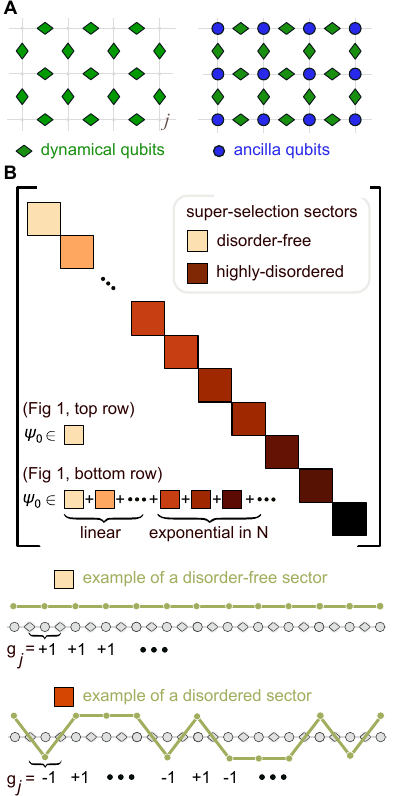}
		\caption{\small \textbf{Qubit lattice and representation of a generator of dynamics respecting local symmetries.} (\textbf{A}) A Hamiltonian with disorder defined on a grid of qubits\,(left) can be mapped into a disorder-free Hamiltonian by introducing ancillary qubits\,(right); the disorder variable is mapped to conserved operators ($\hat \sigma^X$) acting on each ancilla. (\textbf{B}) These local symmetries result in decoupled super-selection sectors\,(colored squares) in Hilbert space, each defined by a unique set of quantum numbers corresponding to the eigenvalues of the conserved operators in Eq.~\eqref{eq:gauss_law}. Most sectors contain disordered background charge configurations; such sectors are the dominant contribution to both generic and uniformly superposed initial states.} 
		\label{fig:block_diagonal}
	\end{figure}

\begin{figure*}[t!]
		\centering
		\includegraphics[width=0.9\textwidth]{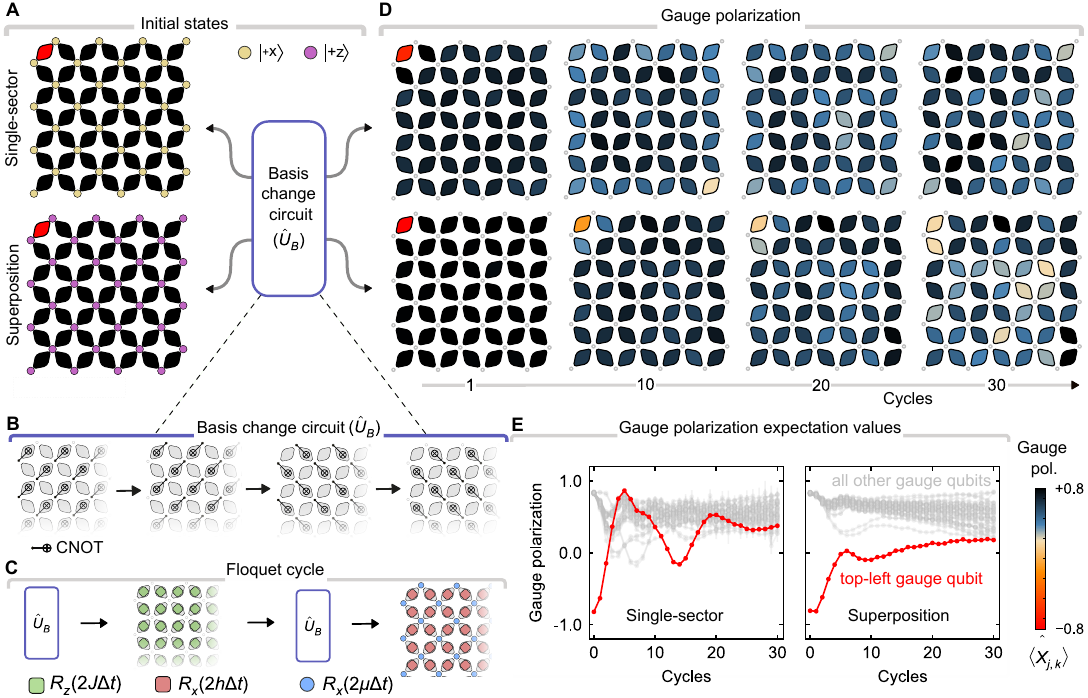}
		\caption{\textbf{Dynamics of a local perturbation in 2d}. (\textbf{A}) Schematic of the quantum circuits to prepare disorder-free initial states of 81 qubits ($N_m=32$, $N_g=49$) with a local perturbation at the top-left link. The perturbation was created by preparing the corresponding gauge qubit Bloch vector in the direction $-\hat{\mathbf{v}}$ (red diamond), opposite to the rest which are prepared in $\hat{\mathbf{v}} \propto J \hat{\mathbf{z}} + h \hat{\mathbf{x}}$ (black diamonds). The matter qubits are prepared in an eigenstate of either Pauli-$X$ (yellow circles) or $Z$ (purple circles). Four layers of CNOTs $(\hat U_B)$ shown in (\textbf{B}) are then applied to these product states to prepare entangled initial states in either a single-sector or a superposition of all sectors. (\textbf{C}) A typical Floquet cycle that implements the second-order Trotter dynamics corresponding to the LGT Hamiltonian $\hat{H}_{\textrm{LGT}}$. In practice, we use CZ and CPhase gates to reduce the circuit depth (see SM). (\textbf{D}) Dynamics of the gauge polarization for the two perturbed initial states. Traces for individual gauge qubits are shown in (\textbf{E}) where the perturbed qubit is shown in red and the rest in gray. The Floquet parameters used were $J=1, \, \Delta t =0.35, \, h=1.5$, and  $\mu=3.5$.}
		\label{fig:2d_energy_dynamics}
	\end{figure*}

	In this work, we implement a protocol that could be leveraged for effective sampling of disorder realizations\,(see discussion on Fig. 5). 
	Leveraging quantum superposition is at the core of quantum algorithms, such as Grover search, and has been discussed in the context of simulating randomness as quantum parallelism\,\cite{Paredes2005PRL}; its relevance for studying disordered systems has been the subject of several pioneering works \,\cite{sirker_purification_and_mbl_2014,sirker_mbl_in_infinite_chains}. Here, we experimentally implement a Trotterized time evolution by augmenting the system with ancillary qubits\,(\cref{fig:block_diagonal}A) for each $\hat{D}_j$ and replacing the disordered coupling $g_j$ with an ancilla operator $\hat{\sigma}^X_j$, i.e., $\hat{H}_{\mathrm{dis}}^Q = \sum_j \hat{\sigma}^X_j \hat{D}_j$. We initialize the system in a product state between the ancillas and dynamical qubits:
	
	\begin{align}
		\label{eq:initial_state}
		\ket{\psi_0} = \ket{\psi_0}_{\mathrm{dynamical}}  \otimes \ket{\psi_0}_{\mathrm{ancillas}}. 
	\end{align}
	Preparing all the ancillas in an eigenstate of $\hat{\sigma}^x_j$ is equivalent to implementing a particular disorder realization. Remarkably, we can also readily prepare the ancillas in $\ket{00\cdots 0}$ which is a superposition over all disorder configurations with equal weights. We then evolve the system under $\hat{H}^Q = \hat{H}_{\mathrm{ord}} \otimes \openone_{\mathrm{ancilla}} + \hat{H}_{\mathrm{dis}}^Q$, where the first term acts only on the dynamical qubits whereas the second acts also on the ancillas. Measuring expectation values after time-evolution is equivalent to the disorder-average \cref{eqn:disorder_averaging}. Furthermore, we can change the disorder density by rotating the ancillas in the $XZ$-plane, and the disorder magnitude by changing $\mu$. By replacing ancilla qubits with qudits we could also study a finer range of disorder amplitudes. It is worth emphasizing that both the Hamiltonian $\hat{H}^Q$ and the initial states of type \cref{eq:initial_state} are \emph{disorder-free}.
	
	According to this framework, we focus on the Hamiltonian\,(before introducing ancillas)
	% \begin{align}
		%     \label{eq:dual_hamiltonian}
		%   \hat{H} &= \frac{1}{2d} \sum_{j} \sum_{k \in N(j)} \left( J \hat{Z}_{j,k} + h \hat{X}_{j,k} \right) +  \sum_{j} g_j \hat{D}_j, \nonumber \\
		%     \hat{D}_j &= \mu \prod_{k \in N(j)} \hat{X}_{j,k},
		% \end{align}
	\begin{align}
		\label{eq:dual_hamiltonian}
		\hat{H} &=  \sum_{\langle j,k\rangle} \left( J \hat{Z}_{j,k} + h \hat{X}_{j,k} \right) +  \sum_{j} g_j \hat{D}_j, \nonumber \\
		\hat{D}_j &= \mu \prod_{k \in N(j)} \hat{X}_{j,k},
	\end{align}
	where the qubits live on the links of either a ring in 1d or a square lattice in 2d. Here, $N(j)$ refers to the set of nearest neighbors of vertex
	$j$, and $\hat Z_{j,k}$ and $\hat X_{j,k}$ are single-qubit Pauli operators on the link qubits between vertices $j$ and $k$. We now implement the above procedure, augmenting the system with
	ancillas and replacing \(g_j\) with ancilla operators \(\hat \sigma^X_j\) that live on the vertices of the lattice. Additionally, in implementing our Trotterized evolution we take advantage of the more local, unitarily equivalent (by the unitary \(\hat U_B\) defined in the SM) Hamiltonian
	\begin{equation}
		\label{eq:lgt-hamiltonian}
		\hat{H}_{\textrm{LGT}} =
		\sum_{\langle j, k \rangle}\left(
		J \hat{\sigma}^{Z}_j \hat{Z}_{j,k} \hat{\sigma}^{Z}_{k}
		+ h \hat{X}_{j,k}\right)
		+ \mu \sum_j \hat{\sigma}^{X}_j.
	\end{equation}
	% \begin{align}
		% \hat{U}_{\mathrm{LGT}} = \hat U_J(\Delta t/2) \; \hat U_h(\Delta t)  \; \hat U_{\mu}(\Delta t) \; \hat U_J(\Delta t/2),
		% \label{eqn:flouqet_unitary}
		% \end{align}
	% where,
	% \begin{align}
		%     \hat U_J(\Delta t) &= \exp\Big(\hspace{-1mm}-\frac{i J\Delta t}{2d}  \sum_{j} \sum_{k \in N(j)} \hat \sigma^Z_j\,\hat Z_{j,k}\,\hat \sigma^Z_{k} \Big), \nonumber\\
		%     \hat U_h(\Delta t) &= \exp\Big(\hspace{-1mm} -\frac{i  h\Delta t}{2d} {\sum_{j} \sum_{k \in N(j)} \hat X_{j,k}}\Big), \nonumber\\
		%     \hat U_\mu (\Delta t) &= \exp\Big(\hspace{-1mm}-i\mu\,\Delta t \hspace{1mm} {\sum_{j} \hat \sigma^X_j }\Big).
		% \end{align}
	
	% \editstrikeout{\begin{equation}
			% %\label{eqn:flouqet_unitary}
			% \begin{split}
				% \hat U_{\mathrm{LGT}}  & = \exp\Big(\hspace{-1mm}-\frac{i J\Delta t}{2d}  \sum_{j} \sum_{k \in N(j)} \hat \sigma^Z_j\,\hat Z_{j,k}\,\hat \sigma^Z_{k} \Big) \\ 
				% & \exp\Big(\hspace{-1mm} -\frac{i  h\Delta t}{2d} {\sum_{j} \sum_{k \in N(j)} \hat X_{j,k}}\Big) \,\,\exp\Big(\hspace{-1mm}-i\mu\,\Delta t \hspace{1mm} {\sum_{j} \hat \sigma^X_j }\Big).  
				% \end{split}
			% \end{equation}}
	\noindent This can be viewed as a lattice gauge theory (LGT) model \cite{Kogut_RMP_1979} with $N_m$ matter qubits on the vertices and $N_g$ gauge qubits on the links, so we borrow the LGT nomenclature. The first term in $\hat{H}_{\textrm{LGT}}$ consists of Ising interactions between neighboring matter sites with coupling strength $J$, mediated by a Pauli-$Z$ on the gauge qubit linking them\,(\,Fig.~\ref{fig:1d_energy_dynamics}A\,). Operators $\hat \sigma_{j}^{X/Y/Z}$ and $\hat X/\hat Y/\hat Z_{j,k}$ refer to the Pauli-$X/Y/Z$ on the matter and gauge qubits respectively. The second term is the sum of local Pauli-$X$ on gauge qubits with strength $h$, referred to as the gauge polarization. Similarly, the last term, which has strength $\mu$, is the matter polarization. 
	In each cycle, the dynamics are advanced by a second-order Trotter unitary $\hat{U}_{\mathrm{LGT}}$ corresponding to \cref{eq:lgt-hamiltonian} with an adjustable Trotter step $\Delta t$. $\hat{U}_{\mathrm{LGT}}$ consists of one-step time evolution for the matter and gauge polarization terms sandwiched between the half-step time evolution for the matter-gauge-matter term (see SM, Sec.~2). A typical Floquet cycle in the middle of the circuit is shown in \cref{fig:1d_energy_dynamics}.
	This LGT model is non-integrable and does not map to Anderson models.
	
	We study the relaxation dynamics of a local perturbation in the energy profile for two distinct disorder-free initial states of $N=38$ qubits placed in a ring on the quantum processor. We choose to study the dynamics of energy perturbations because the total energy is approximately conserved by $\hat U_{\mathrm{LGT}}$ (see SM Fig.~S9). We start with a product state where all the gauge qubits (diamonds in Fig.~\ref{fig:1d_energy_dynamics}B) have their Bloch vectors \(\hat{\mathbf{v}}\) aligned with the direction \(J \hat{\mathbf{z}} + h \hat{\mathbf{x}}\), and apply $\hat \sigma^Y_j$ to the center gauge qubit to flip its Bloch vector to $-\mathbf{\hat v}$ and hence create an energy perturbation. The two initial states are distinguished by the matter qubits (circles in Fig.~\ref{fig:1d_energy_dynamics}B) being prepared in \(\ket{+x}\) or \(\ket{+z}\). The second stage is the application of the entangling unitary, \(\hat U_B\), consisting of two CNOT layers. The first stage can be understood as the specification of an initial state in a convenient basis, where the choice of local symmetry sectors happens to be encoded solely on the matter qubits. The second stage is the transformation of this basis from \cref{eq:dual_hamiltonian} to the one where the Hamiltonian takes the form $\hat H_{\mathrm{LGT}}$ in \cref{eq:lgt-hamiltonian}. The energy profile of these states consists of a center perturbation on a translationally invariant background. Furthermore, the alignment with the local field ensures that we have a finite energy density with respect to a typical disorder realization of the Hamiltonian.
	
	In Fig.~\ref{fig:1d_energy_dynamics}(C-E), we show the results of measuring the local observables that enter the Floquet unitary for the 1d ring, which would comprise the total energy in the continuous-time limit. These observables are the gauge polarization $\langle \hat X_{j,j+1}\rangle$, interaction term per link $\langle \hat \sigma^Z_{j}\,\hat Z_{j,j+1}\,\hat \sigma^Z_{j+1}\rangle$ and the local energy density (per link) $\langle\hat H_{j}\rangle = J \langle  \hat \sigma^Z_{j}\, \hat Z_{j,j+1} \, \hat \sigma^Z_{j+1} \rangle  + h\langle\hat X_{j,j+1} \rangle + \mu \expval{\hat \sigma_j^X + \hat \sigma_{j+1}^X}/2$. We obtain these expectation values by measuring the time-evolved state in the dual basis after the application of $\hat{U}_B$ (See SM, Sec.~2 for further details). The spatiotemporal patterns show distinct behavior that depends on the choice of initial state. For the superposition initial state, the local perturbation (central dark blue stripe in the lower panels) remains relatively localized at the center, whereas a uniform spreading is observed for the single-sector initial state (top panels). Distinct ``ripples'' created by the initial perturbation can be seen in the top panels, indicating relaxation to thermal behavior. However, in the localized dynamics (bottom panels), such ripples are minimized, and we find that the system is largely insensitive to the perturbation and the memory of the initial state is preserved. We emphasize that the only difference between the two experiments is the initial product state, while the circuits applied to the product state are identical.
	
	\begin{figure*}[t!]
		\centering
		\includegraphics[width=0.66\textwidth ]{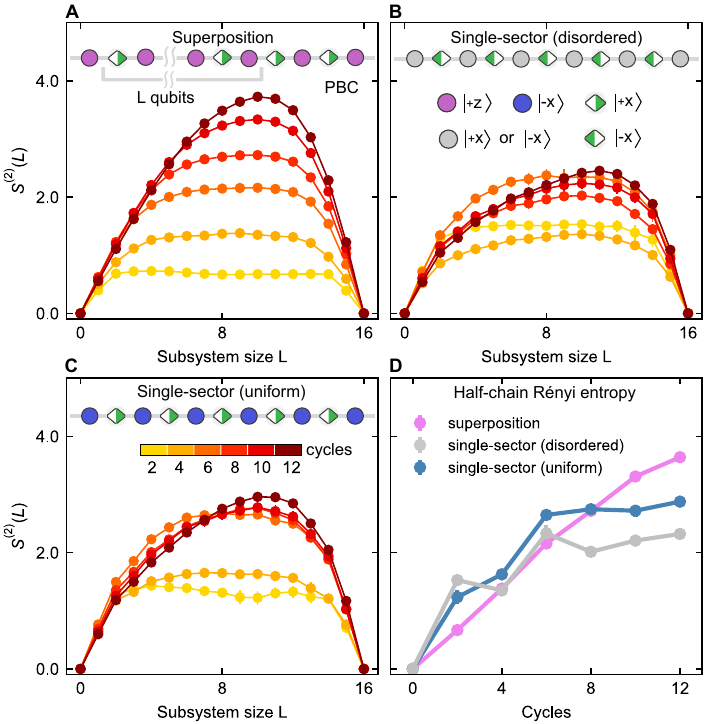}
		\caption{\small {\textbf{Entanglement entropy in a 1d ring.} Second R\'enyi entropy $S^{(2)}(L)$ averaged over all subsystems of size $L$ for a translationally invariant $16$-qubit ring ($N_m=8,\, N_g=8$) measured using randomized single-qubit Clifford measurements for (\textbf{A}) a superposition state with uniform contribution from all the gauge sectors, (\textbf{B}) a single-sector state with disordered gauge configuration ($g_j=\pm 1$), averaged over 100 realizations, and (\textbf{C}) a single-sector state with uniform gauge charges ($g_j=-1$). (\textbf{D})  Half-chain entropy vs.~cycles. We use the fact that the overall entropy of a closed system is $0$ to mitigate error by subtracting the uniform background decoherence\,(see SM). The parameters used were $J=1$, \,$\Delta t =0.4$, \,$h=2.2$, and $\mu=2$.}}
		\label{fig:entropy} 
	\end{figure*} 
	
	There is a complementary way to understand the localization shown in \cref{fig:1d_energy_dynamics}, which emphasizes the roles of an exact and approximate symmetry. In our formulation in terms of ancillas, it was crucial that \(\hat \sigma^X_j\) commutes exactly with the generator of dynamics. After the unitary transformation by \(\hat U_B\), the operator \(\hat \sigma^X_j\) maps to
	\begin{equation}
		\hat G_j= \hat \sigma_j^X \prod_{k \in N(j)} \hat X_{j,k}, \; \mathrm{where}\;\,  [\hat U_{\mathrm{LGT}}, \hat G_j] = 0 \; \forall j;
		\label{eq:gauss_law}
	\end{equation}
	we point out that our choice of Trotterization preserves the symmetry \(\hat G_j\) exactly. This is the structure of a \(\mathbb{Z}_2\) lattice gauge theory (LGT), with integrated Gauss law \(\hat G_j\). In this frame, the eigenvalues $g_j$ can therefore be thought of as realizing a particular disorder configuration by perfectly static \(\mathbb{Z}_2\) background charges, where \(g_j = \pm1\). Since the Floquet unitary commutes with each \(\hat G_j\), it is block diagonal in any basis respecting symmetry, as shown schematically in \cref{fig:block_diagonal}B, where the degree of darkness for each block corresponds to the degree of background disorder in that sector.
	The LGT perspective on disorder-free localization (DFL) has been the subject of many theoretical studies \cite{Smith_PRL_2017_first,Smith_PRL_2017_second,Brenes_PRL_2018,Papaefstathiou_PRB_2020,McClarty_PRB_2020,Russomanno_PhysRevResearch_2020,Karpov-PRL-2021,Heyl_PhysRevResearch_2021,halimeh2021stabilizing,Sarang_PRL_2021,Chakraborty_PRB_2022,Halimeh_PRXQuantum_2022,halimeh2022temperatureinduced,Halimeh2022StarkDFL,Homeier2023realistic,osborne_2D_DFL_2023,sala2024disorder}. The reason to focus on the particular operators shown in \cref{fig:1d_energy_dynamics} is that, in addition to the exact gauge symmetry, the Floquet dynamics approximately conserve the total energy one would obtain in the \(\Delta t \to 0\) limit of $\hat{U}_{LGT}$. We numerically verify this approximate energy conservation for a 24-qubit system (see SM Fig.~S9), which can be thought of as dynamical MBL \cite{dynamical_mbl_abanin_2016}.
	
	Disorder-induced localization depends on dimensionality, tending to be more robust in fewer spatial dimensions; e.g., in interacting systems there is numerical evidence that MBL can become unstable above 1d\,\cite{de2019dynamics}, while for non-interacting fermions, all states are Anderson localized for arbitrarily weak disorder in 1d and 2d\,\cite{50years}. However, several theoretical works on lattice gauge models suggest that DFL is not restricted to 1d but can similarly occur in 2d lattices\,\cite{Chakraborty_PRB_2022,osborne_2D_DFL_2023}. Little is known about the quantitative behavior of DFL in 2d beyond the free-fermion limit\,\cite{Smith_PRB_2018} or for other exactly solvable models\,\cite{Karpov-PRL-2021}. In Fig.~\ref{fig:2d_energy_dynamics}, we show results of measurements in a 2d grid of $N_m=32$\,matter qubits (small circles) and $N_g=49$ gauge qubits\,(diamonds). 
	
	Analogously to the 1d case, state preparation in 2d proceeds in two stages. First, all gauge qubits except one are aligned to \(\hat{\mathbf{v}} \propto J \hat{\mathbf{z}} + h \hat{\mathbf{x}}\), except for the qubit at the top left of the grid (red diamond in \cref{fig:2d_energy_dynamics}A), which points in the \(- \hat{\mathbf{v}}\) direction. The single-sector and superposition states have matter qubits prepared in \(\ket{+x}\) and \(\ket{+z}\), respectively. We then apply the 2d version of the basis change unitary \(\hat U_B\) as shown in \cref{fig:2d_energy_dynamics}B. In Fig.~\ref{fig:2d_energy_dynamics}D and E, we show the spatiotemporal evolution of the local gauge polarization $\langle \hat X_{j,k} \rangle$ measured on the gauge qubits at each link $(j,k)$ for both initial states. The gauge polarization\,(panel D) and the traces of expectation values for all qubits\,(panel E) both suggest that the initial perturbation on the top gauge qubit spreads out within the first few cycles for the single-sector initial state. In contrast, the perturbation stays localized for the superposition initial state. While our data illustrate distinct dynamical behaviors for $\sim 30$ cycles, probing late-time dynamics is beyond what our coherence time scales allow. In SM Fig.~S6, we show both raw and error-mitigated experimental data as well as tensor network simulation results for expectation values of all three terms contributing to the total energy.
	
	Thus far, we have presented observables linear in the total (ancilla and dynamical) system, and commuting with the background charge operators. Such observables correspond directly to disorder-averaged quantities as in \cref{eqn:disorder_averaging}. Quantities non-linear in the density matrix do not have such a direct correspondence. For example, the entropy in a subregion \(A\) does not correspond to the disorder-averaged entropy. Nevertheless, there are non-linear quantities of physical interest because they constrain linear observables; for instance, the mutual information is a linear combination of entropies and bounds connected correlation functions. One advantage of using a quantum simulator is access to such complicated observables. We demonstrate measurements of one of the simplest non-linear quantities, the second R\'enyi entropy, in a subregion \(A\), \(S_A^{(2)} = - \ln \Tr[\hat \rho_A^2]\), for a $N=16$ qubit ring.
	
	In \cref{fig:entropy}, we present our measurements of \(S_A^{(2)}\) as a function of both cycle and subsystem size. As opposed to the local perturbation dynamics experiments, where we used second-order Trotter, we use first-order Trotter here i.e., $\hat{U}_{LGT} =  \hat U_h (\Delta t) \hat U_{\mu}(\Delta t)\hat{U}_J(\Delta t)$. We consider three distinct initial states: (A) superposition over all charges, (B) single-sector state with disordered gauge charges, and (C) single-sector with uniform gauge charges. Keeping all the gauge qubits in an eigenstate of Pauli-X ($\ket{+x}$ for (A,C) and $\ket{-x}$ for (B)), the matter qubits were prepared in either $\ket{+z}$ for (A), randomly chosen $\ket{+x}$ or $\ket{-x}$ and averaged over $100$ disordered configurations for (B), and $\ket{+x}$ for (C). After time-evolving a state with $\hat U_{\mathrm{LGT}}$, we estimate the purity $\Tr[\hat \rho_A^2]$ by measuring each qubit in a random Clifford basis \cite{vanEnk_PRL_2012,Brydges_2019}. Since we are interested in demonstrating generic entropy measurement, each point is an average over all contiguous subregions \(A\) of fixed size (in the LGT frame). 
	
	As mentioned above, the entropy of the superposition state in \cref{fig:entropy}A cannot be directly interpreted as a disorder-averaged entropy. Qualitatively, we expect volume-law contributions from a combination of the background charge disorder and dynamically generated correlations between dynamical and ancilla qubits. By contrast, one expects a slower entropy growth for MBL dynamics. We confirm this expectation in \cref{fig:entropy}B. In fact, the contributions from background disorder dominate; in  \cref{fig:entropy}C we show that even the delocalized uniform-background state has a lower entropy than the localized superposition \cref{fig:entropy}A.
	
	%\vspace{5mm}
	\begin{figure}[t!]
		\centering
		\includegraphics[width=0.34\textwidth ]{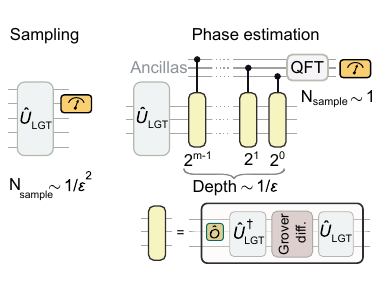}
		\caption{\small {\textbf{Algorithm for efficient sampling.} Quantum processors enable preparing initial states as superposition over all disorder configurations, which can be leveraged to build a phase estimation algorithm (right) to measure the disorder-averaged expectation value of a single-qubit Pauli $\hat O$, offering a polynomial advantage over direct disorder sampling (left). Here, phase estimation is applied to an operator shown as a yellow box, built from Grover's diffusion operator \cite{grover_1996}, using $m=-\log_2 \epsilon $ ancillas.}}
		\label{fig:PhaseEstimation} 
	\end{figure} 
	%\vspace{1mm}
	
	The observation that the entropy of the superposition state is larger compared to the disorder-averaged entropy of a MBL system can be understood in more detail by noting that $\Tr[\hat \rho_A^2]$ has fundamentally different structure from linear observables, as can be seen from 
	\begin{equation}
		\label{eq:entropy_contributions}
		\hat \rho_A \sim  \sum_{\mathbf{g}}  w_{\mathbf{g}} \, \hat \rho^{MBL}_{A,\mathbf{g}} \otimes \ket{\mathbf{g}}\bra{\mathbf{g}} +  \sum_{\mathbf{g_1} \neq \mathbf{g_2}}  \delta \hat \rho_{A, \mathbf{g_1}, \mathbf{g_2}} \otimes \ket{\mathbf{g_1}} \bra{\mathbf{g_2}}.
	\end{equation}
	For linear observables commuting with background charge, the second term does not contribute, so the disorder-averaged expectation value is exactly equal to the expectation value of the superposition state. However, this is no longer the case for non-linear quantities such as entropy, where terms both diagonal (first term) and off-diagonal (second term) in the charge $\mathbf{g}$ give contributions that are not present in disorder averaged $\Tr[\hat \rho_A^2]$.

	The entropy analysis reveals a physically meaningful distinction between initial states formed as superpositions over all disorder configurations and those generated by independently sampling one disorder at a time, a distinction also leveraged in the Grover search algorithm\cite{grover_1996}. We propose a phase estimation algorithm (Fig.\,\ref{fig:PhaseEstimation} and SM, Sec. 6) that can be used to estimate single-qubit Pauli observables to accuracy $\epsilon$ in $O(1)$ experiments, where each is composed of a $\sim 1 / \epsilon$ sequence of controlled applications of $\hat U_{\rm LGT}$. By contrast, naive sampling\,(Fig.~5, left) would require $O(1 / \epsilon^2)$ experiments where time evolution $\hat U$ is applied once per experiment (i.e. disorder realization); in other words, $O(1 / \epsilon^2)$ applications of $\hat U_{\rm LGT}$. This polynomial sampling advantage could boost the search for rare events, which are key to understanding the MBL phase and its stability. \,\cite{Prosen_PRE_2020,Sels2022PRB,Morningstar_PRB_2022}.

	\vspace{8mm}
	\paragraph*{Acknowledgments:} {\footnotesize We acknowledge discussions with Y.~Bahri, U.~E.~Khodaeva, I.~P.~McCulloch, J.~J.~Osborne, A.~Smith, and G.-X.~Su. J.C.H.~acknowledges funding by the Emmy Noether Programme of the German Research Foundation (DFG) under grant no.~HA 8206/1-1, the Max Planck Society, the Deutsche Forschungsgemeinschaft (DFG, German Research Foundation) under Germany’s Excellence Strategy – EXC-2111 – 390814868, and the European Research Council (ERC) under the European Union’s Horizon Europe research and innovation program (Grant Agreement No.~101165667)—ERC Starting Grant QuSiGauge. J.K.~acknowledges support from the Imperial-Technical University of Munich flagship partnership and financial support by the Deutsche Forschungsgemeinschaft (DFG, German Research Foundation) via TRR 360 (Project-ID No. 492547816). The research is part of the Munich Quantum Valley, which is supported by the Bavarian state government with funds from the Hightech Agenda Bayern Plus. D.K.~acknowledges support from Labex MME-DII grant ANR11-LBX-0023, and funding under The Paris Seine Initiative Emergence programme 2019. R.M.~acknowledges support  by the Deutsche Forschungsgemeinschaft under the grant cluster of excellence ct.qmat (EXC 2147, project-id 390858490).}
	
	\vspace{2mm}
	\onecolumngrid
	\vspace{1em}
	\begin{flushleft}
    {\hypertarget{authorlist}{${}^\dagger$}  \small Google Quantum AI and Collaborators}

    \bigskip

    \renewcommand{\author}[2]{#1\textsuperscript{\textrm{\scriptsize #2}}}
    \renewcommand{\affiliation}[2]{\textsuperscript{\textrm{\scriptsize #1} #2} \\}
    \newcommand{\corrauthora}[2]{#1$^{\textrm{\scriptsize #2}, \hyperlink{corra}{\ddagger}}$}
    \newcommand{\corrauthorb}[2]{#1$^{\textrm{\scriptsize #2}, \hyperlink{corrb}{\mathsection}}$}
\newcommand{\xGoogle}{\affiliation{1}{Google Research, Mountain View, CA, USA}}
\newcommand{\xCornell}{\affiliation{2}{Department of Physics, Cornell University, Ithaca, NY, USA}}
\newcommand{\xLASSP}{\affiliation{3}{Laboratory of Solid State and Atomic Physics, Cornell University, Ithaca, NY, USA}}
\newcommand{\xPrincetonEE}{\affiliation{4}{Department of Electrical and Computer Engineering,
Princeton University, Princeton, NJ, USA}}
\newcommand{\xPrinceton}{\affiliation{5}{Department of Physics, Princeton University, Princeton, NJ, USA}}
\newcommand{\xIMM}{\affiliation{6}{Institute of Molecules and Materials, Radboud University, Nijmegen, Netherlands}}
\newcommand{\xUMass}{\affiliation{7}{Department of Electrical and Computer Engineering, University of Massachusetts, Amherst, MA}}
\newcommand{\xYale}{\affiliation{8}{Applied Physics Department, Yale University, New Haven, CT}}
\newcommand{\xUCONN}{\affiliation{9}{Department of Physics, University of Connecticut, Storrs, CT}}
\newcommand{\xAuburnECE}{\affiliation{10}{Department of Electrical and Computer Engineering, Auburn University, Auburn, AL}}
\newcommand{\xUTS}{\affiliation{11}{QSI, Faculty of Engineering \& Information Technology, University of Technology Sydney, NSW, Australia}}
\newcommand{\xUCRECE}{\affiliation{12}{Department of Electrical and Computer Engineering, University of California, Riverside, CA}}
\newcommand{\xHarvard}{\affiliation{13}{Department of Chemistry and Chemical Biology, Harvard University}}
\newcommand{\xUCRPA}{\affiliation{14}{Department of Physics and Astronomy, University of California, Riverside, CA}}
\newcommand{\xParis}{\affiliation{15}{Laboratory of Theoretical Physics and Modelling, CY Cergy Paris Universite, UMR CNRS 8089, Pontoise Cergy-Pontoise Cedex, France}}
\newcommand{\xTUM}{\affiliation{16}{Technical University of Munich, TUM School of Natural Sciences, Physics Department, TQM, Garching, Germany}}
\newcommand{\xMCQST}{\affiliation{17}{Munich Center for Quantum Science and Technology (MCQST), Munich, Germany}}
\newcommand{\xImperial}{\affiliation{18}{Blackett Laboratory, Imperial College London, London SW7 2AZ, United Kingdom}}
\newcommand{\xMPQ}{\affiliation{19}{Max Planck Institute of Quantum Optics, Garching, Germany}}
\newcommand{\xLMU}{\affiliation{20}{Department of Physics and Arnold Sommerfeld Center for Theoretical Physics, Ludwig Maximilian University of Munich, Munich, Germany}}
\newcommand{\xBerlin}{\affiliation{21}{Dahlem Center for Complex Quantum Systems, Free University of Berlin, Berlin, Germany}}
\newcommand{\xMPIPCS}{\affiliation{22}{Max Planck Institute for the Physics of Complex Systems, Dresden, Germany}}

\begin{footnotesize}

\newcommand{\Google}{1}
\newcommand{\Cornell}{2}
\newcommand{\LASSP}{3}
\newcommand{\PrincetonEE}{4}
\newcommand{\Princeton}{5}
\newcommand{\IMM}{6}
\newcommand{\UMass}{7}
\newcommand{\Yale}{8}
\newcommand{\UCONN}{9}
\newcommand{\AuburnECE}{10}
\newcommand{\UTS}{11}
\newcommand{\UCRECE}{12}
\newcommand{\Harvard}{13}
\newcommand{\UCRPA}{14}
\newcommand{\Paris}{15}
\newcommand{\TUM}{16}
\newcommand{\MCQST}{17}
\newcommand{\Imperial}{18}
\newcommand{\MPQ}{19}
\newcommand{\LMU}{20}
\newcommand{\Berlin}{21}
\newcommand{\MPIPCS}{22}

\corrauthora{G.~Gyawali}{\Google, \!\Cornell, \!\LASSP},
\corrauthora{S.~Kumar}{\Google, \!\PrincetonEE},
\corrauthora{Y.~D.~Lensky}{\Google},
\corrauthora{E.~Rosenberg}{\Google},
\author{A.~Szasz}{\Google},
\author{T.~Cochran}{\Google, \!\Princeton},
\author{R.~Chen}{\Google},
\author{A.~H.~Karamlou}{\Google},
\author{K.~Kechedzhi}{\Google},
\author{J.~Berndtsson}{\Google},
\author{T.~Westerhout}{\Google, \!\IMM},
\author{A.~Asfaw}{\Google},
\author{D.~Abanin}{\Google, \Princeton},
\author{R.~Acharya}{\Google},
% ...
\corrauthorb{I. Aleiner}{\Google},
\corrauthorb{R.~Moessner}{\MPIPCS},
\corrauthorb{P.~Roushan}{\Google}

\bigskip

\xGoogle
\xCornell
\xLASSP
\xPrincetonEE
\xPrinceton
\xIMM
\xUMass
\xYale
\xUCONN
\xAuburnECE
\xUTS
\xUCRECE
\xHarvard
\xUCRPA
\xParis
\xTUM
\xMCQST
\xImperial
\xMPQ
\xLMU
\xBerlin
\xMPIPCS

\vspace{2mm}
{\hypertarget{corra}{${}^\ddagger$} These authors contributed equally to this work.}\\

{\hypertarget{corrb}{${}^\mathsection$} Corresponding author: igoraleiner@google.com}\\

{\hypertarget{corrb}{${}^\mathsection$} Corresponding author: moessner@pks.mpg.de}\\

{\hypertarget{corrb}{${}^\mathsection$} Corresponding author: pedramr@google.com}

\end{footnotesize}
\end{flushleft}

	%\onecolumngrid
	\twocolumngrid
	\bibliography{References.bib}

%apsrev4-2.bst 2019-01-14 (MD) hand-edited version of apsrev4-1.bst
%Control: key (0)
%Control: author (8) initials jnrlst
%Control: editor formatted (1) identically to author
%Control: production of article title (0) allowed
%Control: page (0) single
%Control: year (1) truncated
%Control: production of eprint (0) enabled
\begin{thebibliography}{89}%
\makeatletter
\providecommand \@ifxundefined [1]{%
 \@ifx{#1\undefined}
}%
\providecommand \@ifnum [1]{%
 \ifnum #1\expandafter \@firstoftwo
 \else \expandafter \@secondoftwo
 \fi
}%
\providecommand \@ifx [1]{%
 \ifx #1\expandafter \@firstoftwo
 \else \expandafter \@secondoftwo
 \fi
}%
\providecommand \natexlab [1]{#1}%
\providecommand \enquote  [1]{``#1''}%
\providecommand \bibnamefont  [1]{#1}%
\providecommand \bibfnamefont [1]{#1}%
\providecommand \citenamefont [1]{#1}%
\providecommand \href@noop [0]{\@secondoftwo}%
\providecommand \href [0]{\begingroup \@sanitize@url \@href}%
\providecommand \@href[1]{\@@startlink{#1}\@@href}%
\providecommand \@@href[1]{\endgroup#1\@@endlink}%
\providecommand \@sanitize@url [0]{\catcode `\\12\catcode `\$12\catcode
  `\&12\catcode `\#12\catcode `\^12\catcode `\_12\catcode `\%12\relax}%
\providecommand \@@startlink[1]{}%
\providecommand \@@endlink[0]{}%
\providecommand \url  [0]{\begingroup\@sanitize@url \@url }%
\providecommand \@url [1]{\endgroup\@href {#1}{\urlprefix }}%
\providecommand \urlprefix  [0]{URL }%
\providecommand \Eprint [0]{\href }%
\providecommand \doibase [0]{https://doi.org/}%
\providecommand \selectlanguage [0]{\@gobble}%
\providecommand \bibinfo  [0]{\@secondoftwo}%
\providecommand \bibfield  [0]{\@secondoftwo}%
\providecommand \translation [1]{[#1]}%
\providecommand \BibitemOpen [0]{}%
\providecommand \bibitemStop [0]{}%
\providecommand \bibitemNoStop [0]{.\EOS\space}%
\providecommand \EOS [0]{\spacefactor3000\relax}%
\providecommand \BibitemShut  [1]{\csname bibitem#1\endcsname}%
\let\auto@bib@innerbib\@empty
%</preamble>
\bibitem [{\citenamefont {Mott}(1968)}]{Mott1968}%
  \BibitemOpen
  \bibfield  {author} {\bibinfo {author} {\bibfnamefont {N.~F.}\ \bibnamefont
  {Mott}},\ }\bibfield  {title} {\bibinfo {title} {Metal-insulator
  transition},\ }\href {https://doi.org/10.1103/RevModPhys.40.677} {\bibfield
  {journal} {\bibinfo  {journal} {Rev. Mod. Phys.}\ }\textbf {\bibinfo {volume}
  {40}},\ \bibinfo {pages} {677} (\bibinfo {year} {1968})}\BibitemShut
  {NoStop}%
\bibitem [{\citenamefont {Ashcroft}\ and\ \citenamefont
  {Mermin}(1976)}]{AMermin_text}%
  \BibitemOpen
  \bibfield  {author} {\bibinfo {author} {\bibfnamefont {N.~W.}\ \bibnamefont
  {Ashcroft}}\ and\ \bibinfo {author} {\bibfnamefont {N.~D.}\ \bibnamefont
  {Mermin}},\ }\href@noop {} {\emph {\bibinfo {title} {Solid State Physics}}}\
  (\bibinfo  {publisher} {Cengage Learning},\ \bibinfo {year}
  {1976})\BibitemShut {NoStop}%
\bibitem [{\citenamefont {Dobrosavljevic}\ \emph {et~al.}(2012)\citenamefont
  {Dobrosavljevic}, \citenamefont {Trivedi},\ and\ \citenamefont
  {Valles}}]{Dobrosavljevic}%
  \BibitemOpen
  \bibfield  {author} {\bibinfo {author} {\bibfnamefont {V.}~\bibnamefont
  {Dobrosavljevic}}, \bibinfo {author} {\bibfnamefont {N.}~\bibnamefont
  {Trivedi}},\ and\ \bibinfo {author} {\bibfnamefont {J.}~\bibnamefont
  {Valles}},\ }\href@noop {} {\emph {\bibinfo {title} {Conductor-Insulator
  Quantum Phase Transitions}}}\ (\bibinfo  {publisher} {Oxford University
  Press},\ \bibinfo {year} {2012})\BibitemShut {NoStop}%
\bibitem [{\citenamefont {Bahri}\ \emph {et~al.}(2015)\citenamefont {Bahri}
  \emph {et~al.}}]{bahri2015localization}%
  \BibitemOpen
  \bibfield  {author} {\bibinfo {author} {\bibfnamefont {Y.}~\bibnamefont
  {Bahri}} \emph {et~al.},\ }\bibfield  {title} {\bibinfo {title} {Localization
  and topology protected quantum coherence at the edge of hot matter},\ }\href
  {https://doi.org/10.1038/ncomms8341} {\bibfield  {journal} {\bibinfo
  {journal} {Nature Communications}\ }\textbf {\bibinfo {volume} {6}},\
  \bibinfo {pages} {7341} (\bibinfo {year} {2015})}\BibitemShut {NoStop}%
\bibitem [{\citenamefont {Kim}\ and\ \citenamefont
  {Haah}(2016)}]{Kim_PRL_2016}%
  \BibitemOpen
  \bibfield  {author} {\bibinfo {author} {\bibfnamefont {I.~H.}\ \bibnamefont
  {Kim}}\ and\ \bibinfo {author} {\bibfnamefont {J.}~\bibnamefont {Haah}},\
  }\bibfield  {title} {\bibinfo {title} {Localization from superselection rules
  in translationally invariant systems},\ }\href
  {https://doi.org/https://doi.org/10.1103/PhysRevLett.116.027202} {\bibfield
  {journal} {\bibinfo  {journal} {Phys. Rev. Lett.}\ }\textbf {\bibinfo
  {volume} {116}},\ \bibinfo {pages} {027202} (\bibinfo {year}
  {2016})}\BibitemShut {NoStop}%
\bibitem [{\citenamefont {Schulz}\ \emph {et~al.}(2019)\citenamefont {Schulz}
  \emph {et~al.}}]{SchulzPRL2019}%
  \BibitemOpen
  \bibfield  {author} {\bibinfo {author} {\bibfnamefont {M.}~\bibnamefont
  {Schulz}} \emph {et~al.},\ }\bibfield  {title} {\bibinfo {title} {Stark
  many-body localization},\ }\href
  {https://doi.org/https://doi.org/10.1103/PhysRevLett.122.040606} {\bibfield
  {journal} {\bibinfo  {journal} {Phys. Rev. Lett.}\ }\textbf {\bibinfo
  {volume} {122}},\ \bibinfo {pages} {040606} (\bibinfo {year}
  {2019})}\BibitemShut {NoStop}%
\bibitem [{\citenamefont {van Nieuwenburg}\ \emph {et~al.}(2019)\citenamefont
  {van Nieuwenburg}, \citenamefont {Baum},\ and\ \citenamefont
  {Refael}}]{van2019bloch}%
  \BibitemOpen
  \bibfield  {author} {\bibinfo {author} {\bibfnamefont {E.}~\bibnamefont {van
  Nieuwenburg}}, \bibinfo {author} {\bibfnamefont {Y.}~\bibnamefont {Baum}},\
  and\ \bibinfo {author} {\bibfnamefont {G.}~\bibnamefont {Refael}},\
  }\bibfield  {title} {\bibinfo {title} {From {B}loch oscillations to many-body
  localization in clean interacting systems},\ }\href
  {https://doi.org/https://doi.org/10.1073/pnas.1819316116} {\bibfield
  {journal} {\bibinfo  {journal} {Proceedings of the National Academy of
  Sciences}\ }\textbf {\bibinfo {volume} {116}},\ \bibinfo {pages} {9269}
  (\bibinfo {year} {2019})}\BibitemShut {NoStop}%
\bibitem [{\citenamefont {Ribeiro}\ \emph {et~al.}(2020)\citenamefont {Ribeiro}
  \emph {et~al.}}]{ribeiro2020many}%
  \BibitemOpen
  \bibfield  {author} {\bibinfo {author} {\bibfnamefont {P.}~\bibnamefont
  {Ribeiro}} \emph {et~al.},\ }\bibfield  {title} {\bibinfo {title} {Many-body
  quantum dynamics of initially trapped systems due to a {S}tark potential:
  thermalization versus {B}loch oscillations},\ }\href
  {https://doi.org/https://doi.org/10.1103/PhysRevLett.124.110603} {\bibfield
  {journal} {\bibinfo  {journal} {Physical Review Letters}\ }\textbf {\bibinfo
  {volume} {124}},\ \bibinfo {pages} {110603} (\bibinfo {year}
  {2020})}\BibitemShut {NoStop}%
\bibitem [{\citenamefont {Scherg}\ \emph {et~al.}(2021)\citenamefont {Scherg}
  \emph {et~al.}}]{scherg2021observing}%
  \BibitemOpen
  \bibfield  {author} {\bibinfo {author} {\bibfnamefont {S.}~\bibnamefont
  {Scherg}} \emph {et~al.},\ }\bibfield  {title} {\bibinfo {title} {Observing
  non-ergodicity due to kinetic constraints in tilted {F}ermi-{H}ubbard
  chains},\ }\href {https://www.nature.com/articles/s41467-021-24726-0}
  {\bibfield  {journal} {\bibinfo  {journal} {Nature Communications}\ }\textbf
  {\bibinfo {volume} {12}},\ \bibinfo {pages} {4490} (\bibinfo {year}
  {2021})}\BibitemShut {NoStop}%
\bibitem [{\citenamefont {Morong}\ \emph {et~al.}(2021)\citenamefont {Morong}
  \emph {et~al.}}]{morong2021observation}%
  \BibitemOpen
  \bibfield  {author} {\bibinfo {author} {\bibfnamefont {W.}~\bibnamefont
  {Morong}} \emph {et~al.},\ }\bibfield  {title} {\bibinfo {title} {Observation
  of {S}tark many-body localization without disorder},\ }\href
  {https://www.nature.com/articles/s41586-021-03988-0} {\bibfield  {journal}
  {\bibinfo  {journal} {Nature}\ }\textbf {\bibinfo {volume} {599}},\ \bibinfo
  {pages} {393} (\bibinfo {year} {2021})}\BibitemShut {NoStop}%
\bibitem [{\citenamefont {Zisling}(2022)}]{Guy_PRB_2022}%
  \BibitemOpen
  \bibfield  {author} {\bibinfo {author} {\bibfnamefont {G.~a.}\ \bibnamefont
  {Zisling}},\ }\bibfield  {title} {\bibinfo {title} {Transport in {S}tark
  many-body localized systems},\ }\href
  {https://doi.org/10.1103/PhysRevB.105.L140201} {\bibfield  {journal}
  {\bibinfo  {journal} {Phys. Rev. B}\ }\textbf {\bibinfo {volume} {105}},\
  \bibinfo {pages} {L140201} (\bibinfo {year} {2022})}\BibitemShut {NoStop}%
\bibitem [{\citenamefont {Adler}\ \emph {et~al.}(2024)\citenamefont {Adler},
  \citenamefont {Wei}, \citenamefont {Will}, \citenamefont {Srakaew},
  \citenamefont {Agrawal}, \citenamefont {Weckesser}, \citenamefont {Moessner},
  \citenamefont {Pollmann}, \citenamefont {Bloch},\ and\ \citenamefont
  {Zeiher}}]{adler2024observation}%
  \BibitemOpen
  \bibfield  {author} {\bibinfo {author} {\bibfnamefont {D.}~\bibnamefont
  {Adler}}, \bibinfo {author} {\bibfnamefont {D.}~\bibnamefont {Wei}}, \bibinfo
  {author} {\bibfnamefont {M.}~\bibnamefont {Will}}, \bibinfo {author}
  {\bibfnamefont {K.}~\bibnamefont {Srakaew}}, \bibinfo {author} {\bibfnamefont
  {S.}~\bibnamefont {Agrawal}}, \bibinfo {author} {\bibfnamefont
  {P.}~\bibnamefont {Weckesser}}, \bibinfo {author} {\bibfnamefont
  {R.}~\bibnamefont {Moessner}}, \bibinfo {author} {\bibfnamefont
  {F.}~\bibnamefont {Pollmann}}, \bibinfo {author} {\bibfnamefont
  {I.}~\bibnamefont {Bloch}},\ and\ \bibinfo {author} {\bibfnamefont
  {J.}~\bibnamefont {Zeiher}},\ }\bibfield  {title} {\bibinfo {title}
  {Observation of {H}ilbert space fragmentation and fractonic excitations in
  {2D}},\ }\href {https://doi.org/10.1038/s41586-024-08188-0} {\bibfield
  {journal} {\bibinfo  {journal} {Nature}\ }\textbf {\bibinfo {volume} {636}},\
  \bibinfo {pages} {80} (\bibinfo {year} {2024})}\BibitemShut {NoStop}%
\bibitem [{\citenamefont {Anderson}(1958)}]{Anderson1958}%
  \BibitemOpen
  \bibfield  {author} {\bibinfo {author} {\bibfnamefont {P.~W.}\ \bibnamefont
  {Anderson}},\ }\bibfield  {title} {\bibinfo {title} {Absence of diffusion in
  certain random lattices},\ }\href {https://doi.org/10.1103/PhysRev.109.1492}
  {\bibfield  {journal} {\bibinfo  {journal} {Phys. Rev.}\ }\textbf {\bibinfo
  {volume} {109}},\ \bibinfo {pages} {1492} (\bibinfo {year}
  {1958})}\BibitemShut {NoStop}%
\bibitem [{\citenamefont {Wiersma}\ \emph {et~al.}(1997)\citenamefont {Wiersma}
  \emph {et~al.}}]{Wiersma1997nature}%
  \BibitemOpen
  \bibfield  {author} {\bibinfo {author} {\bibfnamefont {D.~S.}\ \bibnamefont
  {Wiersma}} \emph {et~al.},\ }\bibfield  {title} {\bibinfo {title}
  {Localization of light in a disordered medium},\ }\href
  {https://www.nature.com/articles/37757} {\bibfield  {journal} {\bibinfo
  {journal} {Nature}\ }\textbf {\bibinfo {volume} {390}},\ \bibinfo {pages}
  {671–673} (\bibinfo {year} {1997})}\BibitemShut {NoStop}%
\bibitem [{\citenamefont {Schwartz}\ \emph {et~al.}(2007)\citenamefont
  {Schwartz} \emph {et~al.}}]{Schwartz2007}%
  \BibitemOpen
  \bibfield  {author} {\bibinfo {author} {\bibfnamefont {T.}~\bibnamefont
  {Schwartz}} \emph {et~al.},\ }\bibfield  {title} {\bibinfo {title} {Transport
  and {A}nderson localization in disordered two-dimensional photonic
  lattices},\ }\href {https://doi.org/10.1038/nature05623} {\bibfield
  {journal} {\bibinfo  {journal} {Nature}\ }\textbf {\bibinfo {volume} {446}},\
  \bibinfo {pages} {52} (\bibinfo {year} {2007})}\BibitemShut {NoStop}%
\bibitem [{\citenamefont {Billy}\ \emph {et~al.}(2008)\citenamefont {Billy}
  \emph {et~al.}}]{Aspect2008nature}%
  \BibitemOpen
  \bibfield  {author} {\bibinfo {author} {\bibfnamefont {J.}~\bibnamefont
  {Billy}} \emph {et~al.},\ }\bibfield  {title} {\bibinfo {title} {Direct
  observation of {A}nderson localization of matter waves in a controlled
  disorder},\ }\href {https://doi.org/10.1038/nature07000} {\bibfield
  {journal} {\bibinfo  {journal} {Nature}\ }\textbf {\bibinfo {volume} {453}},\
  \bibinfo {pages} {891} (\bibinfo {year} {2008})}\BibitemShut {NoStop}%
\bibitem [{\citenamefont {Kondov}\ \emph {et~al.}(2011)\citenamefont {Kondov}
  \emph {et~al.}}]{kondov2011three}%
  \BibitemOpen
  \bibfield  {author} {\bibinfo {author} {\bibfnamefont {S.}~\bibnamefont
  {Kondov}} \emph {et~al.},\ }\bibfield  {title} {\bibinfo {title}
  {Three-dimensional {A}nderson localization of ultracold matter},\ }\href
  {https://doi.org/https://www.science.org/doi/10.1126/science.1209019}
  {\bibfield  {journal} {\bibinfo  {journal} {Science}\ }\textbf {\bibinfo
  {volume} {334}},\ \bibinfo {pages} {66} (\bibinfo {year} {2011})}\BibitemShut
  {NoStop}%
\bibitem [{\citenamefont {Abrahams}(2010)}]{50years}%
  \BibitemOpen
  \bibfield  {author} {\bibinfo {author} {\bibfnamefont {E.}~\bibnamefont
  {Abrahams}},\ }\href {https://doi.org/https://doi.org/10.1142/7663} {\emph
  {\bibinfo {title} {50 Years Of {A}nderson Localization}}}\ (\bibinfo
  {publisher} {World Scientific Pub Co Inc},\ \bibinfo {year}
  {2010})\BibitemShut {NoStop}%
\bibitem [{\citenamefont {Karamlou}\ \emph {et~al.}(2022)\citenamefont
  {Karamlou} \emph {et~al.}}]{Karamlou2022}%
  \BibitemOpen
  \bibfield  {author} {\bibinfo {author} {\bibfnamefont {A.~H.}\ \bibnamefont
  {Karamlou}} \emph {et~al.},\ }\bibfield  {title} {\bibinfo {title} {Quantum
  transport and localization in 1d and 2d tight-binding lattices},\ }\href
  {https://doi.org/10.1038/s41534-022-00528-0} {\bibfield  {journal} {\bibinfo
  {journal} {npj Quantum Information}\ }\textbf {\bibinfo {volume} {8}},\
  \bibinfo {pages} {35} (\bibinfo {year} {2022})}\BibitemShut {NoStop}%
\bibitem [{\citenamefont {Basko}\ \emph {et~al.}(2006)\citenamefont {Basko},
  \citenamefont {Aleiner},\ and\ \citenamefont {Altshuler}}]{basko2006metal}%
  \BibitemOpen
  \bibfield  {author} {\bibinfo {author} {\bibfnamefont {D.~M.}\ \bibnamefont
  {Basko}}, \bibinfo {author} {\bibfnamefont {I.~L.}\ \bibnamefont {Aleiner}},\
  and\ \bibinfo {author} {\bibfnamefont {B.~L.}\ \bibnamefont {Altshuler}},\
  }\bibfield  {title} {\bibinfo {title} {Metal--insulator transition in a
  weakly interacting many-electron system with localized single-particle
  states},\ }\href {https://doi.org/https://doi.org/10.1016/j.aop.2005.11.014}
  {\bibfield  {journal} {\bibinfo  {journal} {Annals of Physics}\ }\textbf
  {\bibinfo {volume} {321}},\ \bibinfo {pages} {1126} (\bibinfo {year}
  {2006})}\BibitemShut {NoStop}%
\bibitem [{\citenamefont {Serbyn}\ \emph {et~al.}(2013)\citenamefont {Serbyn},
  \citenamefont {Papi\ifmmode~\acute{c}\else \'{c}\fi{}},\ and\ \citenamefont
  {Abanin}}]{Serbyn_PRL_2013}%
  \BibitemOpen
  \bibfield  {author} {\bibinfo {author} {\bibfnamefont {M.}~\bibnamefont
  {Serbyn}}, \bibinfo {author} {\bibfnamefont {Z.}~\bibnamefont
  {Papi\ifmmode~\acute{c}\else \'{c}\fi{}}},\ and\ \bibinfo {author}
  {\bibfnamefont {D.~A.}\ \bibnamefont {Abanin}},\ }\bibfield  {title}
  {\bibinfo {title} {Local conservation laws and the structure of the many-body
  localized states},\ }\href {https://doi.org/10.1103/PhysRevLett.111.127201}
  {\bibfield  {journal} {\bibinfo  {journal} {Phys. Rev. Lett.}\ }\textbf
  {\bibinfo {volume} {111}},\ \bibinfo {pages} {127201} (\bibinfo {year}
  {2013})}\BibitemShut {NoStop}%
\bibitem [{\citenamefont {Huse}\ \emph {et~al.}(2014)\citenamefont {Huse},
  \citenamefont {Nandkishore},\ and\ \citenamefont
  {Oganesyan}}]{Huse_PRB_2014}%
  \BibitemOpen
  \bibfield  {author} {\bibinfo {author} {\bibfnamefont {D.}~\bibnamefont
  {Huse}}, \bibinfo {author} {\bibfnamefont {R.}~\bibnamefont {Nandkishore}},\
  and\ \bibinfo {author} {\bibfnamefont {V.}~\bibnamefont {Oganesyan}},\
  }\bibfield  {title} {\bibinfo {title} {Phenomenology of fully
  many-body-localized systems},\ }\href
  {https://doi.org/10.1103/PhysRevB.90.174202} {\bibfield  {journal} {\bibinfo
  {journal} {Phys. Rev. B}\ }\textbf {\bibinfo {volume} {90}},\ \bibinfo
  {pages} {174202} (\bibinfo {year} {2014})}\BibitemShut {NoStop}%
\bibitem [{\citenamefont {Nandkishore}\ and\ \citenamefont
  {Huse}(2015)}]{nandkishore2015many}%
  \BibitemOpen
  \bibfield  {author} {\bibinfo {author} {\bibfnamefont {R.}~\bibnamefont
  {Nandkishore}}\ and\ \bibinfo {author} {\bibfnamefont {D.~A.}\ \bibnamefont
  {Huse}},\ }\bibfield  {title} {\bibinfo {title} {Many-body localization and
  thermalization in quantum statistical mechanics},\ }\href
  {https://doi.org/https://doi.org/10.1146/annurev-conmatphys-031214-014726}
  {\bibfield  {journal} {\bibinfo  {journal} {Annu. Rev. Condens. Matter
  Phys.}\ }\textbf {\bibinfo {volume} {6}},\ \bibinfo {pages} {15} (\bibinfo
  {year} {2015})}\BibitemShut {NoStop}%
\bibitem [{\citenamefont {Abanin}\ and\ \citenamefont
  {Papi{\'c}}(2017)}]{abanin2017recent}%
  \BibitemOpen
  \bibfield  {author} {\bibinfo {author} {\bibfnamefont {D.~A.}\ \bibnamefont
  {Abanin}}\ and\ \bibinfo {author} {\bibfnamefont {Z.}~\bibnamefont
  {Papi{\'c}}},\ }\bibfield  {title} {\bibinfo {title} {Recent progress in
  many-body localization},\ }\href
  {https://doi.org/https://doi.org/10.1002/andp.201700169} {\bibfield
  {journal} {\bibinfo  {journal} {Annalen der Physik}\ }\textbf {\bibinfo
  {volume} {529}},\ \bibinfo {pages} {1700169} (\bibinfo {year}
  {2017})}\BibitemShut {NoStop}%
\bibitem [{\citenamefont {Schreiber}\ \emph {et~al.}(2015)\citenamefont
  {Schreiber} \emph {et~al.}}]{schreiber2015observation}%
  \BibitemOpen
  \bibfield  {author} {\bibinfo {author} {\bibfnamefont {M.}~\bibnamefont
  {Schreiber}} \emph {et~al.},\ }\bibfield  {title} {\bibinfo {title}
  {Observation of many-body localization of interacting fermions in a
  quasirandom optical lattice},\ }\href
  {https://doi.org/https://doi.org/10.1126/science.aaa7432} {\bibfield
  {journal} {\bibinfo  {journal} {Science}\ }\textbf {\bibinfo {volume}
  {349}},\ \bibinfo {pages} {842} (\bibinfo {year} {2015})}\BibitemShut
  {NoStop}%
\bibitem [{\citenamefont {Grover}(2014)}]{Grover_2014}%
  \BibitemOpen
  \bibfield  {author} {\bibinfo {author} {\bibfnamefont {M.}~\bibnamefont
  {Grover}, \bibfnamefont {T.~Fisher}},\ }\bibfield  {title} {\bibinfo {title}
  {Quantum disentangled liquids},\ }\href
  {https://doi.org/10.1088/1742-5468/2014/10/P10010} {\bibfield  {journal}
  {\bibinfo  {journal} {Journal of Statistical Mechanics: Theory and
  Experiment}\ }\textbf {\bibinfo {volume} {2014}},\ \bibinfo {pages} {P10010}
  (\bibinfo {year} {2014})}\BibitemShut {NoStop}%
\bibitem [{\citenamefont {Schiulaz}\ \emph {et~al.}(2015)\citenamefont
  {Schiulaz} \emph {et~al.}}]{Schiulaz_PRB_2015}%
  \BibitemOpen
  \bibfield  {author} {\bibinfo {author} {\bibfnamefont {M.}~\bibnamefont
  {Schiulaz}} \emph {et~al.},\ }\bibfield  {title} {\bibinfo {title} {Dynamics
  in many-body localized quantum systems without disorder},\ }\href
  {https://doi.org/10.1103/PhysRevB.91.184202} {\bibfield  {journal} {\bibinfo
  {journal} {Phys. Rev. B}\ }\textbf {\bibinfo {volume} {91}},\ \bibinfo
  {pages} {184202} (\bibinfo {year} {2015})}\BibitemShut {NoStop}%
\bibitem [{\citenamefont {Yao}\ \emph {et~al.}(2016)\citenamefont {Yao} \emph
  {et~al.}}]{yao2016quasi}%
  \BibitemOpen
  \bibfield  {author} {\bibinfo {author} {\bibfnamefont {N.}~\bibnamefont
  {Yao}} \emph {et~al.},\ }\bibfield  {title} {\bibinfo {title}
  {Quasi-many-body localization in translation-invariant systems},\ }\href
  {https://doi.org/https://doi.org/10.1103/PhysRevLett.117.240601} {\bibfield
  {journal} {\bibinfo  {journal} {Physical review letters}\ }\textbf {\bibinfo
  {volume} {117}},\ \bibinfo {pages} {240601} (\bibinfo {year}
  {2016})}\BibitemShut {NoStop}%
\bibitem [{\citenamefont {Hickey}\ \emph {et~al.}(2016)\citenamefont {Hickey}
  \emph {et~al.}}]{Hickey_2016}%
  \BibitemOpen
  \bibfield  {author} {\bibinfo {author} {\bibfnamefont {J.}~\bibnamefont
  {Hickey}} \emph {et~al.},\ }\bibfield  {title} {\bibinfo {title} {Signatures
  of many-body localisation in a system without disorder and the relation to a
  glass transition},\ }\href {https://doi.org/10.1088/1742-5468/2016/05/054047}
  {\bibfield  {journal} {\bibinfo  {journal} {Journal of Statistical Mechanics:
  Theory and Experiment}\ }\textbf {\bibinfo {volume} {2016}},\ \bibinfo
  {pages} {054047} (\bibinfo {year} {2016})}\BibitemShut {NoStop}%
\bibitem [{\citenamefont {Mondaini}\ and\ \citenamefont
  {Cai}(2017)}]{Mondaini_PRB_2017}%
  \BibitemOpen
  \bibfield  {author} {\bibinfo {author} {\bibfnamefont {R.}~\bibnamefont
  {Mondaini}}\ and\ \bibinfo {author} {\bibfnamefont {Z.}~\bibnamefont {Cai}},\
  }\bibfield  {title} {\bibinfo {title} {Many-body self-localization in a
  translation-invariant {H}amiltonian},\ }\href
  {https://doi.org/10.1103/PhysRevB.96.035153} {\bibfield  {journal} {\bibinfo
  {journal} {Phys. Rev. B}\ }\textbf {\bibinfo {volume} {96}},\ \bibinfo
  {pages} {035153} (\bibinfo {year} {2017})}\BibitemShut {NoStop}%
\bibitem [{\citenamefont {Mazza}\ \emph {et~al.}(2019)\citenamefont {Mazza}
  \emph {et~al.}}]{Mazza_PRB_2019}%
  \BibitemOpen
  \bibfield  {author} {\bibinfo {author} {\bibfnamefont {P.~P.}\ \bibnamefont
  {Mazza}} \emph {et~al.},\ }\bibfield  {title} {\bibinfo {title} {Suppression
  of transport in nondisordered quantum spin chains due to confined
  excitations},\ }\href {https://doi.org/10.1103/PhysRevB.99.180302} {\bibfield
   {journal} {\bibinfo  {journal} {Phys. Rev. B}\ }\textbf {\bibinfo {volume}
  {99}},\ \bibinfo {pages} {180302} (\bibinfo {year} {2019})}\BibitemShut
  {NoStop}%
\bibitem [{\citenamefont {Bernien}\ \emph {et~al.}(2017)\citenamefont {Bernien}
  \emph {et~al.}}]{Bernien2017Scars}%
  \BibitemOpen
  \bibfield  {author} {\bibinfo {author} {\bibfnamefont {H.}~\bibnamefont
  {Bernien}} \emph {et~al.},\ }\bibfield  {title} {\bibinfo {title} {Probing
  many-body dynamics on a 51-atom quantum simulator},\ }\href
  {https://www.nature.com/articles/nature24622} {\bibfield  {journal} {\bibinfo
   {journal} {Nature}\ }\textbf {\bibinfo {volume} {551}},\ \bibinfo {pages}
  {579} (\bibinfo {year} {2017})}\BibitemShut {NoStop}%
\bibitem [{\citenamefont {Serbyn}\ \emph {et~al.}(2021)\citenamefont {Serbyn},
  \citenamefont {Abanin},\ and\ \citenamefont {Papi{\'c}}}]{Serbyn2021quantum}%
  \BibitemOpen
  \bibfield  {author} {\bibinfo {author} {\bibfnamefont {M.}~\bibnamefont
  {Serbyn}}, \bibinfo {author} {\bibfnamefont {D.}~\bibnamefont {Abanin}},\
  and\ \bibinfo {author} {\bibfnamefont {Z.}~\bibnamefont {Papi{\'c}}},\
  }\bibfield  {title} {\bibinfo {title} {Quantum many-body scars and weak
  breaking of ergodicity},\ }\href
  {https://www.nature.com/articles/s41567-021-01230-2} {\bibfield  {journal}
  {\bibinfo  {journal} {Nature Physics}\ }\textbf {\bibinfo {volume} {17}},\
  \bibinfo {pages} {675} (\bibinfo {year} {2021})}\BibitemShut {NoStop}%
\bibitem [{\citenamefont {Chandran}\ \emph {et~al.}(2023)\citenamefont
  {Chandran} \emph {et~al.}}]{Chandran2023quantum}%
  \BibitemOpen
  \bibfield  {author} {\bibinfo {author} {\bibfnamefont {A.}~\bibnamefont
  {Chandran}} \emph {et~al.},\ }\bibfield  {title} {\bibinfo {title} {Quantum
  many-body scars: A quasiparticle perspective},\ }\href
  {https://doi.org/https://doi.org/10.1146/annurev-conmatphys-031620-101617}
  {\bibfield  {journal} {\bibinfo  {journal} {Annual Review of Condensed Matter
  Physics}\ }\textbf {\bibinfo {volume} {14}},\ \bibinfo {pages} {443}
  (\bibinfo {year} {2023})}\BibitemShut {NoStop}%
\bibitem [{\citenamefont {Paredes}\ \emph {et~al.}(2005)\citenamefont {Paredes}
  \emph {et~al.}}]{Paredes2005PRL}%
  \BibitemOpen
  \bibfield  {author} {\bibinfo {author} {\bibfnamefont {B.}~\bibnamefont
  {Paredes}} \emph {et~al.},\ }\bibfield  {title} {\bibinfo {title} {Exploiting
  quantum parallelism to simulate quantum random many-body systems},\ }\href
  {https://doi.org/10.1103/PhysRevLett.95.140501} {\bibfield  {journal}
  {\bibinfo  {journal} {Phys. Rev. Lett.}\ }\textbf {\bibinfo {volume} {95}},\
  \bibinfo {pages} {140501} (\bibinfo {year} {2005})}\BibitemShut {NoStop}%
\bibitem [{\citenamefont {Andraschko}\ \emph {et~al.}(2014)\citenamefont
  {Andraschko}, \citenamefont {Enss},\ and\ \citenamefont
  {Sirker}}]{sirker_purification_and_mbl_2014}%
  \BibitemOpen
  \bibfield  {author} {\bibinfo {author} {\bibfnamefont {F.}~\bibnamefont
  {Andraschko}}, \bibinfo {author} {\bibfnamefont {T.}~\bibnamefont {Enss}},\
  and\ \bibinfo {author} {\bibfnamefont {J.}~\bibnamefont {Sirker}},\
  }\bibfield  {title} {\bibinfo {title} {Purification and many-body
  localization in cold atomic gases},\ }\href
  {https://doi.org/10.1103/PhysRevLett.113.217201} {\bibfield  {journal}
  {\bibinfo  {journal} {Phys. Rev. Lett.}\ }\textbf {\bibinfo {volume} {113}},\
  \bibinfo {pages} {217201} (\bibinfo {year} {2014})}\BibitemShut {NoStop}%
\bibitem [{\citenamefont {Enss}\ \emph {et~al.}(2017)\citenamefont {Enss},
  \citenamefont {Andraschko},\ and\ \citenamefont
  {Sirker}}]{sirker_mbl_in_infinite_chains}%
  \BibitemOpen
  \bibfield  {author} {\bibinfo {author} {\bibfnamefont {T.}~\bibnamefont
  {Enss}}, \bibinfo {author} {\bibfnamefont {F.}~\bibnamefont {Andraschko}},\
  and\ \bibinfo {author} {\bibfnamefont {J.}~\bibnamefont {Sirker}},\
  }\bibfield  {title} {\bibinfo {title} {Many-body localization in infinite
  chains},\ }\href {https://doi.org/10.1103/PhysRevB.95.045121} {\bibfield
  {journal} {\bibinfo  {journal} {Phys. Rev. B}\ }\textbf {\bibinfo {volume}
  {95}},\ \bibinfo {pages} {045121} (\bibinfo {year} {2017})}\BibitemShut
  {NoStop}%
\bibitem [{\citenamefont {Kogut}(1979)}]{Kogut_RMP_1979}%
  \BibitemOpen
  \bibfield  {author} {\bibinfo {author} {\bibfnamefont {J.~B.}\ \bibnamefont
  {Kogut}},\ }\bibfield  {title} {\bibinfo {title} {An introduction to lattice
  gauge theory and spin systems},\ }\href
  {https://doi.org/10.1103/RevModPhys.51.659} {\bibfield  {journal} {\bibinfo
  {journal} {Rev. Mod. Phys.}\ }\textbf {\bibinfo {volume} {51}},\ \bibinfo
  {pages} {659} (\bibinfo {year} {1979})}\BibitemShut {NoStop}%
\bibitem [{\citenamefont {Smith}\ \emph
  {et~al.}(2017{\natexlab{a}})\citenamefont {Smith} \emph
  {et~al.}}]{Smith_PRL_2017_first}%
  \BibitemOpen
  \bibfield  {author} {\bibinfo {author} {\bibfnamefont {A.}~\bibnamefont
  {Smith}} \emph {et~al.},\ }\bibfield  {title} {\bibinfo {title}
  {Disorder-free localization},\ }\href
  {https://doi.org/10.1103/PhysRevLett.118.266601} {\bibfield  {journal}
  {\bibinfo  {journal} {Phys. Rev. Lett.}\ }\textbf {\bibinfo {volume} {118}},\
  \bibinfo {pages} {266601} (\bibinfo {year} {2017}{\natexlab{a}})}\BibitemShut
  {NoStop}%
\bibitem [{\citenamefont {Smith}\ \emph
  {et~al.}(2017{\natexlab{b}})\citenamefont {Smith} \emph
  {et~al.}}]{Smith_PRL_2017_second}%
  \BibitemOpen
  \bibfield  {author} {\bibinfo {author} {\bibfnamefont {A.}~\bibnamefont
  {Smith}} \emph {et~al.},\ }\bibfield  {title} {\bibinfo {title} {Absence of
  ergodicity without quenched disorder: From quantum disentangled liquids to
  many-body localization},\ }\href
  {https://doi.org/10.1103/PhysRevLett.119.176601} {\bibfield  {journal}
  {\bibinfo  {journal} {Phys. Rev. Lett.}\ }\textbf {\bibinfo {volume} {119}},\
  \bibinfo {pages} {176601} (\bibinfo {year} {2017}{\natexlab{b}})}\BibitemShut
  {NoStop}%
\bibitem [{\citenamefont {Brenes}\ \emph {et~al.}(2018)\citenamefont {Brenes},
  \citenamefont {Dalmonte}, \citenamefont {Heyl},\ and\ \citenamefont
  {Scardicchio}}]{Brenes_PRL_2018}%
  \BibitemOpen
  \bibfield  {author} {\bibinfo {author} {\bibfnamefont {M.}~\bibnamefont
  {Brenes}}, \bibinfo {author} {\bibfnamefont {M.}~\bibnamefont {Dalmonte}},
  \bibinfo {author} {\bibfnamefont {M.}~\bibnamefont {Heyl}},\ and\ \bibinfo
  {author} {\bibfnamefont {A.}~\bibnamefont {Scardicchio}},\ }\bibfield
  {title} {\bibinfo {title} {Many-body localization dynamics from gauge
  invariance},\ }\href {https://doi.org/10.1103/PhysRevLett.120.030601}
  {\bibfield  {journal} {\bibinfo  {journal} {Phys. Rev. Lett.}\ }\textbf
  {\bibinfo {volume} {120}},\ \bibinfo {pages} {030601} (\bibinfo {year}
  {2018})}\BibitemShut {NoStop}%
\bibitem [{\citenamefont {Papaefstathiou}\ \emph {et~al.}(2020)\citenamefont
  {Papaefstathiou} \emph {et~al.}}]{Papaefstathiou_PRB_2020}%
  \BibitemOpen
  \bibfield  {author} {\bibinfo {author} {\bibfnamefont {I.}~\bibnamefont
  {Papaefstathiou}} \emph {et~al.},\ }\bibfield  {title} {\bibinfo {title}
  {Disorder-free localization in a simple $u(1)$ lattice gauge theory},\ }\href
  {https://doi.org/10.1103/PhysRevB.102.165132} {\bibfield  {journal} {\bibinfo
   {journal} {Phys. Rev. B}\ }\textbf {\bibinfo {volume} {102}},\ \bibinfo
  {pages} {165132} (\bibinfo {year} {2020})}\BibitemShut {NoStop}%
\bibitem [{\citenamefont {McClarty}\ \emph {et~al.}(2020)\citenamefont
  {McClarty} \emph {et~al.}}]{McClarty_PRB_2020}%
  \BibitemOpen
  \bibfield  {author} {\bibinfo {author} {\bibfnamefont {P.~A.}\ \bibnamefont
  {McClarty}} \emph {et~al.},\ }\bibfield  {title} {\bibinfo {title}
  {Disorder-free localization and many-body quantum scars from magnetic
  frustration},\ }\href {https://doi.org/10.1103/PhysRevB.102.224303}
  {\bibfield  {journal} {\bibinfo  {journal} {Phys. Rev. B}\ }\textbf {\bibinfo
  {volume} {102}},\ \bibinfo {pages} {224303} (\bibinfo {year}
  {2020})}\BibitemShut {NoStop}%
\bibitem [{\citenamefont {Russomanno}\ \emph {et~al.}(2020)\citenamefont
  {Russomanno} \emph {et~al.}}]{Russomanno_PhysRevResearch_2020}%
  \BibitemOpen
  \bibfield  {author} {\bibinfo {author} {\bibfnamefont {A.}~\bibnamefont
  {Russomanno}} \emph {et~al.},\ }\bibfield  {title} {\bibinfo {title}
  {Homogeneous floquet time crystal protected by gauge invariance},\ }\href
  {https://doi.org/10.1103/PhysRevResearch.2.012003} {\bibfield  {journal}
  {\bibinfo  {journal} {Phys. Rev. Res.}\ }\textbf {\bibinfo {volume} {2}},\
  \bibinfo {pages} {012003} (\bibinfo {year} {2020})}\BibitemShut {NoStop}%
\bibitem [{\citenamefont {Karpov}\ \emph {et~al.}(2021)\citenamefont {Karpov}
  \emph {et~al.}}]{Karpov-PRL-2021}%
  \BibitemOpen
  \bibfield  {author} {\bibinfo {author} {\bibfnamefont {P.}~\bibnamefont
  {Karpov}} \emph {et~al.},\ }\bibfield  {title} {\bibinfo {title}
  {Disorder-free localization in an interacting 2d lattice gauge theory},\
  }\href {https://doi.org/10.1103/PhysRevLett.126.130401} {\bibfield  {journal}
  {\bibinfo  {journal} {Phys. Rev. Lett.}\ }\textbf {\bibinfo {volume} {126}},\
  \bibinfo {pages} {130401} (\bibinfo {year} {2021})}\BibitemShut {NoStop}%
\bibitem [{\citenamefont {Zhu}\ and\ \citenamefont
  {Heyl}(2021)}]{Heyl_PhysRevResearch_2021}%
  \BibitemOpen
  \bibfield  {author} {\bibinfo {author} {\bibfnamefont {G.-Y.}\ \bibnamefont
  {Zhu}}\ and\ \bibinfo {author} {\bibfnamefont {M.}~\bibnamefont {Heyl}},\
  }\bibfield  {title} {\bibinfo {title} {Subdiffusive dynamics and critical
  quantum correlations in a disorder-free localized {K}itaev honeycomb model
  out of equilibrium},\ }\href
  {https://doi.org/10.1103/PhysRevResearch.3.L032069} {\bibfield  {journal}
  {\bibinfo  {journal} {Phys. Rev. Res.}\ }\textbf {\bibinfo {volume} {3}},\
  \bibinfo {pages} {L032069} (\bibinfo {year} {2021})}\BibitemShut {NoStop}%
\bibitem [{\citenamefont {Halimeh}\ \emph {et~al.}(2021)\citenamefont {Halimeh}
  \emph {et~al.}}]{halimeh2021stabilizing}%
  \BibitemOpen
  \bibfield  {author} {\bibinfo {author} {\bibfnamefont {J.~C.}\ \bibnamefont
  {Halimeh}} \emph {et~al.},\ }\bibfield  {title} {\bibinfo {title}
  {Stabilizing disorder-free localization},\ }\href
  {https://doi.org/10.48550/arXiv.2111.02427} {\bibfield  {journal} {\bibinfo
  {journal} {arXiv preprint arXiv:2111.02427}\ } (\bibinfo {year}
  {2021})}\BibitemShut {NoStop}%
\bibitem [{\citenamefont {Hart}\ \emph {et~al.}(2021)\citenamefont {Hart} \emph
  {et~al.}}]{Sarang_PRL_2021}%
  \BibitemOpen
  \bibfield  {author} {\bibinfo {author} {\bibfnamefont {O.}~\bibnamefont
  {Hart}} \emph {et~al.},\ }\bibfield  {title} {\bibinfo {title} {Logarithmic
  entanglement growth from disorder-free localization in the two-leg compass
  ladder},\ }\href {https://doi.org/10.1103/PhysRevLett.126.227202} {\bibfield
  {journal} {\bibinfo  {journal} {Phys. Rev. Lett.}\ }\textbf {\bibinfo
  {volume} {126}},\ \bibinfo {pages} {227202} (\bibinfo {year}
  {2021})}\BibitemShut {NoStop}%
\bibitem [{\citenamefont {Chakraborty}\ \emph {et~al.}(2022)\citenamefont
  {Chakraborty} \emph {et~al.}}]{Chakraborty_PRB_2022}%
  \BibitemOpen
  \bibfield  {author} {\bibinfo {author} {\bibfnamefont {N.}~\bibnamefont
  {Chakraborty}} \emph {et~al.},\ }\bibfield  {title} {\bibinfo {title}
  {Disorder-free localization transition in a two-dimensional lattice gauge
  theory},\ }\href {https://doi.org/10.1103/PhysRevB.106.L060308} {\bibfield
  {journal} {\bibinfo  {journal} {Phys. Rev. B}\ }\textbf {\bibinfo {volume}
  {106}},\ \bibinfo {pages} {L060308} (\bibinfo {year} {2022})}\BibitemShut
  {NoStop}%
\bibitem [{\citenamefont {Halimeh}\ \emph
  {et~al.}(2022{\natexlab{a}})\citenamefont {Halimeh} \emph
  {et~al.}}]{Halimeh_PRXQuantum_2022}%
  \BibitemOpen
  \bibfield  {author} {\bibinfo {author} {\bibfnamefont {J.~C.}\ \bibnamefont
  {Halimeh}} \emph {et~al.},\ }\bibfield  {title} {\bibinfo {title} {Enhancing
  disorder-free localization through dynamically emergent local symmetries},\
  }\href {https://doi.org/10.1103/PRXQuantum.3.020345} {\bibfield  {journal}
  {\bibinfo  {journal} {PRX Quantum}\ }\textbf {\bibinfo {volume} {3}},\
  \bibinfo {pages} {020345} (\bibinfo {year} {2022}{\natexlab{a}})}\BibitemShut
  {NoStop}%
\bibitem [{\citenamefont {Halimeh}\ \emph
  {et~al.}(2022{\natexlab{b}})\citenamefont {Halimeh}, \citenamefont {Hauke},
  \citenamefont {Knolle},\ and\ \citenamefont
  {Grusdt}}]{halimeh2022temperatureinduced}%
  \BibitemOpen
  \bibfield  {author} {\bibinfo {author} {\bibfnamefont {J.~C.}\ \bibnamefont
  {Halimeh}}, \bibinfo {author} {\bibfnamefont {P.}~\bibnamefont {Hauke}},
  \bibinfo {author} {\bibfnamefont {J.}~\bibnamefont {Knolle}},\ and\ \bibinfo
  {author} {\bibfnamefont {F.}~\bibnamefont {Grusdt}},\ }\href
  {https://arxiv.org/abs/2206.11273} {\bibinfo {title} {Temperature-induced
  disorder-free localization}} (\bibinfo {year} {2022}{\natexlab{b}}),\ \Eprint
  {https://arxiv.org/abs/2206.11273} {arXiv:2206.11273 [cond-mat.dis-nn]}
  \BibitemShut {NoStop}%
\bibitem [{\citenamefont {Lang}\ \emph {et~al.}(2022)\citenamefont {Lang},
  \citenamefont {Hauke}, \citenamefont {Knolle}, \citenamefont {Grusdt},\ and\
  \citenamefont {Halimeh}}]{Halimeh2022StarkDFL}%
  \BibitemOpen
  \bibfield  {author} {\bibinfo {author} {\bibfnamefont {H.}~\bibnamefont
  {Lang}}, \bibinfo {author} {\bibfnamefont {P.}~\bibnamefont {Hauke}},
  \bibinfo {author} {\bibfnamefont {J.}~\bibnamefont {Knolle}}, \bibinfo
  {author} {\bibfnamefont {F.}~\bibnamefont {Grusdt}},\ and\ \bibinfo {author}
  {\bibfnamefont {J.~C.}\ \bibnamefont {Halimeh}},\ }\bibfield  {title}
  {\bibinfo {title} {Disorder-free localization with {S}tark gauge
  protection},\ }\href {https://doi.org/10.1103/PhysRevB.106.174305} {\bibfield
   {journal} {\bibinfo  {journal} {Phys. Rev. B}\ }\textbf {\bibinfo {volume}
  {106}},\ \bibinfo {pages} {174305} (\bibinfo {year} {2022})}\BibitemShut
  {NoStop}%
\bibitem [{\citenamefont {Homeier}\ \emph {et~al.}(2023)\citenamefont {Homeier}
  \emph {et~al.}}]{Homeier2023realistic}%
  \BibitemOpen
  \bibfield  {author} {\bibinfo {author} {\bibfnamefont {L.}~\bibnamefont
  {Homeier}} \emph {et~al.},\ }\bibfield  {title} {\bibinfo {title} {Realistic
  scheme for quantum simulation of $\mathbb{Z}_2$ lattice gauge theories with
  dynamical matter in (2 + 1){D}},\ }\href
  {https://doi.org/10.1038/s42005-023-01237-6} {\bibfield  {journal} {\bibinfo
  {journal} {Communications Physics}\ }\textbf {\bibinfo {volume} {6}},\
  \bibinfo {pages} {127} (\bibinfo {year} {2023})}\BibitemShut {NoStop}%
\bibitem [{\citenamefont {Osborne}\ \emph {et~al.}(2023)\citenamefont
  {Osborne}, \citenamefont {McCulloch},\ and\ \citenamefont
  {Halimeh}}]{osborne_2D_DFL_2023}%
  \BibitemOpen
  \bibfield  {author} {\bibinfo {author} {\bibfnamefont {J.}~\bibnamefont
  {Osborne}}, \bibinfo {author} {\bibfnamefont {I.}~\bibnamefont {McCulloch}},\
  and\ \bibinfo {author} {\bibfnamefont {J.~C.}\ \bibnamefont {Halimeh}},\
  }\bibfield  {title} {\bibinfo {title} {Disorder-free localization in $2+
  1${D} lattice gauge theories with dynamical matter},\ }\href
  {https://arxiv.org/abs/2301.07720} {\bibfield  {journal} {\bibinfo  {journal}
  {arXiv preprint arXiv:2301.07720}\ } (\bibinfo {year} {2023})}\BibitemShut
  {NoStop}%
\bibitem [{\citenamefont {Sala}\ \emph {et~al.}(2024)\citenamefont {Sala},
  \citenamefont {Giudici},\ and\ \citenamefont {Halimeh}}]{sala2024disorder}%
  \BibitemOpen
  \bibfield  {author} {\bibinfo {author} {\bibfnamefont {P.}~\bibnamefont
  {Sala}}, \bibinfo {author} {\bibfnamefont {G.}~\bibnamefont {Giudici}},\ and\
  \bibinfo {author} {\bibfnamefont {J.~C.}\ \bibnamefont {Halimeh}},\
  }\bibfield  {title} {\bibinfo {title} {Disorder-free localization as a purely
  classical effect},\ }\href
  {https://doi.org/https://doi.org/10.1103/PhysRevB.109.L060305} {\bibfield
  {journal} {\bibinfo  {journal} {Physical Review B}\ }\textbf {\bibinfo
  {volume} {109}},\ \bibinfo {pages} {L060305} (\bibinfo {year}
  {2024})}\BibitemShut {NoStop}%
\bibitem [{\citenamefont {Abanin}\ \emph {et~al.}(2016)\citenamefont {Abanin},
  \citenamefont {{De Roeck}},\ and\ \citenamefont
  {Huveneers}}]{dynamical_mbl_abanin_2016}%
  \BibitemOpen
  \bibfield  {author} {\bibinfo {author} {\bibfnamefont {D.~A.}\ \bibnamefont
  {Abanin}}, \bibinfo {author} {\bibfnamefont {W.}~\bibnamefont {{De Roeck}}},\
  and\ \bibinfo {author} {\bibfnamefont {F.}~\bibnamefont {Huveneers}},\
  }\bibfield  {title} {\bibinfo {title} {Theory of many-body localization in
  periodically driven systems},\ }\href
  {https://doi.org/https://doi.org/10.1016/j.aop.2016.03.010} {\bibfield
  {journal} {\bibinfo  {journal} {Annals of Physics}\ }\textbf {\bibinfo
  {volume} {372}},\ \bibinfo {pages} {1} (\bibinfo {year} {2016})}\BibitemShut
  {NoStop}%
\bibitem [{\citenamefont {De~Tomasi}\ \emph {et~al.}(2019)\citenamefont
  {De~Tomasi} \emph {et~al.}}]{de2019dynamics}%
  \BibitemOpen
  \bibfield  {author} {\bibinfo {author} {\bibfnamefont {G.}~\bibnamefont
  {De~Tomasi}} \emph {et~al.},\ }\bibfield  {title} {\bibinfo {title} {Dynamics
  of strongly interacting systems: From {F}ock-space fragmentation to many-body
  localization},\ }\href
  {https://doi.org/https://doi.org/10.1103/PhysRevB.100.214313} {\bibfield
  {journal} {\bibinfo  {journal} {Physical Review B}\ }\textbf {\bibinfo
  {volume} {100}},\ \bibinfo {pages} {214313} (\bibinfo {year}
  {2019})}\BibitemShut {NoStop}%
\bibitem [{\citenamefont {Smith}\ \emph {et~al.}(2018)\citenamefont {Smith}
  \emph {et~al.}}]{Smith_PRB_2018}%
  \BibitemOpen
  \bibfield  {author} {\bibinfo {author} {\bibfnamefont {A.}~\bibnamefont
  {Smith}} \emph {et~al.},\ }\bibfield  {title} {\bibinfo {title} {Dynamical
  localization in {${\ensuremath{\mathbb{Z}}}_{2}$} lattice gauge theories},\
  }\href {https://doi.org/10.1103/PhysRevB.97.245137} {\bibfield  {journal}
  {\bibinfo  {journal} {Phys. Rev. B}\ }\textbf {\bibinfo {volume} {97}},\
  \bibinfo {pages} {245137} (\bibinfo {year} {2018})}\BibitemShut {NoStop}%
\bibitem [{\citenamefont {van Enk}\ and\ \citenamefont
  {Beenakker}(2012{\natexlab{a}})}]{vanEnk_PRL_2012}%
  \BibitemOpen
  \bibfield  {author} {\bibinfo {author} {\bibfnamefont {S.~J.}\ \bibnamefont
  {van Enk}}\ and\ \bibinfo {author} {\bibfnamefont {C.~W.~J.}\ \bibnamefont
  {Beenakker}},\ }\bibfield  {title} {\bibinfo {title} {Measuring
  $\mathrm{Tr}{\ensuremath{\rho}}^{n}$ on single copies of $\ensuremath{\rho}$
  using random measurements},\ }\href
  {https://doi.org/10.1103/PhysRevLett.108.110503} {\bibfield  {journal}
  {\bibinfo  {journal} {Phys. Rev. Lett.}\ }\textbf {\bibinfo {volume} {108}},\
  \bibinfo {pages} {110503} (\bibinfo {year} {2012}{\natexlab{a}})}\BibitemShut
  {NoStop}%
\bibitem [{\citenamefont {Brydges}\ \emph {et~al.}(2019)\citenamefont {Brydges}
  \emph {et~al.}}]{Brydges_2019}%
  \BibitemOpen
  \bibfield  {author} {\bibinfo {author} {\bibfnamefont {T.}~\bibnamefont
  {Brydges}} \emph {et~al.},\ }\bibfield  {title} {\bibinfo {title} {Probing
  {R}\'enyi entanglement entropy via randomized measurements},\ }\href
  {https://doi.org/10.1126/science.aau4963} {\bibfield  {journal} {\bibinfo
  {journal} {Science}\ }\textbf {\bibinfo {volume} {364}},\ \bibinfo {pages}
  {260} (\bibinfo {year} {2019})}\BibitemShut {NoStop}%
\bibitem [{\citenamefont {Grover}(1996)}]{grover_1996}%
  \BibitemOpen
  \bibfield  {author} {\bibinfo {author} {\bibfnamefont {L.~K.}\ \bibnamefont
  {Grover}},\ }\bibfield  {title} {\bibinfo {title} {A fast quantum mechanical
  algorithm for database search},\ }in\ \href
  {https://doi.org/10.1145/237814.237866} {\emph {\bibinfo {booktitle}
  {Proceedings of the Twenty-Eighth Annual ACM Symposium on Theory of
  Computing}}},\ \bibinfo {series and number} {STOC '96}\ (\bibinfo
  {publisher} {Association for Computing Machinery},\ \bibinfo {address} {New
  York, NY, USA},\ \bibinfo {year} {1996})\ p.\ \bibinfo {pages}
  {212–219}\BibitemShut {NoStop}%
\bibitem [{\citenamefont {\ifmmode~\check{S}\else \v{S}\fi{}untajs}\ \emph
  {et~al.}(2020)\citenamefont {\ifmmode~\check{S}\else \v{S}\fi{}untajs} \emph
  {et~al.}}]{Prosen_PRE_2020}%
  \BibitemOpen
  \bibfield  {author} {\bibinfo {author} {\bibfnamefont {J.}~\bibnamefont
  {\ifmmode~\check{S}\else \v{S}\fi{}untajs}} \emph {et~al.},\ }\bibfield
  {title} {\bibinfo {title} {Quantum chaos challenges many-body localization},\
  }\href {https://doi.org/10.1103/PhysRevE.102.062144} {\bibfield  {journal}
  {\bibinfo  {journal} {Phys. Rev. E}\ }\textbf {\bibinfo {volume} {102}},\
  \bibinfo {pages} {062144} (\bibinfo {year} {2020})}\BibitemShut {NoStop}%
\bibitem [{\citenamefont {Sels}(2022)}]{Sels2022PRB}%
  \BibitemOpen
  \bibfield  {author} {\bibinfo {author} {\bibfnamefont {D.}~\bibnamefont
  {Sels}},\ }\bibfield  {title} {\bibinfo {title} {Bath-induced delocalization
  in interacting disordered spin chains},\ }\href
  {https://doi.org/10.1103/PhysRevB.106.L020202} {\bibfield  {journal}
  {\bibinfo  {journal} {Phys. Rev. B}\ }\textbf {\bibinfo {volume} {106}},\
  \bibinfo {pages} {L020202} (\bibinfo {year} {2022})}\BibitemShut {NoStop}%
\bibitem [{\citenamefont {Morningstar}\ \emph {et~al.}(2022)\citenamefont
  {Morningstar} \emph {et~al.}}]{Morningstar_PRB_2022}%
  \BibitemOpen
  \bibfield  {author} {\bibinfo {author} {\bibfnamefont {A.}~\bibnamefont
  {Morningstar}} \emph {et~al.},\ }\bibfield  {title} {\bibinfo {title}
  {Avalanches and many-body resonances in many-body localized systems},\ }\href
  {https://doi.org/10.1103/PhysRevB.105.174205} {\bibfield  {journal} {\bibinfo
   {journal} {Phys. Rev. B}\ }\textbf {\bibinfo {volume} {105}},\ \bibinfo
  {pages} {174205} (\bibinfo {year} {2022})}\BibitemShut {NoStop}%
\bibitem [{\citenamefont {Acharya}\ \emph {et~al.}(2024)\citenamefont {Acharya}
  \emph {et~al.}}]{google_y1}%
  \BibitemOpen
  \bibfield  {author} {\bibinfo {author} {\bibfnamefont {R.}~\bibnamefont
  {Acharya}} \emph {et~al.},\ }\href {https://arxiv.org/abs/2408.13687}
  {\bibinfo {title} {Quantum error correction below the surface code
  threshold}} (\bibinfo {year} {2024}),\ \Eprint
  {https://arxiv.org/abs/2408.13687} {arXiv:2408.13687 [quant-ph]} \BibitemShut
  {NoStop}%
\bibitem [{\citenamefont {Will}\ \emph {et~al.}(2025)\citenamefont {Will},
  \citenamefont {Cochran}, \citenamefont {Rosenberg}, \citenamefont {Jobst},
  \citenamefont {Eassa}, \citenamefont {Roushan}, \citenamefont {Knap},
  \citenamefont {Gammon-Smith},\ and\ \citenamefont
  {Pollmann}}]{will2025_probing_nonequilibrium_topologicalorder}%
  \BibitemOpen
  \bibfield  {author} {\bibinfo {author} {\bibfnamefont {M.}~\bibnamefont
  {Will}}, \bibinfo {author} {\bibfnamefont {T.~A.}\ \bibnamefont {Cochran}},
  \bibinfo {author} {\bibfnamefont {E.}~\bibnamefont {Rosenberg}}, \bibinfo
  {author} {\bibfnamefont {B.}~\bibnamefont {Jobst}}, \bibinfo {author}
  {\bibfnamefont {N.~M.}\ \bibnamefont {Eassa}}, \bibinfo {author}
  {\bibfnamefont {P.}~\bibnamefont {Roushan}}, \bibinfo {author} {\bibfnamefont
  {M.}~\bibnamefont {Knap}}, \bibinfo {author} {\bibfnamefont {A.}~\bibnamefont
  {Gammon-Smith}},\ and\ \bibinfo {author} {\bibfnamefont {F.}~\bibnamefont
  {Pollmann}},\ }\href {https://arxiv.org/abs/2501.18461} {\bibinfo {title}
  {Probing non-equilibrium topological order on a quantum processor}} (\bibinfo
  {year} {2025}),\ \Eprint {https://arxiv.org/abs/2501.18461} {arXiv:2501.18461
  [quant-ph]} \BibitemShut {NoStop}%
\bibitem [{\citenamefont {Acharya}\ \emph {et~al.}(2023)\citenamefont {Acharya}
  \emph {et~al.}}]{Google_M2_2023}%
  \BibitemOpen
  \bibfield  {author} {\bibinfo {author} {\bibfnamefont {R.}~\bibnamefont
  {Acharya}} \emph {et~al.},\ }\bibfield  {title} {\bibinfo {title}
  {Suppressing quantum errors by scaling a surface code logical qubit},\ }\href
  {https://doi.org/10.1038/s41586-022-05434-1} {\bibfield  {journal} {\bibinfo
  {journal} {Nature}\ }\textbf {\bibinfo {volume} {614}},\ \bibinfo {pages}
  {676} (\bibinfo {year} {2023})}\BibitemShut {NoStop}%
\bibitem [{\citenamefont {Foxen}\ \emph {et~al.}(2020)\citenamefont {Foxen},
  \citenamefont {Neill}, \citenamefont {Dunsworth}, \citenamefont {Roushan},
  \citenamefont {Chiaro}, \citenamefont {Megrant}, \citenamefont {Kelly},
  \citenamefont {Chen}, \citenamefont {Satzinger}, \citenamefont {Barends},
  \citenamefont {Arute}, \citenamefont {Arya}, \citenamefont {Babbush},
  \citenamefont {Bacon}, \citenamefont {Bardin}, \citenamefont {Boixo},
  \citenamefont {Buell}, \citenamefont {Burkett}, \citenamefont {Chen},
  \citenamefont {Collins}, \citenamefont {Farhi}, \citenamefont {Fowler},
  \citenamefont {Gidney}, \citenamefont {Giustina}, \citenamefont {Graff},
  \citenamefont {Harrigan}, \citenamefont {Huang}, \citenamefont {Isakov},
  \citenamefont {Jeffrey}, \citenamefont {Jiang}, \citenamefont {Kafri},
  \citenamefont {Kechedzhi}, \citenamefont {Klimov}, \citenamefont {Korotkov},
  \citenamefont {Kostritsa}, \citenamefont {Landhuis}, \citenamefont {Lucero},
  \citenamefont {McClean}, \citenamefont {McEwen}, \citenamefont {Mi},
  \citenamefont {Mohseni}, \citenamefont {Mutus}, \citenamefont {Naaman},
  \citenamefont {Neeley}, \citenamefont {Niu}, \citenamefont {Petukhov},
  \citenamefont {Quintana}, \citenamefont {Rubin}, \citenamefont {Sank},
  \citenamefont {Smelyanskiy}, \citenamefont {Vainsencher}, \citenamefont
  {White}, \citenamefont {Yao}, \citenamefont {Yeh}, \citenamefont {Zalcman},
  \citenamefont {Neven},\ and\ \citenamefont
  {Martinis}}]{Foxen_2020_demonstrating_twoq_gates}%
  \BibitemOpen
  \bibfield  {author} {\bibinfo {author} {\bibfnamefont {B.}~\bibnamefont
  {Foxen}}, \bibinfo {author} {\bibfnamefont {C.}~\bibnamefont {Neill}},
  \bibinfo {author} {\bibfnamefont {A.}~\bibnamefont {Dunsworth}}, \bibinfo
  {author} {\bibfnamefont {P.}~\bibnamefont {Roushan}}, \bibinfo {author}
  {\bibfnamefont {B.}~\bibnamefont {Chiaro}}, \bibinfo {author} {\bibfnamefont
  {A.}~\bibnamefont {Megrant}}, \bibinfo {author} {\bibfnamefont
  {J.}~\bibnamefont {Kelly}}, \bibinfo {author} {\bibfnamefont
  {Z.}~\bibnamefont {Chen}}, \bibinfo {author} {\bibfnamefont {K.}~\bibnamefont
  {Satzinger}}, \bibinfo {author} {\bibfnamefont {R.}~\bibnamefont {Barends}},
  \bibinfo {author} {\bibfnamefont {F.}~\bibnamefont {Arute}}, \bibinfo
  {author} {\bibfnamefont {K.}~\bibnamefont {Arya}}, \bibinfo {author}
  {\bibfnamefont {R.}~\bibnamefont {Babbush}}, \bibinfo {author} {\bibfnamefont
  {D.}~\bibnamefont {Bacon}}, \bibinfo {author} {\bibfnamefont
  {J.}~\bibnamefont {Bardin}}, \bibinfo {author} {\bibfnamefont
  {S.}~\bibnamefont {Boixo}}, \bibinfo {author} {\bibfnamefont
  {D.}~\bibnamefont {Buell}}, \bibinfo {author} {\bibfnamefont
  {B.}~\bibnamefont {Burkett}}, \bibinfo {author} {\bibfnamefont
  {Y.}~\bibnamefont {Chen}}, \bibinfo {author} {\bibfnamefont {R.}~\bibnamefont
  {Collins}}, \bibinfo {author} {\bibfnamefont {E.}~\bibnamefont {Farhi}},
  \bibinfo {author} {\bibfnamefont {A.}~\bibnamefont {Fowler}}, \bibinfo
  {author} {\bibfnamefont {C.}~\bibnamefont {Gidney}}, \bibinfo {author}
  {\bibfnamefont {M.}~\bibnamefont {Giustina}}, \bibinfo {author}
  {\bibfnamefont {R.}~\bibnamefont {Graff}}, \bibinfo {author} {\bibfnamefont
  {M.}~\bibnamefont {Harrigan}}, \bibinfo {author} {\bibfnamefont
  {T.}~\bibnamefont {Huang}}, \bibinfo {author} {\bibfnamefont
  {S.}~\bibnamefont {Isakov}}, \bibinfo {author} {\bibfnamefont
  {E.}~\bibnamefont {Jeffrey}}, \bibinfo {author} {\bibfnamefont
  {Z.}~\bibnamefont {Jiang}}, \bibinfo {author} {\bibfnamefont
  {D.}~\bibnamefont {Kafri}}, \bibinfo {author} {\bibfnamefont
  {K.}~\bibnamefont {Kechedzhi}}, \bibinfo {author} {\bibfnamefont
  {P.}~\bibnamefont {Klimov}}, \bibinfo {author} {\bibfnamefont
  {A.}~\bibnamefont {Korotkov}}, \bibinfo {author} {\bibfnamefont
  {F.}~\bibnamefont {Kostritsa}}, \bibinfo {author} {\bibfnamefont
  {D.}~\bibnamefont {Landhuis}}, \bibinfo {author} {\bibfnamefont
  {E.}~\bibnamefont {Lucero}}, \bibinfo {author} {\bibfnamefont
  {J.}~\bibnamefont {McClean}}, \bibinfo {author} {\bibfnamefont
  {M.}~\bibnamefont {McEwen}}, \bibinfo {author} {\bibfnamefont
  {X.}~\bibnamefont {Mi}}, \bibinfo {author} {\bibfnamefont {M.}~\bibnamefont
  {Mohseni}}, \bibinfo {author} {\bibfnamefont {J.}~\bibnamefont {Mutus}},
  \bibinfo {author} {\bibfnamefont {O.}~\bibnamefont {Naaman}}, \bibinfo
  {author} {\bibfnamefont {M.}~\bibnamefont {Neeley}}, \bibinfo {author}
  {\bibfnamefont {M.}~\bibnamefont {Niu}}, \bibinfo {author} {\bibfnamefont
  {A.}~\bibnamefont {Petukhov}}, \bibinfo {author} {\bibfnamefont
  {C.}~\bibnamefont {Quintana}}, \bibinfo {author} {\bibfnamefont
  {N.}~\bibnamefont {Rubin}}, \bibinfo {author} {\bibfnamefont
  {D.}~\bibnamefont {Sank}}, \bibinfo {author} {\bibfnamefont {V.}~\bibnamefont
  {Smelyanskiy}}, \bibinfo {author} {\bibfnamefont {A.}~\bibnamefont
  {Vainsencher}}, \bibinfo {author} {\bibfnamefont {T.}~\bibnamefont {White}},
  \bibinfo {author} {\bibfnamefont {Z.}~\bibnamefont {Yao}}, \bibinfo {author}
  {\bibfnamefont {P.}~\bibnamefont {Yeh}}, \bibinfo {author} {\bibfnamefont
  {A.}~\bibnamefont {Zalcman}}, \bibinfo {author} {\bibfnamefont
  {H.}~\bibnamefont {Neven}},\ and\ \bibinfo {author} {\bibfnamefont
  {J.}~\bibnamefont {Martinis}},\ }\bibfield  {title} {\bibinfo {title}
  {Demonstrating a continuous set of two-qubit gates for near-term quantum
  algorithms},\ }\bibfield  {journal} {\bibinfo  {journal} {Physical Review
  Letters}\ }\textbf {\bibinfo {volume} {125}},\ \href
  {https://doi.org/10.1103/physrevlett.125.120504}
  {10.1103/physrevlett.125.120504} (\bibinfo {year} {2020})\BibitemShut
  {NoStop}%
\bibitem [{\citenamefont {Cochran}\ \emph {et~al.}(2025)\citenamefont {Cochran}
  \emph {et~al.}}]{cochran_2dlgt}%
  \BibitemOpen
  \bibfield  {author} {\bibinfo {author} {\bibfnamefont {T.}~\bibnamefont
  {Cochran}} \emph {et~al.},\ }\bibfield  {title} {\bibinfo {title}
  {Visualizing dynamics of charges and strings in (2 + 1){D} lattice gauge
  theories},\ }\href {https://doi.org/10.1038/s41586-025-08999-9} {\bibfield
  {journal} {\bibinfo  {journal} {Nature}\ }\textbf {\bibinfo {volume} {642}},\
  \bibinfo {pages} {315} (\bibinfo {year} {2025})}\BibitemShut {NoStop}%
\bibitem [{\citenamefont {Klimov}\ \emph {et~al.}(2024)\citenamefont {Klimov},
  \citenamefont {Bengtsson}, \citenamefont {Quintana}, \citenamefont
  {Bourassa}, \citenamefont {Hong}, \citenamefont {Dunsworth}, \citenamefont
  {Satzinger}, \citenamefont {Livingston}, \citenamefont {Sivak}, \citenamefont
  {Niu}, \citenamefont {Andersen}, \citenamefont {Zhang}, \citenamefont {Chik},
  \citenamefont {Chen}, \citenamefont {Neill}, \citenamefont {Erickson},
  \citenamefont {Grajales~Dau}, \citenamefont {Megrant}, \citenamefont
  {Roushan}, \citenamefont {Korotkov}, \citenamefont {Kelly}, \citenamefont
  {Smelyanskiy}, \citenamefont {Chen},\ and\ \citenamefont
  {Neven}}]{optimizing_klimov_2024}%
  \BibitemOpen
  \bibfield  {author} {\bibinfo {author} {\bibfnamefont {P.~V.}\ \bibnamefont
  {Klimov}}, \bibinfo {author} {\bibfnamefont {A.}~\bibnamefont {Bengtsson}},
  \bibinfo {author} {\bibfnamefont {C.}~\bibnamefont {Quintana}}, \bibinfo
  {author} {\bibfnamefont {A.}~\bibnamefont {Bourassa}}, \bibinfo {author}
  {\bibfnamefont {S.}~\bibnamefont {Hong}}, \bibinfo {author} {\bibfnamefont
  {A.}~\bibnamefont {Dunsworth}}, \bibinfo {author} {\bibfnamefont {K.~J.}\
  \bibnamefont {Satzinger}}, \bibinfo {author} {\bibfnamefont {W.~P.}\
  \bibnamefont {Livingston}}, \bibinfo {author} {\bibfnamefont
  {V.}~\bibnamefont {Sivak}}, \bibinfo {author} {\bibfnamefont {M.~Y.}\
  \bibnamefont {Niu}}, \bibinfo {author} {\bibfnamefont {T.~I.}\ \bibnamefont
  {Andersen}}, \bibinfo {author} {\bibfnamefont {Y.}~\bibnamefont {Zhang}},
  \bibinfo {author} {\bibfnamefont {D.}~\bibnamefont {Chik}}, \bibinfo {author}
  {\bibfnamefont {Z.}~\bibnamefont {Chen}}, \bibinfo {author} {\bibfnamefont
  {C.}~\bibnamefont {Neill}}, \bibinfo {author} {\bibfnamefont
  {C.}~\bibnamefont {Erickson}}, \bibinfo {author} {\bibfnamefont
  {A.}~\bibnamefont {Grajales~Dau}}, \bibinfo {author} {\bibfnamefont
  {A.}~\bibnamefont {Megrant}}, \bibinfo {author} {\bibfnamefont
  {P.}~\bibnamefont {Roushan}}, \bibinfo {author} {\bibfnamefont {A.~N.}\
  \bibnamefont {Korotkov}}, \bibinfo {author} {\bibfnamefont {J.}~\bibnamefont
  {Kelly}}, \bibinfo {author} {\bibfnamefont {V.}~\bibnamefont {Smelyanskiy}},
  \bibinfo {author} {\bibfnamefont {Y.}~\bibnamefont {Chen}},\ and\ \bibinfo
  {author} {\bibfnamefont {H.}~\bibnamefont {Neven}},\ }\bibfield  {title}
  {\bibinfo {title} {Optimizing quantum gates towards the scale of logical
  qubits},\ }\href {https://doi.org/10.1038/s41467-024-46623-y} {\bibfield
  {journal} {\bibinfo  {journal} {Nature Communications}\ }\textbf {\bibinfo
  {volume} {15}},\ \bibinfo {pages} {2442} (\bibinfo {year}
  {2024})}\BibitemShut {NoStop}%
\bibitem [{\citenamefont {Bengtsson}\ \emph {et~al.}(2024)\citenamefont
  {Bengtsson}, \citenamefont {Opremcak}, \citenamefont {Khezri}, \citenamefont
  {Sank}, \citenamefont {Bourassa}, \citenamefont {Satzinger}, \citenamefont
  {Hong}, \citenamefont {Erickson}, \citenamefont {Lester}, \citenamefont
  {Miao}, \citenamefont {Korotkov}, \citenamefont {Kelly}, \citenamefont
  {Chen},\ and\ \citenamefont
  {Klimov}}]{model-based_optimization_bengtsson_2024}%
  \BibitemOpen
  \bibfield  {author} {\bibinfo {author} {\bibfnamefont {A.}~\bibnamefont
  {Bengtsson}}, \bibinfo {author} {\bibfnamefont {A.}~\bibnamefont {Opremcak}},
  \bibinfo {author} {\bibfnamefont {M.}~\bibnamefont {Khezri}}, \bibinfo
  {author} {\bibfnamefont {D.}~\bibnamefont {Sank}}, \bibinfo {author}
  {\bibfnamefont {A.}~\bibnamefont {Bourassa}}, \bibinfo {author}
  {\bibfnamefont {K.~J.}\ \bibnamefont {Satzinger}}, \bibinfo {author}
  {\bibfnamefont {S.}~\bibnamefont {Hong}}, \bibinfo {author} {\bibfnamefont
  {C.}~\bibnamefont {Erickson}}, \bibinfo {author} {\bibfnamefont {B.~J.}\
  \bibnamefont {Lester}}, \bibinfo {author} {\bibfnamefont {K.~C.}\
  \bibnamefont {Miao}}, \bibinfo {author} {\bibfnamefont {A.~N.}\ \bibnamefont
  {Korotkov}}, \bibinfo {author} {\bibfnamefont {J.}~\bibnamefont {Kelly}},
  \bibinfo {author} {\bibfnamefont {Z.}~\bibnamefont {Chen}},\ and\ \bibinfo
  {author} {\bibfnamefont {P.~V.}\ \bibnamefont {Klimov}},\ }\bibfield  {title}
  {\bibinfo {title} {Model-based optimization of superconducting qubit
  readout},\ }\href {https://doi.org/10.1103/PhysRevLett.132.100603} {\bibfield
   {journal} {\bibinfo  {journal} {Phys. Rev. Lett.}\ }\textbf {\bibinfo
  {volume} {132}},\ \bibinfo {pages} {100603} (\bibinfo {year}
  {2024})}\BibitemShut {NoStop}%
\bibitem [{\citenamefont {Wallman}\ and\ \citenamefont
  {Emerson}(2016)}]{randomized_compiling_wallman_2016}%
  \BibitemOpen
  \bibfield  {author} {\bibinfo {author} {\bibfnamefont {J.~J.}\ \bibnamefont
  {Wallman}}\ and\ \bibinfo {author} {\bibfnamefont {J.}~\bibnamefont
  {Emerson}},\ }\bibfield  {title} {\bibinfo {title} {Noise tailoring for
  scalable quantum computation via randomized compiling},\ }\href
  {https://doi.org/10.1103/PhysRevA.94.052325} {\bibfield  {journal} {\bibinfo
  {journal} {Phys. Rev. A}\ }\textbf {\bibinfo {volume} {94}},\ \bibinfo
  {pages} {052325} (\bibinfo {year} {2016})}\BibitemShut {NoStop}%
\bibitem [{\citenamefont {{Google Quantum AI team}}\ and\ \citenamefont
  {collaborators}(2024{\natexlab{a}})}]{quantumlib_cirq_gauge_compiling_2024}%
  \BibitemOpen
  \bibfield  {author} {\bibinfo {author} {\bibnamefont {{Google Quantum AI
  team}}}\ and\ \bibinfo {author} {\bibnamefont {collaborators}},\ }\href
  {https://github.com/quantumlib/Cirq/tree/main/cirq-core/cirq/transformers/gauge_compiling}
  {\bibinfo {title} {Gauge compiling in cirq}} (\bibinfo {year}
  {2024}{\natexlab{a}})\BibitemShut {NoStop}%
\bibitem [{\citenamefont {Hashim}\ \emph {et~al.}(2021)\citenamefont {Hashim},
  \citenamefont {Naik}, \citenamefont {Morvan}, \citenamefont {Ville},
  \citenamefont {Mitchell}, \citenamefont {Kreikebaum}, \citenamefont {Davis},
  \citenamefont {Smith}, \citenamefont {Iancu}, \citenamefont {O'Brien},
  \citenamefont {Hincks}, \citenamefont {Wallman}, \citenamefont {Emerson},\
  and\ \citenamefont {Siddiqi}}]{hashim_randomized_compiling_2021}%
  \BibitemOpen
  \bibfield  {author} {\bibinfo {author} {\bibfnamefont {A.}~\bibnamefont
  {Hashim}}, \bibinfo {author} {\bibfnamefont {R.~K.}\ \bibnamefont {Naik}},
  \bibinfo {author} {\bibfnamefont {A.}~\bibnamefont {Morvan}}, \bibinfo
  {author} {\bibfnamefont {J.-L.}\ \bibnamefont {Ville}}, \bibinfo {author}
  {\bibfnamefont {B.}~\bibnamefont {Mitchell}}, \bibinfo {author}
  {\bibfnamefont {J.~M.}\ \bibnamefont {Kreikebaum}}, \bibinfo {author}
  {\bibfnamefont {M.}~\bibnamefont {Davis}}, \bibinfo {author} {\bibfnamefont
  {E.}~\bibnamefont {Smith}}, \bibinfo {author} {\bibfnamefont
  {C.}~\bibnamefont {Iancu}}, \bibinfo {author} {\bibfnamefont {K.~P.}\
  \bibnamefont {O'Brien}}, \bibinfo {author} {\bibfnamefont {I.}~\bibnamefont
  {Hincks}}, \bibinfo {author} {\bibfnamefont {J.~J.}\ \bibnamefont {Wallman}},
  \bibinfo {author} {\bibfnamefont {J.}~\bibnamefont {Emerson}},\ and\ \bibinfo
  {author} {\bibfnamefont {I.}~\bibnamefont {Siddiqi}},\ }\bibfield  {title}
  {\bibinfo {title} {Randomized compiling for scalable quantum computing on a
  noisy superconducting quantum processor},\ }\href
  {https://doi.org/10.1103/PhysRevX.11.041039} {\bibfield  {journal} {\bibinfo
  {journal} {Phys. Rev. X}\ }\textbf {\bibinfo {volume} {11}},\ \bibinfo
  {pages} {041039} (\bibinfo {year} {2021})}\BibitemShut {NoStop}%
\bibitem [{\citenamefont {{Google Quantum AI team}}\ and\ \citenamefont
  {collaborators}(2024{\natexlab{b}})}]{cirq_readout_mitigation}%
  \BibitemOpen
  \bibfield  {author} {\bibinfo {author} {\bibnamefont {{Google Quantum AI
  team}}}\ and\ \bibinfo {author} {\bibnamefont {collaborators}},\ }\href
  {https://github.com/quantumlib/Cirq/blob/95204668565ae56e482b9f9bd9c9f36032580b59/cirq-core/cirq/experiments/readout_confusion_matrix.py}
  {\bibinfo {title} {Readout mitigation in cirq}} (\bibinfo {year}
  {2024}{\natexlab{b}})\BibitemShut {NoStop}%
\bibitem [{\citenamefont {Cai}\ \emph {et~al.}(2023)\citenamefont {Cai},
  \citenamefont {Babbush}, \citenamefont {Benjamin}, \citenamefont {Endo},
  \citenamefont {Huggins}, \citenamefont {Li}, \citenamefont {McClean},\ and\
  \citenamefont {O'Brien}}]{quantum_error_mitigation}%
  \BibitemOpen
  \bibfield  {author} {\bibinfo {author} {\bibfnamefont {Z.}~\bibnamefont
  {Cai}}, \bibinfo {author} {\bibfnamefont {R.}~\bibnamefont {Babbush}},
  \bibinfo {author} {\bibfnamefont {S.~C.}\ \bibnamefont {Benjamin}}, \bibinfo
  {author} {\bibfnamefont {S.}~\bibnamefont {Endo}}, \bibinfo {author}
  {\bibfnamefont {W.~J.}\ \bibnamefont {Huggins}}, \bibinfo {author}
  {\bibfnamefont {Y.}~\bibnamefont {Li}}, \bibinfo {author} {\bibfnamefont
  {J.~R.}\ \bibnamefont {McClean}},\ and\ \bibinfo {author} {\bibfnamefont
  {T.~E.}\ \bibnamefont {O'Brien}},\ }\bibfield  {title} {\bibinfo {title}
  {Quantum error mitigation},\ }\href
  {https://doi.org/10.1103/RevModPhys.95.045005} {\bibfield  {journal}
  {\bibinfo  {journal} {Rev. Mod. Phys.}\ }\textbf {\bibinfo {volume} {95}},\
  \bibinfo {pages} {045005} (\bibinfo {year} {2023})}\BibitemShut {NoStop}%
\bibitem [{\citenamefont {Weinberg}\ and\ \citenamefont
  {Bukov}(2019)}]{quspin}%
  \BibitemOpen
  \bibfield  {author} {\bibinfo {author} {\bibfnamefont {P.}~\bibnamefont
  {Weinberg}}\ and\ \bibinfo {author} {\bibfnamefont {M.}~\bibnamefont
  {Bukov}},\ }\bibfield  {title} {\bibinfo {title} {{QuSpin: a Python package
  for dynamics and exact diagonalisation of quantum many body systems. Part II:
  bosons, fermions and higher spins}},\ }\href
  {https://doi.org/10.21468/SciPostPhys.7.2.020} {\bibfield  {journal}
  {\bibinfo  {journal} {SciPost Phys.}\ }\textbf {\bibinfo {volume} {7}},\
  \bibinfo {pages} {020} (\bibinfo {year} {2019})}\BibitemShut {NoStop}%
\bibitem [{\citenamefont {{Google Quantum AI team}}\ and\ \citenamefont
  {collaborators}(2020)}]{quantum_ai_team_qsim_2020}%
  \BibitemOpen
  \bibfield  {author} {\bibinfo {author} {\bibnamefont {{Google Quantum AI
  team}}}\ and\ \bibinfo {author} {\bibnamefont {collaborators}},\ }\href
  {https://doi.org/10.5281/zenodo.4110662} {\bibinfo {title} {{qsim}}}
  (\bibinfo {year} {2020})\BibitemShut {NoStop}%
\bibitem [{\citenamefont {{Google Quantum AI team}}\ and\ \citenamefont
  {collaborators}(2024{\natexlab{c}})}]{cirq_developers_cirq_2024}%
  \BibitemOpen
  \bibfield  {author} {\bibinfo {author} {\bibnamefont {{Google Quantum AI
  team}}}\ and\ \bibinfo {author} {\bibnamefont {collaborators}},\ }\href
  {https://doi.org/10.5281/zenodo.11398048} {\bibinfo {title} {{Cirq}}}
  (\bibinfo {year} {2024}{\natexlab{c}})\BibitemShut {NoStop}%
\bibitem [{\citenamefont {Fannes}\ \emph {et~al.}(1992)\citenamefont {Fannes},
  \citenamefont {Nachtergaele},\ and\ \citenamefont {Werner}}]{mps1}%
  \BibitemOpen
  \bibfield  {author} {\bibinfo {author} {\bibfnamefont {M.}~\bibnamefont
  {Fannes}}, \bibinfo {author} {\bibfnamefont {B.}~\bibnamefont
  {Nachtergaele}},\ and\ \bibinfo {author} {\bibfnamefont {R.~F.}\ \bibnamefont
  {Werner}},\ }\bibfield  {title} {\bibinfo {title} {{Finitely correlated
  states on quantum spin chains}},\ }\href {https://doi.org/10.1007/BF02099178}
  {\bibfield  {journal} {\bibinfo  {journal} {Communications in Mathematical
  Physics}\ }\textbf {\bibinfo {volume} {144}},\ \bibinfo {pages} {443 }
  (\bibinfo {year} {1992})}\BibitemShut {NoStop}%
\bibitem [{\citenamefont {\"Ostlund}\ and\ \citenamefont
  {Rommer}(1995)}]{mps2}%
  \BibitemOpen
  \bibfield  {author} {\bibinfo {author} {\bibfnamefont {S.}~\bibnamefont
  {\"Ostlund}}\ and\ \bibinfo {author} {\bibfnamefont {S.}~\bibnamefont
  {Rommer}},\ }\bibfield  {title} {\bibinfo {title} {Thermodynamic limit of
  density matrix renormalization},\ }\href
  {https://doi.org/10.1103/PhysRevLett.75.3537} {\bibfield  {journal} {\bibinfo
   {journal} {Phys. Rev. Lett.}\ }\textbf {\bibinfo {volume} {75}},\ \bibinfo
  {pages} {3537} (\bibinfo {year} {1995})}\BibitemShut {NoStop}%
\bibitem [{\citenamefont {Schollw\"ock}(2011)}]{mps3}%
  \BibitemOpen
  \bibfield  {author} {\bibinfo {author} {\bibfnamefont {U.}~\bibnamefont
  {Schollw\"ock}},\ }\bibfield  {title} {\bibinfo {title} {The density-matrix
  renormalization group in the age of matrix product states},\ }\href
  {https://doi.org/10.1016/j.aop.2010.09.012} {\bibfield  {journal} {\bibinfo
  {journal} {Annals of Physics}\ }\textbf {\bibinfo {volume} {326}},\ \bibinfo
  {pages} {96} (\bibinfo {year} {2011})}\BibitemShut {NoStop}%
\bibitem [{\citenamefont {Verstraete}\ and\ \citenamefont
  {Cirac}(2004)}]{peps}%
  \BibitemOpen
  \bibfield  {author} {\bibinfo {author} {\bibfnamefont {F.}~\bibnamefont
  {Verstraete}}\ and\ \bibinfo {author} {\bibfnamefont {J.~I.}\ \bibnamefont
  {Cirac}},\ }\href@noop {} {\bibinfo {title} {Renormalization algorithms for
  quantum-many body systems in two and higher dimensions}} (\bibinfo {year}
  {2004}),\ \Eprint {https://arxiv.org/abs/cond-mat/0407066}
  {arXiv:cond-mat/0407066 [cond-mat.str-el]} \BibitemShut {NoStop}%
\bibitem [{\citenamefont {Hauschild}\ \emph {et~al.}(2024)\citenamefont
  {Hauschild}, \citenamefont {Unfried}, \citenamefont {Anand}, \citenamefont
  {Andrews}, \citenamefont {Bintz}, \citenamefont {Borla}, \citenamefont
  {Divic}, \citenamefont {Drescher}, \citenamefont {Geiger}, \citenamefont
  {Hefel}, \citenamefont {H\'{e}mery}, \citenamefont {Kadow}, \citenamefont
  {Kemp}, \citenamefont {Kirchner}, \citenamefont {Liu}, \citenamefont
  {M\"{o}ller}, \citenamefont {Parker}, \citenamefont {Rader}, \citenamefont
  {Romen}, \citenamefont {Scalet}, \citenamefont {Schoonderwoerd},
  \citenamefont {Schulz}, \citenamefont {Soejima}, \citenamefont {Thoma},
  \citenamefont {Wu}, \citenamefont {Zechmann}, \citenamefont {Zweng},
  \citenamefont {Mong}, \citenamefont {Zaletel},\ and\ \citenamefont
  {Pollmann}}]{tenpy}%
  \BibitemOpen
  \bibfield  {author} {\bibinfo {author} {\bibfnamefont {J.}~\bibnamefont
  {Hauschild}}, \bibinfo {author} {\bibfnamefont {J.}~\bibnamefont {Unfried}},
  \bibinfo {author} {\bibfnamefont {S.}~\bibnamefont {Anand}}, \bibinfo
  {author} {\bibfnamefont {B.}~\bibnamefont {Andrews}}, \bibinfo {author}
  {\bibfnamefont {M.}~\bibnamefont {Bintz}}, \bibinfo {author} {\bibfnamefont
  {U.}~\bibnamefont {Borla}}, \bibinfo {author} {\bibfnamefont
  {S.}~\bibnamefont {Divic}}, \bibinfo {author} {\bibfnamefont
  {M.}~\bibnamefont {Drescher}}, \bibinfo {author} {\bibfnamefont
  {J.}~\bibnamefont {Geiger}}, \bibinfo {author} {\bibfnamefont
  {M.}~\bibnamefont {Hefel}}, \bibinfo {author} {\bibfnamefont
  {K.}~\bibnamefont {H\'{e}mery}}, \bibinfo {author} {\bibfnamefont
  {W.}~\bibnamefont {Kadow}}, \bibinfo {author} {\bibfnamefont
  {J.}~\bibnamefont {Kemp}}, \bibinfo {author} {\bibfnamefont {N.}~\bibnamefont
  {Kirchner}}, \bibinfo {author} {\bibfnamefont {V.~S.}\ \bibnamefont {Liu}},
  \bibinfo {author} {\bibfnamefont {G.}~\bibnamefont {M\"{o}ller}}, \bibinfo
  {author} {\bibfnamefont {D.}~\bibnamefont {Parker}}, \bibinfo {author}
  {\bibfnamefont {M.}~\bibnamefont {Rader}}, \bibinfo {author} {\bibfnamefont
  {A.}~\bibnamefont {Romen}}, \bibinfo {author} {\bibfnamefont
  {S.}~\bibnamefont {Scalet}}, \bibinfo {author} {\bibfnamefont
  {L.}~\bibnamefont {Schoonderwoerd}}, \bibinfo {author} {\bibfnamefont
  {M.}~\bibnamefont {Schulz}}, \bibinfo {author} {\bibfnamefont
  {T.}~\bibnamefont {Soejima}}, \bibinfo {author} {\bibfnamefont
  {P.}~\bibnamefont {Thoma}}, \bibinfo {author} {\bibfnamefont
  {Y.}~\bibnamefont {Wu}}, \bibinfo {author} {\bibfnamefont {P.}~\bibnamefont
  {Zechmann}}, \bibinfo {author} {\bibfnamefont {L.}~\bibnamefont {Zweng}},
  \bibinfo {author} {\bibfnamefont {R.~S.~K.}\ \bibnamefont {Mong}}, \bibinfo
  {author} {\bibfnamefont {M.~P.}\ \bibnamefont {Zaletel}},\ and\ \bibinfo
  {author} {\bibfnamefont {F.}~\bibnamefont {Pollmann}},\ }\bibfield  {title}
  {\bibinfo {title} {{Tensor network Python (TeNPy) version 1}},\ }\href
  {https://doi.org/10.21468/SciPostPhysCodeb.41} {\bibfield  {journal}
  {\bibinfo  {journal} {SciPost Phys. Codebases}\ ,\ \bibinfo {pages} {41}}
  (\bibinfo {year} {2024})}\BibitemShut {NoStop}%
\bibitem [{\citenamefont
  {Kitaev}(1995)}]{kitaev_1995_quantum_measurements_abelianstabilizer}%
  \BibitemOpen
  \bibfield  {author} {\bibinfo {author} {\bibfnamefont {A.~Y.}\ \bibnamefont
  {Kitaev}},\ }\href {https://arxiv.org/abs/quant-ph/9511026} {\bibinfo {title}
  {Quantum measurements and the {A}belian stabilizer problem}} (\bibinfo {year}
  {1995}),\ \Eprint {https://arxiv.org/abs/quant-ph/9511026}
  {arXiv:quant-ph/9511026 [quant-ph]} \BibitemShut {NoStop}%
\bibitem [{\citenamefont {van Enk}\ and\ \citenamefont
  {Beenakker}(2012{\natexlab{b}})}]{randomized_measurements_beenakker}%
  \BibitemOpen
  \bibfield  {author} {\bibinfo {author} {\bibfnamefont {S.~J.}\ \bibnamefont
  {van Enk}}\ and\ \bibinfo {author} {\bibfnamefont {C.~W.~J.}\ \bibnamefont
  {Beenakker}},\ }\bibfield  {title} {\bibinfo {title} {Measuring
  $\mathrm{Tr}{\ensuremath{\rho}}^{n}$ on single copies of $\ensuremath{\rho}$
  using random measurements},\ }\href
  {https://doi.org/10.1103/PhysRevLett.108.110503} {\bibfield  {journal}
  {\bibinfo  {journal} {Phys. Rev. Lett.}\ }\textbf {\bibinfo {volume} {108}},\
  \bibinfo {pages} {110503} (\bibinfo {year} {2012}{\natexlab{b}})}\BibitemShut
  {NoStop}%
\bibitem [{\citenamefont {Johri}\ \emph {et~al.}(2017)\citenamefont {Johri},
  \citenamefont {Steiger},\ and\ \citenamefont
  {Troyer}}]{entanglement_spectroscopy_troyer}%
  \BibitemOpen
  \bibfield  {author} {\bibinfo {author} {\bibfnamefont {S.}~\bibnamefont
  {Johri}}, \bibinfo {author} {\bibfnamefont {D.~S.}\ \bibnamefont {Steiger}},\
  and\ \bibinfo {author} {\bibfnamefont {M.}~\bibnamefont {Troyer}},\
  }\bibfield  {title} {\bibinfo {title} {Entanglement spectroscopy on a quantum
  computer},\ }\href {https://doi.org/10.1103/PhysRevB.96.195136} {\bibfield
  {journal} {\bibinfo  {journal} {Phys. Rev. B}\ }\textbf {\bibinfo {volume}
  {96}},\ \bibinfo {pages} {195136} (\bibinfo {year} {2017})}\BibitemShut
  {NoStop}%
\bibitem [{\citenamefont {Efron}\ and\ \citenamefont
  {Stein}(1981)}]{jacknife_resampling}%
  \BibitemOpen
  \bibfield  {author} {\bibinfo {author} {\bibfnamefont {B.}~\bibnamefont
  {Efron}}\ and\ \bibinfo {author} {\bibfnamefont {C.}~\bibnamefont {Stein}},\
  }\bibfield  {title} {\bibinfo {title} {{The Jackknife Estimate of
  Variance}},\ }\href {https://doi.org/10.1214/aos/1176345462} {\bibfield
  {journal} {\bibinfo  {journal} {The Annals of Statistics}\ }\textbf {\bibinfo
  {volume} {9}},\ \bibinfo {pages} {586 } (\bibinfo {year} {1981})}\BibitemShut
  {NoStop}%
\bibitem [{\citenamefont {Hoke}\ \emph {et~al.}(2023)\citenamefont {Hoke} \emph
  {et~al.}}]{mipt_hoke_2023}%
  \BibitemOpen
  \bibfield  {author} {\bibinfo {author} {\bibfnamefont {J.~C.}\ \bibnamefont
  {Hoke}} \emph {et~al.},\ }\bibfield  {title} {\bibinfo {title}
  {Measurement-induced entanglement and teleportation on a noisy quantum
  processor},\ }\href {https://doi.org/10.1038/s41586-023-06505-7} {\bibfield
  {journal} {\bibinfo  {journal} {Nature}\ }\textbf {\bibinfo {volume} {622}},\
  \bibinfo {pages} {481} (\bibinfo {year} {2023})}\BibitemShut {NoStop}%
\end{thebibliography}%
	
	\newpage
	\onecolumngrid
	\setcounter{equation}{0}
	\setcounter{figure}{0}
	\setcounter{table}{0}
	
	\renewcommand{\thefigure}{S\arabic{figure}}
	
	\renewcommand{\abstractname}{\vspace{+\baselineskip}}
	\makeatletter
	\renewcommand{\thesection}{\arabic{section}}
	\renewcommand{\thesubsection}{\thesection.\Alph{subsection}}
	\renewcommand{\thesubsubsection}{\alph{subsubsection}}
	\renewcommand{\thefigure}{S\@arabic\c@figure}
	\renewcommand{\theequation}{S\@arabic\c@equation}
	\renewcommand{\thetable}{S\@arabic\c@table}

	\begin{center}
		\textbf{Supplementary Materials for\\[4mm]
			\Large Observation of disorder-free localization and efficient disorder averaging \\ on a quantum processor} \\
		\vspace{5pt}
		Google Quantum AI and Collaborators\hyperlink{authorlist}{$\,^\dagger$}
	\end{center}
	%\vspace{3pt}
	
	\begin{center}
		\textbf{\large Contents}
	\end{center}
	%\vspace{1em}
	
	% Define commands for TOC entries to keep the code clean and aligned.
	\newcommand{\tocitem}[3]{\noindent\hyperref[#3]{#1 #2}\dotfill\pageref{#3}\par}
	\newcommand{\subtocitem}[3]{\noindent\hspace{1.5em}\hyperref[#3]{#1 #2}\dotfill\pageref{#3}\par}
	\newcommand{\subsubtocitem}[3]{\noindent\hspace{3em}\hyperref[#3]{#1 #2}\dotfill\pageref{#3}\par}
	
	\tocitem{1.}{List of symbols}{sec:symbols}
	\tocitem{2.}{Experimental techniques and device characterization}{sec:exp_tech}
	\subtocitem{A.}{Overview}{ssec:overview}
	\subtocitem{B.}{Error Suppression and mitigation}{ssec:mitigation}
	\subsubtocitem{a.}{Measurement in the dual basis and postselection}{sssec:postselection}
	\subsubtocitem{b.}{Second-order Trotter circuits and compilation with CPhase gates}{sssec:cphase}
	\subsubtocitem{c.}{Multi-layer gauge compiling}{sssec:gauge_compiling}
	\subsubtocitem{d.}{Uncorrelated readout mitigation}{sssec:readout}
	\subsubtocitem{e.}{Global depolarizing channel correction}{sssec:depolarizing}
	\tocitem{3.}{Further experimental data}{sec:further_data}
	\subtocitem{A.}{Dynamics of energy perturbation in 1d}{ssec:data_1d}
	\subtocitem{B.}{Dynamics of energy perturbation in 2d}{ssec:data_2d}
	\subtocitem{C.}{Matter imbalance}{ssec:imbalance}
	\tocitem{4.}{Design of the circuit and mapping to fixed disorder}{sec:design-circ-mapp}
	\subtocitem{A.}{\(\mathbb{Z}_2\) gauge theory with \(\mathbb{Z}_2\) matter}{ssec:z2_gauge}
	\subtocitem{B.}{A device-efficient Hamiltonian}{ssec:device_hamiltonian}
	\subtocitem{C.}{No-flux sectors and models with symmetry}{ssec:no_flux}
	\tocitem{5.}{Simulation techniques and numerical results}{sec:simulations}
	\subtocitem{A.}{Exact statevector simulation}{ssec:sim_statevector}
	\subtocitem{B.}{ED results}{ssec:sim_ed}
	\subsubtocitem{a.}{ED using disordered mixed-field Ising model}{sssec:ed_ising}
	\subsubtocitem{b.}{Tunable DFL in the matter imbalance}{sssec:ed_imbalance}
	\subtocitem{C.}{Matrix Product States (MPS) simulations}{ssec:mps_simulations}
	\subsubtocitem{a.}{Approach to MPS simulations of digital circuits}{sssec:MPS_sim_method}
	\subsubtocitem{b.}{Scaling of MPS simulation cost with system size}{sssec:MPS_scaling_cost}
	\tocitem{6.}{Grover's algorithm and DFL}{sec:grover}
	\tocitem{7.}{Second R\'enyi entropy}{sec:renyi_entropy}
	\subtocitem{A.}{Measurement}{ssec:renyi_measurement}
	\subtocitem{B.}{Numerical results}{ssec:renyi_numerical}
	
	\vspace{2em}

	\section{List of symbols}
	\label{sec:symbols}
	\renewcommand{\arraystretch}{1.5}
	\begin{center}
		\begin{tabular}{ |l|l| } 
			\hline
			Symbol & Description\\ 
			\hline
			$N$ & Total number of qubits \\
			$d$ & Spatial dimension \\
			$\hat H_{\mathrm{LGT}}$ & Lattice Gauge Theory (LGT) Hamiltonian \\
			$\hat H_{\textrm{ord}}$ & Ordered part of a Hamiltonian $\hat{H}$ \\
			$\hat H_{\textrm{dis}}$ & Disordered part of a Hamiltonian $\hat{H}$  given by $\hat H_{\mathrm{dis}} = \sum_j g_j \hat D_j$\\
			$g_j$ & Binary disorder at site $j$ \\
			$\mathbf{g}$ & $\{g_j\}$, a particular disorder realization\\
			$\hat{O}$ & An observable whose disorder-average we are interested in measuring \\
			$\hat H^Q$ & Hamiltonian $\hat{H}$ obtained by promoting disorder variables $g_j$'s to ancillary qubits \\
			$J$ & Coefficient of the $\hat{\sigma}^Z_j\, \hat{Z}_{j,k}\, \hat{\sigma}^Z_{k}$ term in the LGT \\
			$h$ & Coefficient of the $\hat X_{j,k}$ term in the LGT \\
			$\mu$ & Coefficient of the $\hat{\sigma}^{X}_j$ term in the LGT \\
			$\hat X_{j,k}$, $\hat Y_{j,k}$, $\hat Z_{j,k}$ & Spin-1/2 Pauli operators for the qubit on link $(j,k)$ \\
			$\hat \sigma^X_j$, $\hat \sigma^Y_j$, $\hat \sigma^Z_j$ & Spin-1/2 Pauli operators for the qubit on vertex $j$ \\
			$\hat{\mathbf{x}},\hat{\mathbf{y}},\hat{\mathbf{z}}$ & Unit vectors in the Bloch sphere \\
			$\ket{\pm x},\ket{\pm y}, \ket{\pm z}$ &$\pm 1$ eigenstates of Pauli $X$,$Y$,$Z$ matrices \\
			$\ket{\psi}$ & A generic state in the Hilbert space \\
			$\hat U_{\textrm{LGT}}$ & Trotterized time evolution operator for the lattice gauge theory (LGT) model \\
			$\hat U_{J}$ & Time evolution operator corresponding to the  $\hat \sigma^Z_j\,\hat Z_{j,k}\,\hat \sigma^Z_{k}$ term in the LGT\\
			$\hat U_{h}$ & Time evolution operator corresponding to $\hat{X}_{j,k}$ term in the LGT \\
			$\hat U_{\mu}$ & Time evolution corresponding to $\hat \sigma^X_j$ in the LGT \\
			$N_m$ & Number of matter qubits, equal to the number of gauge symmetry operators \\
			$N_g$ & Number of gauge qubits \\
			$\hat G_j$ & The generator of gauge invariant transformation,  conserved operator centered at matter qubit $j$ \\
			$\hat U_B$ &  Unitary that transforms a state in the original representation (with static ancillas) to the LGT  \\
			$\Delta t$ & Trotter step size\\
			$\hat H_{j,k}$ & Term contributing to the total Hamiltonian corresponding to link $(j,k)$ \\
			$\epsilon_{j,k}$ & Energy per link; expectation value of $\hat H_{j,k}$ \\
			$\ket{\hat{\mathbf{v}}}$  & Unit vector in the Bloch sphere that points along direction $\mathbf{v} =   J\hat{\mathbf{z}} + h \hat{\mathbf{x}}$\\
			$L$ & Size of a subsystem consisting of $L$ contiguous qubits in 1d\\
			$S^{(2)}(L)$ & Second R\'enyi Entropy of a subsystem with $L$ qubits\\
			$\ket{\theta}$ & Single-qubit state in the $XZ$-plane of the Bloch sphere that makes an angle $\theta$ with the X-axis \\
			$\mathcal{I}(\theta, \text{cycle})$ & Matter imbalance at a given cycle after evolving an initial state parametrized by angle $\theta$ \\
			$\mathcal{R}_{q}$ & Confusion/Response matrix for qubit $q$ obtained by sampling random bitstrings 
			\\
			$\chi$ & Maximum bond dimension of a matrix product state (MPS)\\ 
			\hline
		\end{tabular}
	\end{center}
	
	\newpage
	\section{Experimental techniques and device characterization}
	\label{sec:exp_tech}
	\subsection{Overview}
	\label{ssec:overview}
	
\begin{figure*}[h!]
		\centering
		\includegraphics{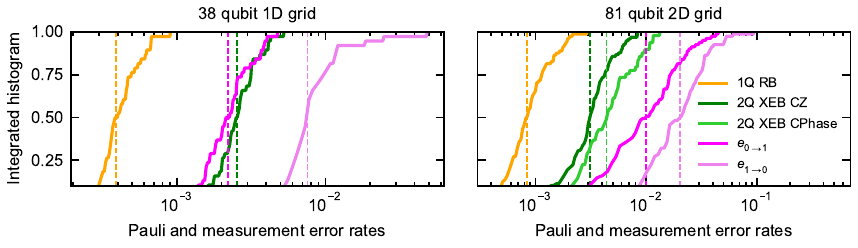}
		\caption{Typical error rates achieved on a 38-qubit 1d grid on the 72-qubit Willow chip and  81-qubit 2d grid on the 105-qubit Willow chip, showing readout (measurement) errors, single-qubit randomized benchmarking (1Q RB), and two-qubit cross-entropy benchmarking (2Q XEB) Pauli errors for both CZ and CPhase gates. The corresponding medians are shown with dashed lines. $e_{0\to 1}$ and $e_{1\to 0}$ are the measurement errors qubits prepared in 0 and 1, respectively. }
		\label{sfig:error_rates}
	\end{figure*}

	The experimental results for energy perturbation dynamics reported in this paper were obtained using 72- and 105-qubit Willow chips, similar in architecture to the chips used to demonstrate logical error suppression below the surface code threshold \cite{ google_y1}. We employed either CZ  (for 1d experiments) or a combination of CZ and CPhase gates (for 2d experiments, similar to Ref. \cite{will2025_probing_nonequilibrium_topologicalorder}) as the native entangling gates \cite{Google_M2_2023, google_y1,Foxen_2020_demonstrating_twoq_gates}. Figure~\ref{sfig:error_rates} shows the typical error rates for a 38-qubit 1d grid with CZ gate and an 81-qubit 2d grid that implements both CZ and CPhase gates. Further, the entropy (Fig.~4) and matter imbalance (Fig.~\ref{sfig:imbalance_exp_vs_sim}) measurements were performed on a different 72-qubit Sycamore chip \cite{Google_M2_2023} with CZ gates (error rates similar to Ref.~\cite{cochran_2dlgt}).  Additional information about gate and readout optimization protocols is provided in Refs.~\cite {optimizing_klimov_2024, model-based_optimization_bengtsson_2024}.  
	
	\subsection{Error Suppression and mitigation}
	\label{ssec:mitigation}
	
	\subsubsection{Measurement in the dual basis and postselection}
	\label{sssec:postselection}
	The Floquet unitary operator in our LGT model has $N_m$ local conserved operators $\hat G_j$. Thus, if we start in a fixed sector labeled by the charges $\mathbf{g} = \{ g_1, g_2, \cdots, g_{N_m}\}$, ideal unitary dynamics should keep the state in this charge sector. We exploit these local symmetries to postselect the measured bitstrings for the single-sector initial state in both 1d (Fig.~1) and 2d (Fig.~3).

\begin{figure*}[t!]
		\centering
		\includegraphics{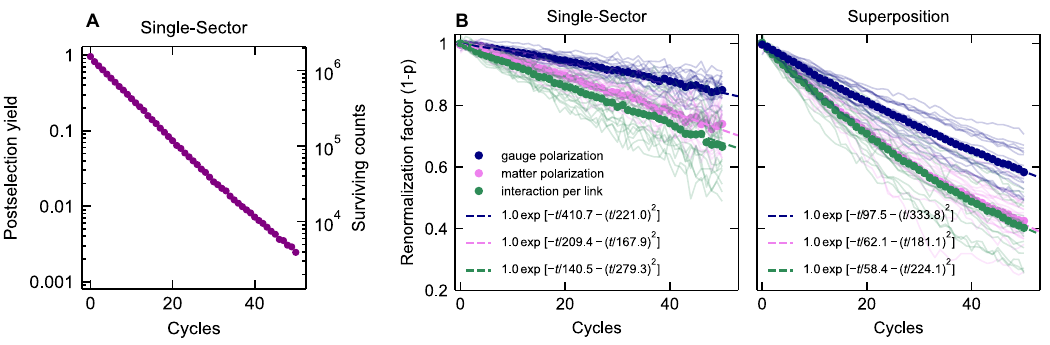}
		\caption{(\textbf{A}) Total postselection yield for the locally-perturbed 1d single-sector initial state shown in Fig.~1B of the main text, where the measurements were performed in the dual basis with the matter qubits always measured in the X-basis, whereas the gauge qubits were measured in X- or Z- basis, and postselected such that each of the $N_m=19$ matter qubits is measured in the $\ket{+}$ state, as expected for an ideal circuit. (\textbf{B})  Rescaling factors ($1-p$) that were used to error-mitigate the gauge polarization $\expval{\hat X_{j,j+1}} $, matter polarization $\expval{\hat \sigma^X_j} $  and interaction per link $\expval{\hat \sigma^Z_{j} \hat Z_{j,j+1} \hat \sigma^Z_{j+1}}$, obtained by measuring the respective quantities for a similar circuit, but with Trotter step $\Delta t$ set to 0. Site-resolved traces are shown in faint colors; means are shown with circles, then fitted to the function $y(t) = \exp[-t/a -(t/b)^2]$ as a function of cycle number $t$, shown by dashed lines.  The matter polarization in the superposition initial state is initially zero, so the single-sector values without postselection are used for rescaling.}
		\label{sfig:error_mitigation_1d}
	\end{figure*}

	We noted in the main text that a unitary transformation ($\hat{U}_B$) relates the LGT model to the disordered model augmented with ancillas, which we refer to as the dual model. We will provide further details about the connection between these two models in \cref{sec:design-circ-mapp}, but here we note the following mapping between the local observables
	\begin{align}
		&\hat{G}_j \rightarrow \sigma^X_j \nonumber \\
		&\hat{X}_{j,k} \rightarrow \hat{X}_{j,k} \nonumber \\
		&\sigma^{X}_j \rightarrow \hat{G}_j =  \hat \sigma^X_j \prod_{j\in N(j)} \hat X_{j,k} = g_j\prod_{j
			\in N(j)} \hat X_{j,k} \nonumber \\
		&\hat \sigma^Z_j \hat Z_{j,k} \hat \sigma^Z_k \rightarrow \hat Z_{j,k} 
	\end{align}
	To estimate the linear observables of interest in the LGT model, namely the gauge polarization $\hat X_{j,k}$, matter polarization $\hat \sigma^X_j$ and the interaction per link  $\hat \sigma^Z_j \hat Z_{j,k} \hat \sigma^Z_k$, intuitively, one would measure the time-evolved wavefunction directly in the X or the Z basis. Since the conserved operators $\hat G_j$ are products of Pauli-X operators, we would be able to postselect only the $X$-basis measurements. However, here, we measure in the dual basis (related to the LGT basis by the basis change $\hat U_B$). This maps the gauge-symmetry operators onto the matter qubits, i.e., $\hat G_j \rightarrow \hat \sigma^X_j$ and all the Hamiltonian terms onto the gauge qubits, so we can measure all conserved charges simultaneously with all the Hamiltonian terms. This allows us to postselect for all the observables shown in Figs.~\ref{sfig:all_energy_data_1d} and \ref{sfig:all_energy_data_2d} for the single-sector initial states. We show the total postselection yield for the single-sector initial state in 1d in \cref{sfig:error_mitigation_1d}A.
	
\begin{figure*}[b!]
		\centering
		\includegraphics{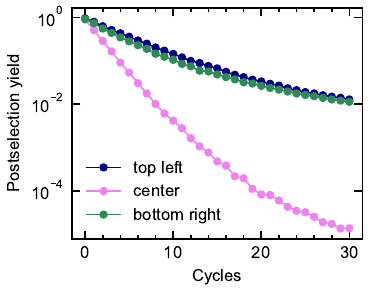}
		\caption{Postselection yields for the locally-perturbed 2d single-sector initial state shown in Fig. 3A of the main text, where the measurements were performed in the Pauli-$X$ basis, and postselected such that the matter qubits less than a Manhattan distance of 7 from the indicated gauge site are measured in the $\ket{+}$ state, as expected in the ideal circuit. The postselection yield for the center qubit is lower than that of the corner qubits since the center qubit has more matter qubits within a given Manhattan distance to postselect on.}
		\label{sfig:error_mitigation_2d}
	\end{figure*}

	For the 2d experiment, because the system has $N_m = 32$ gauge symmetry operators, we are not able to take enough shots to postselect on all of the $\hat G_j$. Instead, we throw out shots where an error occurs less than a Manhattan distance of 7 away from the operator that we are measuring. In \cref{sfig:error_mitigation_2d}, we show the postselection yield for measuring gauge polarization $\hat X_{j,k}$ corresponding to the gauge qubits on the top-left corner, bottom-right corner, and the center of the grid. As expected, our Manhattan-distance-based local postselection protocol yields a higher rate for the corner qubits than the center qubit, since the center qubit has more gauge symmetry operators to postselect on within a given distance.

	\subsubsection{Second-order Trotter circuits and compilation with CPhase gates}
	\label{sssec:cphase}
\begin{figure*}[h!]
		\centering
		\includegraphics{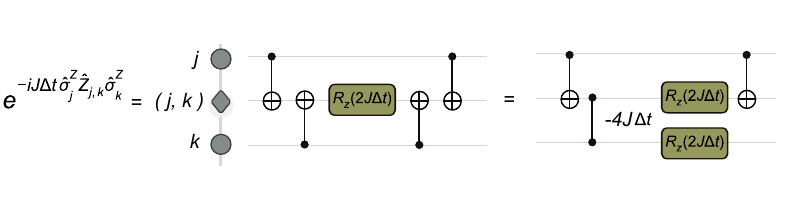}
		\caption{Implementation of the matter-gauge-matter interaction in the LGT model via CZ and CPhase gates. CNOT-Rz-CNOT layers can be written as CPhase-Rz, reducing the depth of the circuit from 4 to 3 entangling gates for implementing this term.}
		\label{sfig:cz_and_cphase}
	\end{figure*}

	The Floquet circuits shown in Fig.~1 and Fig.~3 of the main text can be written efficiently by having both CZ and CPhase entangling gates in the same grid. In \cref{sfig:cz_and_cphase}, we show the implementation of the three-qubit interaction term in terms of CZ and CPhase, with the CPhase implementation requiring one fewer entangling gate. For the 1d case, CPhase implementation has 3 instead of 4 entangling layers per Floquet unitary. Likewise, for the 2d case, CPhase implementation has 6 instead of 8 entangling layers per Floquet unitary. Although the median 2-qubit XEB of the CZ gate is lower than that of the CPhase, as shown in \cref{sfig:error_rates}, the shorter depth of the circuit gives an advantage, enabling coherent simulation until about 30 cycles for the 2d grid consisting of 81 qubits.
	
	We further note that we implement the second-order Trotter cycle at no additional cost in terms of circuit depth. As described in the main text, the second-order Trotter unitary is given by:
	\begin{align}
		\hat{U}_{\mathrm{LGT}}(\Delta t) = \hat U_J(\Delta t/2) \; \hat U_h(\Delta t)  \; \hat U_{\mu}(\Delta t) \; \hat U_J(\Delta t/2),
		\label{eqn:flouqet_unitary}
	\end{align}
	where,
	\begin{align}
		\hat U_J(\Delta t) &= \exp\Big(\hspace{-1mm}-i J\Delta t \sum_{\langle j, k \rangle} \hat \sigma^Z_j\,\hat Z_{j,k}\,\hat \sigma^Z_{k} \Big), \nonumber\\
		\hat U_h(\Delta t) &= \exp\Big(\hspace{-1mm} -i  h\Delta t {\sum_{\langle j, k \rangle} \hat X_{j,k}}\Big), \nonumber\\
		\hat U_\mu (\Delta t) &= \exp\Big(\hspace{-1mm}-i\mu\,\Delta t \hspace{1mm} {\sum_{j} \hat \sigma^X_j }\Big).
	\end{align}
	Furthermore, we express $\hat U_J(\Delta t/2)$ in terms of $\hat U_B$ and single qubit rotations as 
	\begin{equation}
		\hat U_J(\Delta t/2) = \hat{U}_B  \mathcal{R}_z  \hat{U}_B,
	\end{equation}
	where $\mathcal{R}_z$ is the product of Z-rotations, i.e., $R_z(J\Delta t)$ on all the gauge qubits. Consecutive layers of $\hat U_{LGT}$ can thus be written as
	\begin{align}
		\cdots {\color{blue}\hat{U}_{LGT}} {\color{red}\hat{U}_{LGT}} \cdots &= \cdots {\color{blue} \hat{U}_B  \mathcal{R}_z  \hat{U}_B U_h  \; \hat U_{\mu} \hat{U}_B  \mathcal{R}_z}  \underbrace{{\color{blue} \hat{U}_B} {\color{red} \hat{U}_B}}_{\hat \openone}  {\color{red} \mathcal{R}_z  \hat{U}_B U_h  \; \hat U_{\mu}  \hat{U}_B  \mathcal{R}_z  \hat{U}_B} \cdots \\
		&= \cdots {\color{blue} \hat{U}_B  \mathcal{R}_z  \hat{U}_B \underbrace{U_h  \; \hat U_{\mu}  {\color{blue}\hat{U}_B  \mathcal{R}_z} {\color{red}   \mathcal{R}_z  \hat{U}_B}}_{\color{violet} \textrm{repeating unit}} U_h  \; \hat U_{\mu}}  {\color{red}\hat{U}_B  \mathcal{R}_z  \hat{U}_B} \cdots \\
		&= \cdots {\color{violet} (\hat{U}_h \hat{U}_{\mu} \hat{U}_B \mathcal{R}_Z^2 \hat{U}_B }) {\color{violet} (\hat{U}_h \hat{U}_{\mu}\hat{U}_B \mathcal{R}_Z^2 \hat{U}_B })\cdots 
	\end{align}
	The repeating unit in the middle of the circuit, shown in purple, is the first-order Trotter unitary.
	Thus, the typical Floquet layer is the same for both first and second-order Trotter, and the two circuits differ only at the beginning and the end. The first layer of $\hat{U}_B$ in the second-order Trotter cancels the initial state preparation circuit's $\hat{U}_B$. Since we perform the energy spreading measurements in the dual basis, which can be done by applying $\hat U_B$ at the end of the circuit, this further cancels the last $\hat U_B$ occurring in the last layer of the second-order Trotter.

	\subsubsection{Multi-layer gauge compiling}
	\label{sssec:gauge_compiling}
	We utilize gauge compiling \cite{randomized_compiling_wallman_2016,quantumlib_cirq_gauge_compiling_2024, hashim_randomized_compiling_2021} for all of the experimental results presented in this paper. For Figs.~1 and 3, we implement multi-layer gauge compiling, meaning that when there are consecutive CZ layers, we twirl these layers together instead of inserting single-qubit layers between them. 
	
	For the results shown in Fig.~1 of the main text, we collected 20000 and 5000 shots per randomized instance for single-sector and superposition initial states, respectively. For the results corresponding to the superposition initial state in Fig.~3, where no postselection was performed, we measured 1000 shots per instance. However, we used postselection for the single-sector initial state. The number of shots for the postselected results depends exponentially on the number of cycles- ranging from $1000$ at cycle 0 to $682846$ at cycle 30. Likewise, $20000$ shots per instance were measured for the matter imbalance results shown in \cref{sfig:imbalance_exp_vs_sim}. We find that using gauge compiling leads to a smooth exponential decay of Pauli observables under the Floquet unitary corresponding to Trotter step $\Delta t$ equal to 0, which is an identity operation in the absence of noise (shown in \cref{sfig:error_mitigation_1d}, \cref{sfig:error_mitigation_2d}). It also lowers the raw value of the experimentally measured second R\'enyi entropy, allowing for error mitigation by subtracting the uniform background entropy\,(see Sec.~\ref{sec:renyi_entropy}).

	\subsubsection{Uncorrelated readout mitigation}
	\label{sssec:readout}
	Assuming uncorrelated readout errors, the error model is parametrized by two probabilities for each qubit \(q\): \(\epsilon_{q,0}\) is the probability that \(1\) is read out for a qubit in the \(0\) state (written \(0 \to 1\)), and \(\epsilon_{q,1}\) is the probability for a \(1 \to 0\) error. To estimate these probabilities, as we run circuits for a given experiment we interleave circuits that prepare random known bitstrings (by randomly choosing whether to apply \(X\) on a given bit). This assumes that the effect of errors from single-qubit gates is much smaller than readout error, as confirmed by randomized benchmarking. From these measurements, we construct the $N$-qubit confusion matrix (also known as the response matrix):
	\begin{equation}
		\mathcal{R} = \mathcal{R}_{q_1} \otimes \cdots \otimes \mathcal{R}_{q_N}, \quad \mathcal{R}_{q} = \begin{pmatrix} 1 - \epsilon_{q,0} & \epsilon_{q,1} \\ \epsilon_{q, 0} & 1 - \epsilon_{q,1},
		\end{pmatrix}
	\end{equation}
	which satisfies $\mathcal{R} \vec p_{\rm ideal} = \vec p_{\rm noisy}$, where $\vec p_{\rm ideal}$ are the probabilities of measurement outcomes in the absence of readout errors and $\vec p_{\rm noisy}$ are the corresponding probabilities in the presence of readout errors. Because of the tensor product structure, one can efficiently invert $\mathcal{R}$ and apply the inverse to measured probabilities on small subsets of qubits (in this work, we apply readout error mitigation to single- and three-site observables). In addition, for all observables that share this tensor product structure, we can perform efficient readout mitigation that scales only linearly in the number of shots and number of qubits. We use the implementation in Ref.~\cite{cirq_readout_mitigation}. We perform uncorrelated readout error mitigation for the local quantities in Figures 1 and 3 and the corresponding SM figures when we do not perform postselection.
	
	\subsubsection{Global depolarizing channel correction}
	\label{sssec:depolarizing}
	After the application of randomized compiling, we find that the errors in the expectation values of Pauli strings such as $\hat X_{j,k}$, $\hat \sigma^X_{j}$, $\hat \sigma^Z_{j} \hat Z_{j,k} \hat \sigma^Z_{k}$ can largely be modeled as incoherent errors. This can be seen through the exponential decay of gauge polarization $\langle\hat X_{j,j+1}\rangle$, matter polarization $\langle\hat \sigma_j^X\rangle$, and the interaction term $\langle\hat \sigma^Z_{j} \hat Z_{j,j+1} \hat \sigma^Z_{j+1}\rangle$ shown in Fig.~\ref{sfig:error_mitigation_1d}B. We model these incoherent errors with a global depolarizing channel given by:
	\begin{equation}
		\label{seqn:depolarizing_channel}
		\mathcal{D}(\hat \rho) = (1-p) \hat \rho + \frac{p}{2^N} \mathbb{\hat \openone},
	\end{equation}
	where $p$ is the depolarization probability, and $\hat \openone$ is the $2^N$ dimensional identity matrix in the Hilbert space of $N$ qubits. We do not know \textit{a priori} the depolarization probability $p$, hence we characterize it by measuring the expectation value of the corresponding Pauli observable when there are supposed to be no dynamics by setting the Trotter step $\Delta t$ to $0$. Further discussion of this and related error mitigation methods can be found in Ref.~\cite{quantum_error_mitigation}. The expectation value of a Pauli observable $\hat O$ under the action of depolarizing channel $\mathcal{D}$ is given by:
	\begin{align}
		\langle\hat O\rangle_{\text{noisy}} &= \Tr[\hat O {\hat \rho_{\text{noisy}}}]  \nonumber \\
		&= \Tr[\hat O \mathcal{D}\left( \hat \rho_{\text{exact}} \right)] \nonumber \\
		&= (1-p) \Tr[\hat O \hat \rho_{\text{exact}}] \nonumber \\
		&= (1-p)\langle\hat O\rangle_{\text{exact}} \\
		\implies \langle\hat O\rangle_{\text{exact}} &= \frac{\langle\hat O\rangle_{\text{noisy}}}{\left(1-p \right)},
	\end{align}
	In the third line, we use the fact that Pauli operators are traceless. Thus, we obtain the error-mitigated signal by rescaling the experimental signal by $1/(1-p)$. A comparison between error-mitigated experiment vs.~noiseless simulation for the dynamics of a local perturbation in Fig.~\ref{sfig:all_energy_data_1d} and Fig.~\ref{sfig:all_energy_data_2d} shows that the error-mitigated values closely match the noiseless simulation values. This technique was applied to both initial states: single-sector and superposition. For single-sector results, we first used postselection and then rescaled utilizing this technique. Furthermore, the matter polarization $\expval{\hat \sigma^X_j}$ in the superposition state is initially zero, making the characterization of the depolarizing channel impossible with this method. Thus, we use the matter polarization of the single-sector initial state but without postselection to compute the renormalization factor for this case in 1d. In contrast, we found that matter-polarization rescaling for the superposition initial state with the single-sector measurements did not work well in the 2d case, possibly because, when using CPhase gates, the error channels are more different between the circuit of interest and the one with $\Delta t = 0$, so the superposition matter polarization result in \cref{sfig:all_energy_data_2d} is not rescaled.
	
	\newpage
	\section{Further experimental data}
	\label{sec:further_data}
	\subsection{Dynamics of energy perturbation in 1d}
	\label{ssec:data_1d}
	\begin{figure*}[!h]
		\centering
		\includegraphics{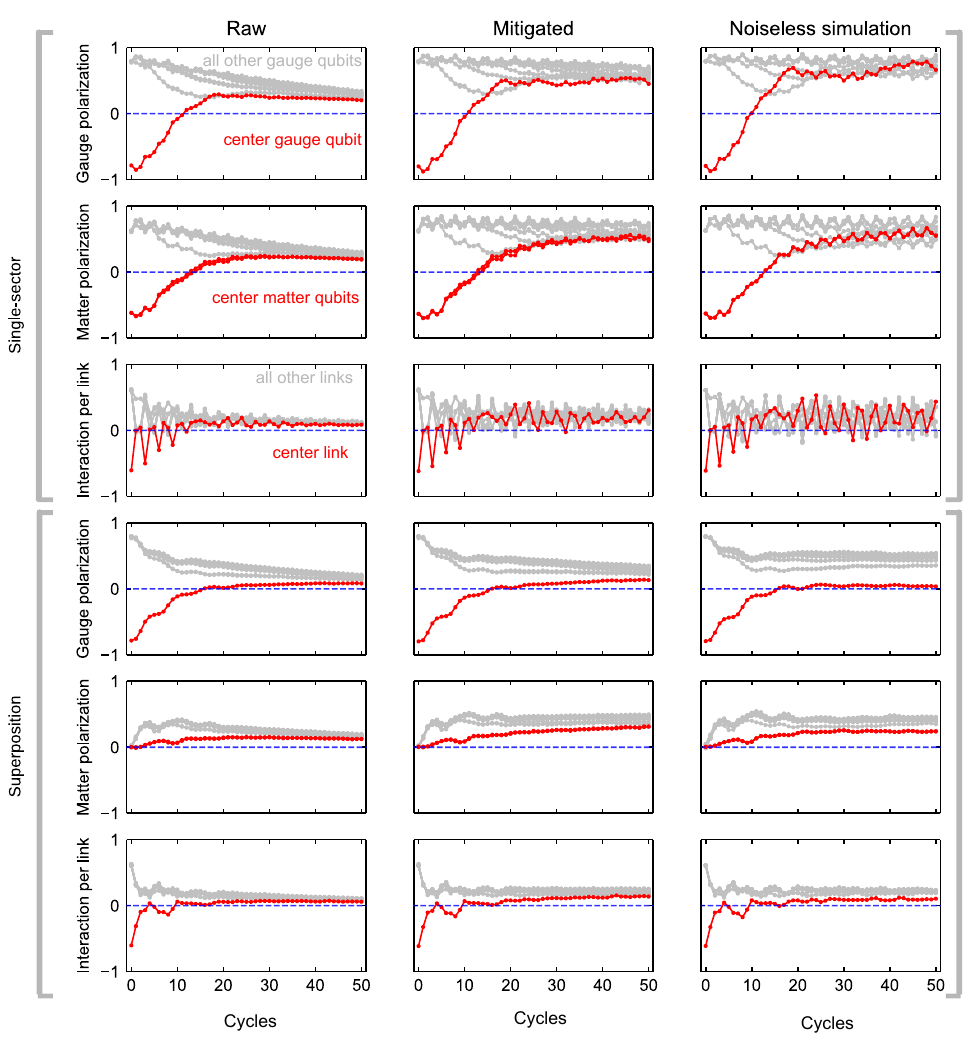}
		\caption{(\textbf{A}) Raw, (\textbf{B}) error-mitigated experiment, and (\textbf{C}) noiseless simulation data for the dynamics of the perturbed initial state in 1d, as shown in Fig.~1 of the main text. For each initial state, we show the expectation values of three terms in the Floquet unitary (and the underlying Hamiltonian), i.e., gauge polarization $\langle\hat X_{j,j+1}\rangle$, matter polarization $\langle\hat \sigma_j^X\rangle$, and the interaction term $\langle\hat \sigma^Z_{j} \hat Z_{j,j+1} \hat \sigma^Z_{j+1}\rangle$. Since the initial states are perturbed at the center, the center qubits (red) have negative values compared to the rest (gray). The simulation was done in the dual picture by fixing $N_m=19$ gauge charges and working only with the gauge qubits. The simulation results for the superposition initial state were obtained by sampling over 2000 gauge charges $\mathbf{g}$, chosen randomly from a uniform distribution.}
		\label{sfig:all_energy_data_1d}
	\end{figure*}
	
	\subsection{Dynamics of energy perturbation in 2d}
	\label{ssec:data_2d}
	\begin{figure*}[!h]
		\centering
		\includegraphics{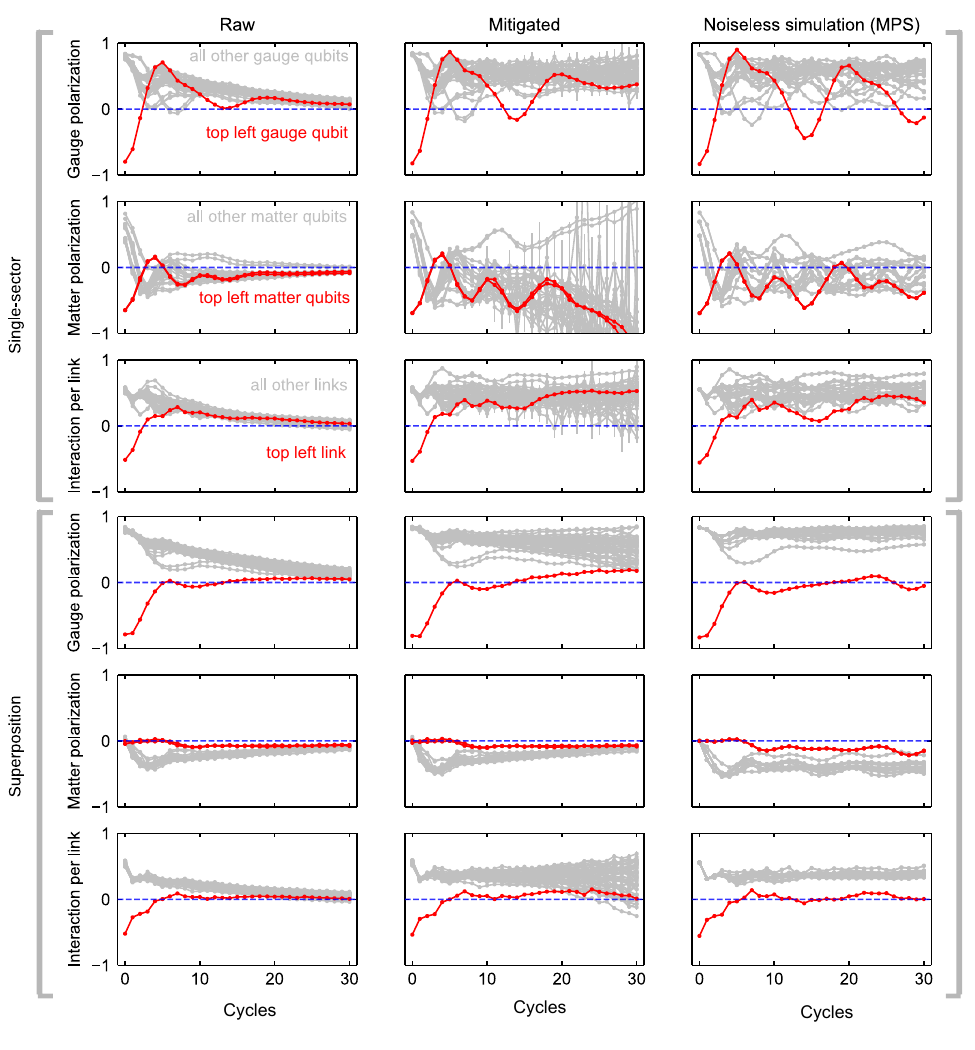}
		\caption{(\textbf{A}) Raw, (\textbf{B}) error-mitigated experiment, and (\textbf{C}) noiseless simulation data for the dynamics of the perturbed initial state in 2d, as shown in Fig.~3 of the main text. For each initial state, we show the expectation values of three terms in the Floquet unitary (and the underlying Hamiltonian), i.e., gauge polarization $\langle\hat X_{j,j+1}\rangle$, matter polarization $\langle\hat \sigma_j^X\rangle$, and the interaction term $\langle\hat \sigma^Z_{j} \hat Z_{j,j+1} \hat \sigma^Z_{j+1}\rangle$. Since the initial states are perturbed at the top left link, the top left qubits (red) have negative values compared to the rest (gray). The simulations were performed using matrix product states (MPS) with a maximum bond dimension $\chi = 1024$; the single-sector initial state was simulated in the dual basis with 49 qubits, while the superposition initial state was simulated in the full LGT basis with 81 qubits.  Further details about the MPS simulations are provided in \cref{sec:simulations}C.}
		\label{sfig:all_energy_data_2d}
	\end{figure*}
	
	\newpage
	In Fig.~\ref{sfig:all_energy_data_1d} and Fig.~\ref{sfig:all_energy_data_2d}, we show a comparison between raw (A), error mitigated (B), and  (C) noiseless simulation results for the dynamics of the perturbed initial states shown in Fig.~1 and Fig.~3 of the main text, respectively. We utilized the dual representation of the LGT in the form of a disordered mixed field Ising model, with the disorder set by gauge charges, as discussed in \cref{sec:design-circ-mapp} for the noiseless simulation. This allows us to simulate the 38 qubit 1d ring with only $N_g=19$ qubits by fixing $N_m=19$ gauge charges $\mathbf{g}$. Since the uniform superposition initial state consists of $2^{N_m}$ gauge configurations, we sample over $2000$ random samples. Similarly, for the 2d simulations of the single-sector initial state, we fixed $N_m=32$ matter charges, simulated only $N_g=49$ qubits on a dual basis using matrix product states (MPS). In contrast, we simulated the full LGT model for the superposition initial state.  An exponential decay of the signal strength shown in Fig.~\ref{sfig:error_mitigation_1d}B, Fig.~\ref{sfig:error_mitigation_2d}, and a good agreement of the error-mitigated experimental results with noiseless simulation in Fig.~\ref{sfig:all_energy_data_1d} and Fig.~\ref{sfig:all_energy_data_2d} suggest the reasonableness of the error mitigation techniques used in this paper.

	\subsection{Matter imbalance}
	\label{ssec:imbalance}
	\begin{figure*}[!h]
		\centering
		\includegraphics{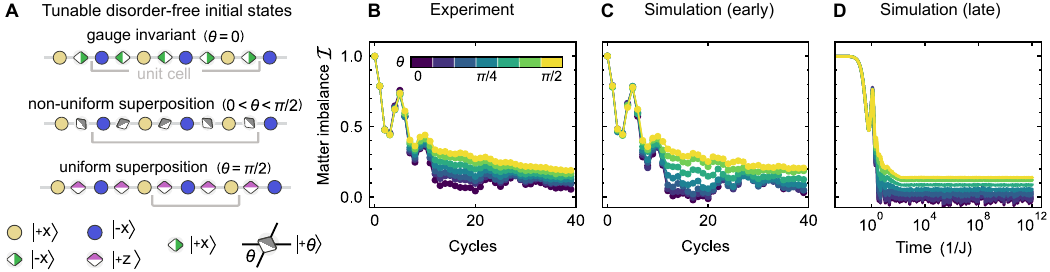}
		\caption{(\textbf{A}) Parameterized disorder-free initial product states. Keeping the matter sites in a staggered configuration, the gauge qubits are prepared in an eigenstate of Pauli-X (single-sector), Pauli-Z (uniform superposition) or some non-uniform superposition state parameterized by angle $\theta$. The unit cell indicator lines highlight the pattern that is repeated. (\textbf{B}) Experimentally measured matter imbalance, $\mathcal{I}$, with respect to the initial staggered configuration plotted as a function of the initial state parameter $\theta$, for a $24$ qubit open chain with Floquet parameters $J=1, \, \Delta t =0.25, \, h=2.2$ and  $\mu=2.0 $ and using first-order Trotterization. (\textbf{C}) Corresponding results obtained from exact statevector simulation. (\textbf{D}) Results for exponentially long times obtained using exact diagonalization (ED) in the Hamiltonian limit ($\Delta t \rightarrow 0$).}
		\label{sfig:imbalance_exp_vs_sim}
	\end{figure*}
	
	Having provided evidence for DFL in 1d and 2d for two distinct, yet both translationally invariant, initial states, we next show signatures of DFL for a tunable family of initial states in the dynamics of matter imbalance. In Fig.~\ref{sfig:imbalance_exp_vs_sim}A, we describe a class of parameterized product states that are disorder-free for all values of a parameter $\theta$:
	\vspace{-1mm}
	\begin{equation}
		\ket{\psi_0(\theta)}{=} \bigotimes_{j=1} ^ {N_m} \ket{({-}1)^j x} \bigotimes_{j=1}^{N_g} \ket{\cos\theta\,\hat{\mathbf{x}}_{j,j+1}{+}\sin\theta\,\hat{\mathbf{z}}}.
	\end{equation}
	
	\noindent These states have a staggered pattern on the matter sites (alternating pattern of blue and yellow circles in Fig.~\ref{sfig:imbalance_exp_vs_sim}A), $\hat{\bold{x}}_{j,j+1}=\sqrt{2}\sin(\pi/4+\pi j/2)\hat{\bold{x}}$, and the angle that the Bloch vector of each of the gauge qubits makes with the equatorial plane of the Bloch sphere is $\theta$. Here, we choose to measure the imbalance $\mathcal{I}$:
	\vspace{-1mm}
	\begin{equation}
		\mathcal{I}\, (\theta, \text{cycle}) = \frac{1}{N_m} \sum_{j=1}^{N_m}  \expval{\sigma_j^X (\text{cycle})}  \expval{\sigma_j^X (\text{0})}, 
	\end{equation}
	\noindent which has played a significant role in studying disorder-induced MBL\,\cite{nandkishore2015many,abanin2017recent,schreiber2015observation}. Results of measuring $\mathcal{I}$ on an $N=23$ qubit chain ($N_m=12, N_g=11$) are shown in Fig.~\ref{sfig:imbalance_exp_vs_sim}B and corresponding early-time simulation in Fig.~\ref{sfig:imbalance_exp_vs_sim}C. The matter-imbalance as well as the entropy experiments were performed using first-order Trotter circuits instead of second-order for the energy spreading experiments. When $\theta=0$, $\mathcal{I}$ decays to zero faster than for $\theta=\pi/2$, with a uniform trend for the $\theta$ values in between. 
	The decay of $\mathcal{I}$ to zero results from the loss of the initial state memory and suggests ergodic dynamics. In contrast, the asymptotic nonzero value of $\mathcal{I}$ implies localization. As $\theta$ is varied from 0 to $\pi/2$, the translational invariance of the initial state is preserved, yet distinct dynamical responses are observed. Since \(\mathcal{I}\) is not related to a conserved or nearly conserved quantity, it is a much subtler indicator of localization than, for example, energy density. It also suffers from finite-size effects more strongly than the conserved quantities since the total value can decay arbitrarily. Nevertheless, we find that at this system size, localization persists until exponentially long times (Fig.~\ref{sfig:imbalance_exp_vs_sim}D). 
	
	\section{Design of the circuit and mapping to fixed disorder}
	\label{sec:design-circ-mapp}
	
	Most well-known examples of localization require disorder, either in the Hamiltonian or initial state. One way to get a random background potential from an initial state is to \emph{simulate} a gauge theory. The idea is that violations of a Gauss law (for example \(\nabla \cdot E = \rho\) in a \(U(1)\) gauge theory) can easily be introduced in the initial state, and have the effect of static background charges (they have no dynamics since gauge-invariant dynamics preserve the Gauss law constraints). Averaging over background charge configurations can
	re-introduce translation symmetry to the initial state, but dynamics may still display signatures of localization. Crucially, this relies on having access to a simulation of a gauge theory, and therefore unphysical configurations of the gauge and matter fields.
	
	There is also a more direct perspective. Consider a model where randomizing
	certain couplings leads to localization. For example, say an interaction
	includes a term \(\sum_j g_j \hat{\mathcal{O}}_j\), where initially
	\(\hat{\mathcal{O}}_j\) is a Hermitian operator on some site \(j\) and \(g_j \in
	\mathbb{R}\); suppose that randomizing the \emph{sign} of \(g_j\) gives rise to
	localization. We augment the system with an ancilla for each coupling, and
	replace the coupling by an operator on that ancilla. In our example, we add
	an ancilla qubit at each site \(j\) (labelled by, say, \(j'\)), and replace
	\(g_j \to g_j \hat X_{j'}\). We can obtain averages over disorder realizations
	from single-state measurements by initializing the ancillas in superpositions
	of coupling values. For the example above, we would initialize each ancilla
	qubit in the \(\ket{0}\) state. Quantities, like the energy density, that
	involve the couplings will involve ancilla operators. To connect back to the
	discussion of gauge theory, to replace couplings by operators we need these
	operators to commute amongst each other and with the Hamiltonian, a situation
	which is automatic in the case that the operators are gauge constraints and
	the Hamiltonian is gauge-invariant.
	\subsection{\(\mathbb{Z}_2\) gauge theory with \(\mathbb{Z}_2\) matter}
	\label{ssec:z2_gauge}
	A convenient model for quantum simulators is \(\mathbb{Z}_2\) gauge theory with
	\(\mathbb{Z}_2\) matter. This is a system defined on a graph, with qubits on the
	vertices representing matter and qubits on the links representing gauge
	fields. We write the matter Pauli operators as $\hat \sigma^{X/Y/Z}$, and label them by
	a vertex, i.e. \(\hat \sigma^{X}_j\). Gauge field operators are written as $\hat X/\hat Y/\hat Z$ and
	labelled by vertices connected by their corresponding edge, i.e.
	\(\hat X_{j,k} = \hat X_{k,j}\). We write \(j \vee k\) to mean there is an edge between
	vertices \(j\) and \(k\). For each vertex we define an integrated Gauss law
	\begin{equation}
		\label{eq:z2-z2-gauss-law}
		\hat G_j = \hat \sigma^X_j \prod_{j \vee k} \hat X_{j,k} = (-1)^{q_j};
	\end{equation}
	the meaning of gauge invariance for dynamics is that our Hamiltonian (or
	Floquet unitary) commutes with \(\hat G_j\), and \(\hat G_j = + 1\) defines the
	gauge-invariant subspace. As usual, the low-order terms that commute with all
	such constraints are
	\begin{enumerate}
		\item arbitrary functions of \(\hat X_{j,k}\) (``gauge polarizations'') and \(\hat \sigma^X_j\)
		(``matter polarizations'')
		\item ``non-conserving hopping'' terms \(\hat \sigma^Z_j \hat Z_{j,k} \hat \sigma^Z_k\) that do not preserve a
		\(U(1)\) symmetry for matter consistent with the gauge constraint
		\item and therefore \(\hat \sigma^Y_j \hat Z_{j,k} \hat \sigma^Z_k\) and \(\hat \sigma^Y_j \hat Z_{j,k} \hat \sigma^Y_k\), which can be
		combined to the ``number-conserving hopping'' terms \((\hat \sigma^Z_j - i \hat \sigma^Y_j) \hat Z_{j,k}
		(\hat \sigma^Z_k + i \hat \sigma^Y_k)\) that preserve a full \(U(1)\) matter symmetry (number
		conservation in the \(x\)-basis) consistent with the gauge constraint
		\item ``magnetic flux'' loops \(\hat Z_{j_1, j_2} \cdots \hat Z_{j_{n-1}, j_n} \hat Z_{j_n,
			j_1}\).
	\end{enumerate}
	We can thus consider both the model with and without \(U(1)\) matter
	conservation (in our convention in the \(x\)-basis) using the same Gauss law
	\cref{eq:z2-z2-gauss-law}.
	
	To understand models in this class, it is useful to consider the operator
	\begin{equation}
		\label{eq:z2-bg-charge-decoupler}
		\hat U_B = \prod_j \left( \prod_{k \vee j} \CNOT{j}{j,k} \right).
	\end{equation}
	Note that all the operators in the product commute, so for graphs of bounded
	degree \(\hat U_B\) can be implemented by a finite depth circuit. Conjugation by
	this operator enacts (note \(\hat U^2 = \openone \))
	\begin{align}
		\hat U_B \hat G_j \hat U_B^{\dagger} &= \hat \sigma^X_j, \; \hat U_B \hat \sigma^Z_j \hat Z_{j,k} \hat \sigma^Z_k \hat U_B^{\dagger} = \hat Z_{j,k}, \; \hat U_B \hat X_{j,k} \hat U_B^{\dagger} = \hat X_{j,k} \\
		\hat U_B \hat Z_{j_1, j_2} &\cdots \hat Z_{j_{n-1}, j_n} \hat Z_{j_n, j_1} \hat U_B^{\dagger}
		= \hat Z_{j_1, j_2} \cdots \hat Z_{j_{n-1}, j_n} \hat Z_{j_n, j_1}.
	\end{align}
	Since the gauge constraint has become simply \(\hat \sigma^X_j\), this is a concrete
	implementation of the idea discussed above: the matter sites can simply be
	considered ancillary degrees of freedom that play the role of couplings, or
	equivalently background charges. This perspective is exact because, by virtue
	of being local gauge constraints, the \(\hat \sigma^X_j\) commute with the
	Hamiltonian. For example, say that in the original gauge theory there was a
	term \(\hat \sigma^X_j\) in the Hamiltonian; this becomes \(\hat G_j = \hat \sigma^X_j \prod_{j
		\vee k} \hat X_{j,k}\) after the unitary transformation. Then in each charge
	sector, this term contributes \(\pm \prod_{j \vee k} \hat X_{j,k}\), depending
	on the eigenvalue of \(\hat \sigma^X_j\).
	\subsection{A device-efficient Hamiltonian}
	\label{ssec:device_hamiltonian}
	We focus on a simple family of Hamiltonians defined by the energy density
	(per link)
	\begin{equation}
		\hat H_{j,k} =  J \hat \sigma^Z_j \hat Z_{j,k} \hat \sigma^Z_k
		+ \mu \left( \frac{\hat \sigma^X_j}{\mathrm{deg}(j)} + \frac{\hat \sigma^X_k}{\mathrm{deg}(k)} \right)
		+ h \hat X_{j,k}
		+ Q \left( \hat \sigma^X_j \frac{\prod_{\ell, j \vee \ell} \hat X_{j,\ell}}{\mathrm{deg}(j)} + \hat \sigma^X_k \frac{\prod_{\ell, k \vee \ell} \hat X_{k, \ell}}{\mathrm{deg}(k)} \right)
	\end{equation}
	The degree of a vertex \(\mathrm{\deg}(j)\) is the number of edges it
	touches. On a general graph, this model has no symmetries apart from the
	gauge constraint.
	
	To analyze the model by charge sectors, it is helpful to conjugate by
	\cref{eq:z2-bg-charge-decoupler}, obtaining
	\begin{equation}
		\hat H_{j,k} = J \hat Z_{j,k}
		+ \mu \left( (-1)^{q_j} \frac{\prod_{\ell, j \vee \ell} \hat X_{j,\ell}}{\mathrm{deg}(j)} + (-1)^{q_k} \frac{\prod_{\ell, k \vee \ell} \hat X_{k, \ell}}{\mathrm{deg}(k)} \right)
		+ h \hat X_{j,k}
		+ Q \left( \frac{(-1)^{q_j}}{\mathrm{deg}(j)} + \frac{(-1)^{q_k}}{\mathrm{deg}(k)} \right)
	\end{equation}
	where we have replaced \(\hat \sigma^X_j \to (-1)^{q_j}\) to emphasize the new role of
	matter sites as labelling background charge. This mapping gives a simple
	strategy for short-depth circuits that prepare states with desired energy
	density profiles, roughly independent of the particular charge sector: align
	(anti-align) link qubits with the effective single-site field \(J \hat Z + h \hat X\)
	for higher (lower) energy density.
	
	In a fixed charge sector, we can imagine there is a charge at each site where \(q_j = 1\). To understand dynamics in a fixed sector, it can be convenient to move the dependence on background charge from the multi-qubit term (proportional to \(\mu\)) to the single-qubit ``longitudinal'' field
	(proportional to \(h\)). In systems with appropriate boundaries this can always be done (in general, there can remain a single charge affecting the multi-qubit term if the total number of initial charges is odd) by
	conjugating on the appropriate links with \(\hat Z^{j,k}\); looking just at the effect of charges on the multi-qubit term, we note this operation just ``hops'' charges between \(\{j,k\}\), or pair annihilates them if there is a charge on both sites. The resulting energy density is
	\begin{equation}
		\hat H_{j,k}' =
		\mu \left( \frac{\prod_{\ell, j \vee \ell} \hat X_{j,\ell}}{\mathrm{deg}(j)} + \frac{\prod_{\ell, k \vee \ell} \hat X_{k, \ell}}{\mathrm{deg}(k)} \right)
		+J \hat Z^{j,k}
		+ h (-1)^{\kappa_{j,k}(q)} \hat X_{j,k}
		+ Q \left( \frac{(-1)^{q_j}}{\mathrm{deg}(j)} + \frac{(-1)^{q_k}}{\mathrm{deg}(k)} \right),
	\end{equation}
	where \(\kappa_{j,k}(q)\) is the sign accumulated on \(\hat X_{j,k}\) from the
	conjugations used to eliminate background charges \(q\) on the hopping terms.
	\subsection{No-flux sectors and models with symmetry}
	\label{ssec:no_flux}
	In a special case of such models, which can be interpreted as those where all
	dynamics can take place in a no-flux sector, there is another way to isolate
	the background charges. One can think of this as solving \(\oint A = 0\) by
	writing \(A = d \phi\) for some scalar field \(\phi\). In the
	\(\mathbb{Z}_2\) case, we look for a transformation which ``splits'' each link
	to two spins, satisfying \(\hat Z_{j,k} \sim \hat \tau^Z_{j} \hat \tau^Z_{k}\). Clearly, such
	a mapping could only work for the sector where all loops \(\hat Z_{j_1, j_2}
	\cdots \hat Z_{j_n, j_1} \sim 1\). Note that \(\hat \tau^X_j\) would flip all links
	touching \(j\). This means we have the correspondence
	\begin{equation*}
		\hat Z_{j,k} \sim \hat \tau^Z_{j} \hat \tau^Z_{k}, \; \hat \tau^X_{j} \sim \prod_{k \vee j} \hat X_{j,k}.
	\end{equation*}
	Reading the correspondences as going from \(\hat \tau\) to \(\hat Z,\hat X\) instead, this
	is how a \(\mathbb{Z}_2\) gauge theory arises from an Ising model. In any
	case, the gauge constraint now lives on a single site,
	\begin{equation*}
		\hat G_j \sim \hat \sigma^{X}_j \hat \tau^X_{j},
	\end{equation*}
	and a \(\mathsf{CNOT}\) gate between the new gauge sites and the matter will
	again reduce the gauge constraint to a single site, \(\hat \tau^X_{j}\).
	
	Compared to the mapping of the previous section, this transformation is not local, and only appears local (i.e. local operators mapping to local operators) for operators which cannot create flux (it can, though, be modified to work in \emph{fixed} flux sectors). In particular, it is always possible to make the transformation for 1d graphs or trees, though not all terms map to local terms; for example, in systems with a boundary single \(\hat X\) operators will become strings of \(\hat \tau^X\) stretching to a boundary. The main advantage is that the non-trivial part of the
	transformation has only acted on the gauge fields. In particular, the only
	non-trivial gauge-invariant single-site operator, \(\hat \sigma^X_j\), is unchanged by
	any of the transformations. Any on-site structure, including a possible
	\(U(1)\) symmetry generator would have to be built from the \(\hat \sigma^X_j\) in a
	gauge theory, and this structure will be preserved by this transformation.
	
	To be more explicit, let us call \(\hat \sigma^{\pm}_j = \hat \sigma^Z_j \mp i \hat \sigma^Y_j\). The
	\(U(1)\)-conserving hopping has a simple transformation
	\begin{equation}
		\hat \sigma^+_j \hat Z_{j,k} \hat \sigma^-_k \sim \hat \sigma^+_j \hat \sigma^-_k, \quad \hat \sigma^X_j \sim \hat \sigma^X_j.
	\end{equation}
	An operator which has a local image under the mapping is
	\begin{align}
		&\frac{1}{2} \left(\sum_{j \vee k} J' \hat \sigma^+_{j} \hat Z_{j,k} \hat \sigma^{-}_{k} + \Delta \hat \sigma^X_{j} \hat \sigma^X_{k}
		+ J \hat \sigma^Z_j \hat Z_{j,k} \hat \sigma^Z_k \right)
		+ h \hat \sigma^X_j + u \prod_{j \vee k} \hat X_{j,k} \\
		&\sim \frac{1}{2} \left( \sum_{j \vee k} J' \hat \sigma^{+}_j \hat \sigma^{-}_k + \Delta \hat \sigma^X_{j} \hat \sigma^X_{k} + J \hat \sigma^Z_j \hat \sigma^Z_k \right) + (h + u \hat \tau^{x}_{j}) \hat \sigma^X_j.
	\end{align}
	Setting \(J = 0\) gives a Heisenberg ``XXZ''-type model with a \(U(1)\)
	symmetry and local potential dictated by background charge \(\hat \tau^x_j\),
	while setting \(J' = \Delta = 0\) gives an Ising model with transverse field
	dictated by the same background charge.
	
	\newpage
	\section{Simulation techniques and numerical results}
	\label{sec:simulations}
	The numerical results presented in this paper were obtained primarily using four methods. (A) The short-time dynamics for 1d were simulated using an exact statevector simulation of the quantum circuits. (B) The long-time dynamics for 1d were simulated using exact diagonalization.  (C) The short-time dynamics for 2d were simulated using matrix product states (MPS). (D) Additionally, we used the Lanczos method available in the quspin \cite{quspin} library for computing the entropy of a 16-qubit system at intermediate time scales. 
	\begin{figure*}[b!]
		\centering
		\includegraphics{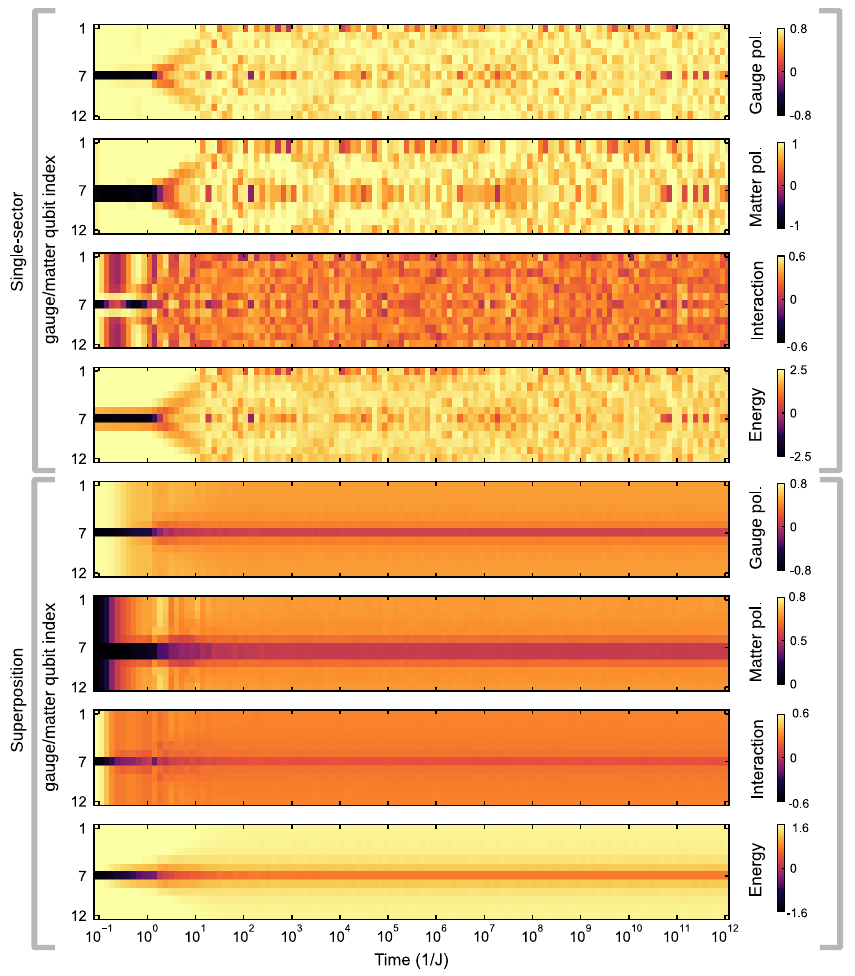}
		\caption{Long-time Hamiltonian dynamics ($\Delta t = 0$) of a $24$ qubit perturbed initial state in 1d obtained from ED in the dual basis. We examined localization in energy per link as well as the contributions from the respective Hamiltonian terms: gauge polarization $\langle\hat X_{j,j+1}\rangle$, matter polarization $\langle\hat \sigma^X_j \rangle$, and the interaction term $\langle\hat \sigma^Z_{j} \hat Z_{j,j+1} \hat \sigma^Z_{j+1}\rangle$. The $24$-qubit single-sector state with periodic boundary conditions (PBC) maps to a $12$-qubit mixed-field Ising model with the sign of Ising coupling determined by the background charges. Hence, the uniform superposition initial state observables can be obtained by summing over $2^{12}$ gauge configurations. Hamiltonian parameters used were $J=1$, $h=1.3$, and $\mu=1.5$.}
		\label{sfig:energy_long_times}
	\end{figure*}
	
	\subsection{Exact statevector simulation}
	\label{ssec:sim_statevector}
	We used the qsim library \cite{quantum_ai_team_qsim_2020} to simulate the Cirq quantum circuits \cite{cirq_developers_cirq_2024}  for comparison against the experimental results. Since the experimental results for energy perturbation dynamics in 1d and 2d were obtained for 38 and 81 qubits, they are challenging to simulate with exact statevector in the original LGT basis.  For the 1d results, we used the dual (gauge-symmetry) basis to carry out the statevector dynamics of $19$ gauge qubits separately in each sector. Thus, the superposition initial state results were obtained by averaging over disorder realizations of the background charges.  For further discussion of the dual mapping, see \cref{sec:design-circ-mapp}. The gauge charges map to $19$ matter sites in the dual basis, allowing us to work with only the $19$ gauge qubits. The 2d simulations were performed using matrix product states (MPS), and the details are provided in \cref{sec:simulations}C.

	% In Fig.~\ref{sfig:2d_energy_dynamics}, we show the noiseless Floquet dynamics of gauge polarization as well as energy per link on a 2d grid consisting of 39 qubits. We observe localization of both quantities for the superposition initial state and delocalization for the single-sector initial states. We used the Floquet parameters $J=1, \, \Delta t=0.35, \, h = 1.5$, and $\mu=3.5$. Since the simulation of 39 qubits using exact state-vector methods is challenging, we utilized the dual mapping to fix $n_c=16$ gauge charges and work with an effective $23$-qubit system. For the uniform superposition case, we averaged the results over $100$ randomly chosen gauge charges $\mathbf{g}$.
	
	% \begin{figure*}[b!]
		% \centering
		% \includegraphics{Figures/FigS5.pdf}
		% \caption{Dynamics of gauge polarization (top panels) and energy (bottom panels) in a noiseless simulation for the superposition initial state (right panels), compared against the single-sector initial state (left panels). The initial state was prepared for both cases with an energy imbalance on the top gauge qubit. The dynamics of the 39-qubit 2d grid were simulated in the dual picture by fixing $n_c=16$ gauge charges, resulting in an effective $23$-qubit system.}
		% \label{sfig:2d_energy_dynamics}
		% \end{figure*}
	\subsection{ED results}
	\label{ssec:sim_ed}
	\begin{figure*}[b!]
		\centering
		\includegraphics{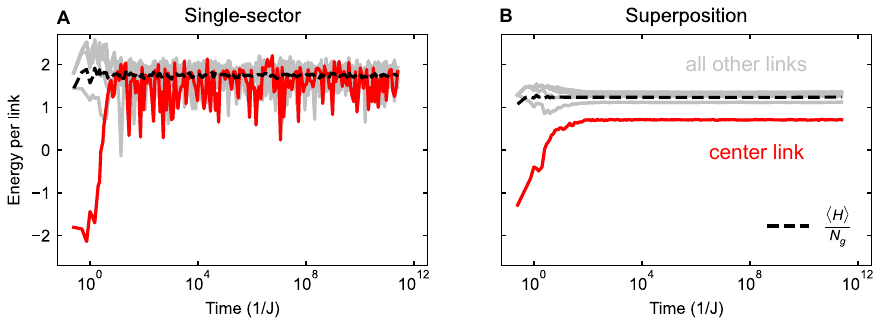}
		\caption{Long-time Floquet dynamics ($J=1$, $h=1.3$, and $\mu=1.5$, $\Delta t =0.25$) of the energy per link in a noiseless ED simulation for the single-sector (left panel) and superposition initial states (right panel). In contrast to the second-order Trotter dynamics for our energy spreading experiments, these numerics use a first-order Trotter approximation. For both cases, the initial state was prepared with an energy imbalance on the center gauge qubit (shown in red). The average energy per link is shown in the dashed black line.}
		\label{sfig:1d_energy_dynamics_trotter_long}
	\end{figure*}
	
	We further employ our exact diagonalization (ED) code for continuous and Floquet time evolution performed using matrix exponentiation. In ED, very long evolution times are accessible for moderate system sizes, which can provide insights into the long-time dynamical behavior currently inaccessible on quantum hardware for small system sizes. Although the experimental results presented are for a $38$-qubit ring,  due to the difficulty of diagonalizing an effective $ 19$-qubit Hamiltonian in the dual mapping (Sec.~\ref{sec:design-circ-mapp}), we present results for a 24-qubit ring here. 
	
	\subsubsection{ED using disordered mixed-field Ising model}
	\label{sssec:ed_ising}
	Let us first consider the dual disordered mixed-field Ising model for $N=12$ qubits (corresponding to $N=24$ qubits in the LGT model) with periodic boundary conditions, given by the Hamiltonian
	\begin{align}
		H=\sum_{j=1}^N\big(\mu g_j \hat X_j \hat X_{j+1}+h\hat X_j+J \hat Z_j\big),
	\end{align}
	where $g_j$ are the background charges. There are $2^L$ unique configurations of $g_j$ over the entire lattice. The initial state $\ket{\psi_0}$ is a product state where each of the gauge qubits has its Bloch vector \(\hat{\mathbf{v}}\) aligned with the direction \(J \hat{\mathbf{z}} + h \hat{\mathbf{x}}\). $\hat \sigma^Y_j$ is applied to the center gauge qubit to flip its Bloch vector to $-\mathbf{\hat v}$ and create an energy perturbation. This is the setup of the experiment summarized in Fig.~1 of the main text.
	
	We consider two cases: (i) the single-sector case where $g_j=+1,\,\forall j$, and (ii) the superposition case where the dynamics starting in $\ket{\psi_0}$ will be carried out for each single configuration of $g_j$ over the lattice, and afterwards averaged over all configurations. The corresponding ED results for the Hamiltonian dynamics ($\Delta t = 0$) up to evolution time $t=10^{12}/J$ are shown in Fig.~\ref{sfig:energy_long_times} for the local gauge polarization, matter polarization, matter-gauge-matter interaction, and energy per link. We see a stark contrast between the single-sector and superposition cases. Whereas in the former, the dynamics become locally indistinguishable between sites already at times $\sim\mathcal{O}(10/J)$, in the latter there are distinct robust localization dynamics up to arbitrarily large times. Similar observations were made also for the Floquet dynamics with Trotter step ($\Delta t = 0.25$) as shown in Fig.~\ref{sfig:1d_energy_dynamics_trotter_long}, where we focus only on the energy density (per link). Furthermore, we note that the total energy is approximately conserved by the Floquet dynamics.

	\subsubsection{Tunable DFL in the matter imbalance}
	\label{sssec:ed_imbalance}
	We employ our ED code to also look at the ``infinite-time'' dynamics of the imbalance for a $23$-qubit chain ($N_m=12$ and $N_g=11$) with open boundary conditions, as is done experimentally for finite times in Fig.~\ref{sfig:imbalance_exp_vs_sim}. In particular, we consider the initial state 
	
	\begin{subequations}
		\begin{align}
			\ket{\psi_0(\theta)}&=\ket{\psi^m_0}\otimes\ket{\psi^g_0(\theta)},\\
			\ket{\psi^m_0}&=\bigotimes_{j=1}^{N_m}\ket{(-1)^j\hat{\mathbf{x}}},\\
			\ket{\psi^g_0(\theta)}&=\bigotimes_{j=1}^{N_g}\ket{\cos\theta\,\,\hat{\mathbf{x}}_{j,j+1}+\sin\theta\,\,\hat{\mathbf{z}}},\\
			\hat{\mathbf{x}}_{j,j+1}&=\sqrt{2}\sin\frac{\pi\big(2j+1\big)}{4}\,\,\hat{\mathbf{x}}.
		\end{align}
	\end{subequations}
	When $\theta=0$ or $\pi$, the initial state $\ket{\psi_0(\theta)}$ is in the local-symmetry sector $g_j=+1,\,\forall j$. For $\theta=\pi/2$, $\ket{\psi_0(\theta)}$ is in an equal superposition of an extensive number of local-symmetry sectors. In general, we have to consider $2^{N_g}$ different states, each with a probability amplitude $c_n=\cos^{N_g-M}(\theta/2)\sin^M(\theta/2)$, where $M$ is the number of ``flips'' with respect to $\hat{\mathbf{x}}_{j,j+1}$. Accordingly, we can rewrite the initial state as
	
	\begin{align}
		\ket{\psi_0(\theta)}&=\sum_{n=1}^{2^{N_g}}c_n(\theta)\ket{\psi^m_0}\otimes\ket{\psi^g_n}=\sum_{n=1}^{2^{N_g}}c_n(\theta)\ket{n},
	\end{align}
	where $\ket{\psi^g_n}$ are the $2^{N_g}$ product states in the computational basis of the $X_{j,j+1}$. For notational brevity, we have denoted $\ket{n}=\ket{\psi^m_0}\otimes\ket{\psi^g_n}$. We now calculate the time-evolved wave function
	\begin{align}\label{eq:TimeEvolution}
		\ket{\psi(\theta,t)}=e^{-iHt}\ket{\psi_0(\theta)}=\sum_{n=1}^{2^{N_g}}c_n(\theta)\ket{n(t)},
	\end{align}
	We can then calculate the matter imbalance as follows
	\begin{align}
		\mathcal{I}(\theta,t)&=\frac{1}{N_m}\sum_{j=1}^{N_m}(-1)^j\bra{\psi(\theta,t)}\hat \sigma^X_j\ket{\psi(\theta,t)}\\\nonumber
		&=\frac{1}{N_m}\sum_{j=1}^{N_m}\sum_{m,n=1}^{2^{N_g}}(-1)^jc_n(\theta)c_m(\theta)\bra{m(t)}\hat \sigma^X_j\ket{n(t)}\\\nonumber
		&=\frac{1}{N_m}\sum_{j=1}^{N_m}\sum_{\mathbf{g}}^{2^{N_g-1}}\sum_{m_{\mathbf{g}},n_{\mathbf{g}}=1}^{2}(-1)^jc_{n_{\mathbf{g}}}(\theta)c_{m_{\mathbf{g}}}(\theta)\bra{m_{\mathbf{g}}(t)}\hat \sigma^X_j\ket{n_{\mathbf{g}}(t)}\nonumber \\
		&=\frac{1}{N_m} \sum_{j=1}^{N_m}\sum_{\mathbf{g}}^{2^{N_g-1}} \sum_{n_{\mathbf{g}}=1}^{2}(-1)^jc^2_{n_{\mathbf{g}}}(\theta)\bra{n_{\mathbf{g}}(t)}\hat \sigma^X_j\ket{n_{\mathbf{g}}(t)},
		\label{eq:imbalance}
	\end{align}
	where we have resolved the local-symmetry sectors $\mathbf{g}$ in the last two lines of Eq.~\eqref{eq:imbalance}, and made use of the fact that a given matter configuration allows two opposite gauge polarizations in the same local-symmetry sector. As shown in Fig.~\ref{sfig:imbalance_exp_vs_sim}, we see a tunable robust signature of DFL that persists up to all considered arbitrarily long evolution times, with stronger localization the larger $\theta$ is in $[0,\pi/2]$.
	
	\subsection{Matrix Product States (MPS) simulations}
	\label{ssec:mps_simulations}
	\begin{figure*}[!t]
		\centering
		\includegraphics{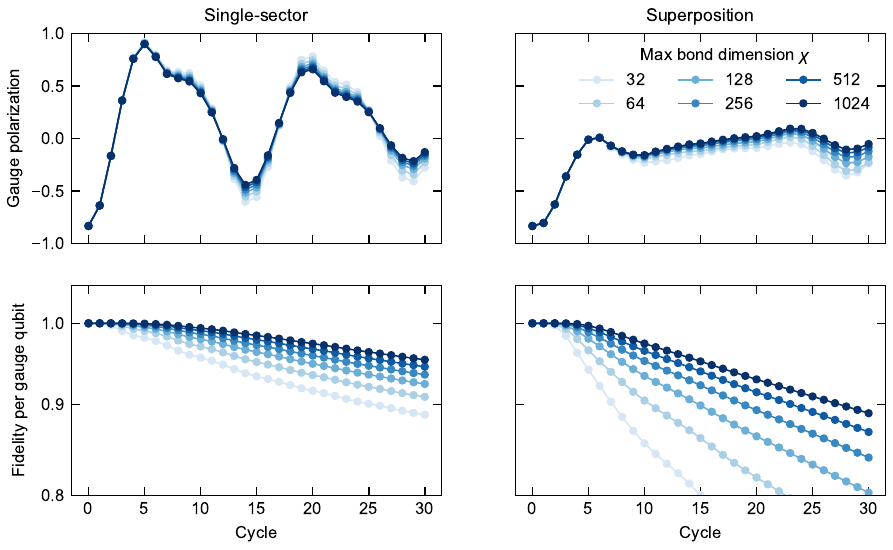}
		\caption{(Top panels) Gauge polarization, i.e., the expectation value of $\hat X$, for the perturbed top left qubit on the 81 qubit 2d grid showing convergence of a local observable with increasing bond dimension $\chi$. (Bottom panels) Exponential decay of the fidelity per gauge qubit of the time-evolved MPS with the number of cycles as a function of $\chi$. The y-axis is on a log scale. The single-sector initial state (left panels) was simulated by fixing the gauge charges and working with only the gauge qubits in the dual model. In contrast, the superposition initial state (right panels) was simulated in an LGT basis with both matter and gauge qubits.}
		\label{fig::mps_simulations}
	\end{figure*}
	
	The 2d experimental grid contains 81 qubits, consisting of 32 matter qubits and 49 gauge qubits.  The Hamiltonian is block diagonal, with each block labeled by a flux in $\{+1,-1\}$ on each matter qubit, so in principle the dynamics can be simulated independently for each flux sector, with an effective Hilbert space of dimension $2^{49}$.  Despite the substantial reduction from the full Hilbert space of size $2^{81}$, the Hamiltonian blocks for each flux sector remain too large to be simulated by exact diagonalization.
	
	We therefore turn to tensor networks to simulate the dynamics of the system.  Tensor networks can efficiently simulate very large quantum systems so long as the amount of entanglement remains moderate.  Especially in 1d systems, Hamiltonian or Trotterized/circuit evolution of hundreds of spins can be simulated using matrix product states (MPS)~\cite{mps1,mps2,mps3} at low computational cost.  For ergodic dynamics, the entanglement between two halves of the system grows linearly over time, and consequently the tensor network bond dimension (and hence computational cost) grows exponentially, so even in 1d the simulations (with fixed computational resources) will eventually fail.  
	
	In 2d, the entanglement structure of the system would be most naturally represented by an inherently 2d tensor network such as the projected entangled-pair state (PEPS) ansatz~\cite{peps}.  However, PEPS-based simulations suffer from a number of problems, including expensive and approximate contraction and calculation of observables, and consequently a lack of reliable time-evolution algorithms.  Therefore in this work we use MPS-based simulations.  For a 2d system with area law entanglement (for example, the single sector initial state in our experiments), the bond dimension required for a faithful MPS representation scales exponentially in the smaller of the two linear dimensions. 
	
	Despite this exponential scaling, at the system size used in the current experiment, our MPS simulations achieve greater fidelity with the ideal time evolution than do the experiments on the device.  The decay of MPS fidelity per gauge qubit with Trotter step, for bond dimensions up to 1024, is shown in Fig.~\ref{fig::mps_simulations} above.  Evidently, the fidelity does decay exponentially, but the decay is slower than that of the experimental fidelity (given, roughly, by $(1-p)^{6 C}$ where $p$ is the error rate per two qubit gate and $C$ is the number of cycles).
	
	\subsubsection{Approach to MPS simulations of digital circuits\label{sssec:MPS_sim_method}}
	
	Here we give some details of our MPS simulations.  When simulating the full model in the LGT basis, including both matter and gauge qubits, there are two types of Trotter circuit layers: single-qubit $X$ rotations on all qubits (with different angles for matter and gauge qubits), and three-qubit $ZZZ$ rotations on links connecting two matter qubits.  The single-qubit rotations can be applied directly on each MPS tensor, with negligible computational cost and no error.  
	
	For the entangling layers, we group the $ZZZ$ rotation gates so that each group of gates can be represented by a single matrix product operator (MPO) of bond dimension $D=2$.  In practice for the grid used in the experiment, we do this by choosing an MPS snake that (referring to the layout in Fig.~3 of the main text) starts in the upper left and runs up and to the right along each anti-diagonal before going to the start of the next anti-diagonal.  Then a single MPO of $D=2$ captures all $ZZZ$ rotations on links along the anti-diagonals, while we need one additional MPO with $D=2$ for each diagonal (apart from the lower left and upper right, where there is no link along the downwards diagonal).  Thus in practice, each $ZZZ$ layer is implemented by successively applying seven $D=2$ MPOs.  We take a similar approach for other qubit layouts; for example, each $L\times L$ grid used to study scaling of MPS cost with system size, below, requires $(L+1)$ bond-dimension 2 MPOs per $ZZZ$ layer.
	
	In each time evolution simulation, we fix a maximum MPS bond dimension $\chi$.  Then for each MPO application, we first apply the MPO with no truncation, temporarily resulting in an MPS of bond dimension $2\chi$ that is not in canonical form.  We then sweep from left to right making no truncations to put the MPS into canonical form, then right to left making optimal truncations back to bond dimension $\chi$ using singular value decompositions; this truncation is the only source of error in the simulations.  We record the truncated weight of Schmidt values, $\epsilon$, at each bond, and we use $\prod (1-\epsilon)$ as a proxy for the fidelity of the resulting state.  To confirm that this is a good estimate for fidelity, we have also used a small grid of $3\times3$ matter qubits, which has just 21 total qubits and thus can be simulated exactly, and we directly compute the fidelity $|\langle \psi_\chi|\psi_{\text{exact}}\rangle|^2$; the estimate described above based on accumulated truncation error is, as expected, a lower bound on the true fidelity, but the values are also quite close.  
	
	When simulating the model in the dual basis, with only gauge qubits, we take a nearly identical approach.  The Trotter evolution layers consist of single-qubit $X$ and $Z$ rotations, which can be implemented exactly, and ``$XXXX$'' rotations with $X$ operators on all gauge qubits around a particular (removed) matter qubit.  (``  '' are used because at the edges of the system each matter qubit has fewer than four gauge qubit neighbors, so in practice these operators can be as small as a single-qubit rotation.)  The latter layer can again be implemented by grouping the rotation gates into $D=2$ MPOs.  Specifically, we need one MPO for each diagonal, which includes the rotation gates for the $XXXX$ stars around each matter qubit along that diagonal.  The lower left and upper right are implemented each with just a single-qubit rotation, giving six MPOs per Trotter step for the experimental grid (and $L$ MPOs per step for the $L\times L$ square grids used below).
	
	For the dual-basis simulations, we simulate the single-sector initial state by setting a coefficient $g_j=+1$ for all ``$XXXX$'' terms.  We can also learn about the behavior of the superposition initial state by separately simulating a variety of flux sectors with different configurations of $g_j\in\{+1,-1\}$.
	
	To match the experiment, the MPS simulations on the experimental grid are performed with second-order Trotter, meaning that each Trotter step consists of a half-layer of $Z$ rotations (the $ZZZ$ layer for the full model, the single-qubit $Z$ layer for the dual basis simulation), a full layer of $X$ rotations (including the ``$XXXX$'' rotations in the dual basis simulations), and another half-layer of $Z$ rotations.  When simulating multiple steps, the $Z$ half-layers from the end of one step and the start of the next are combined, so that the simulation cost is only marginally higher than for first-order Trotter simulations.
	
	Our simulations make use of some functionality from the TeNPy tensor network library~\cite{tenpy}.
	
	\subsubsection{Scaling of MPS simulation cost with system size\label{sssec:MPS_scaling_cost}}
	
\begin{figure}
		\centering
		\includegraphics[width = 0.7\textwidth]{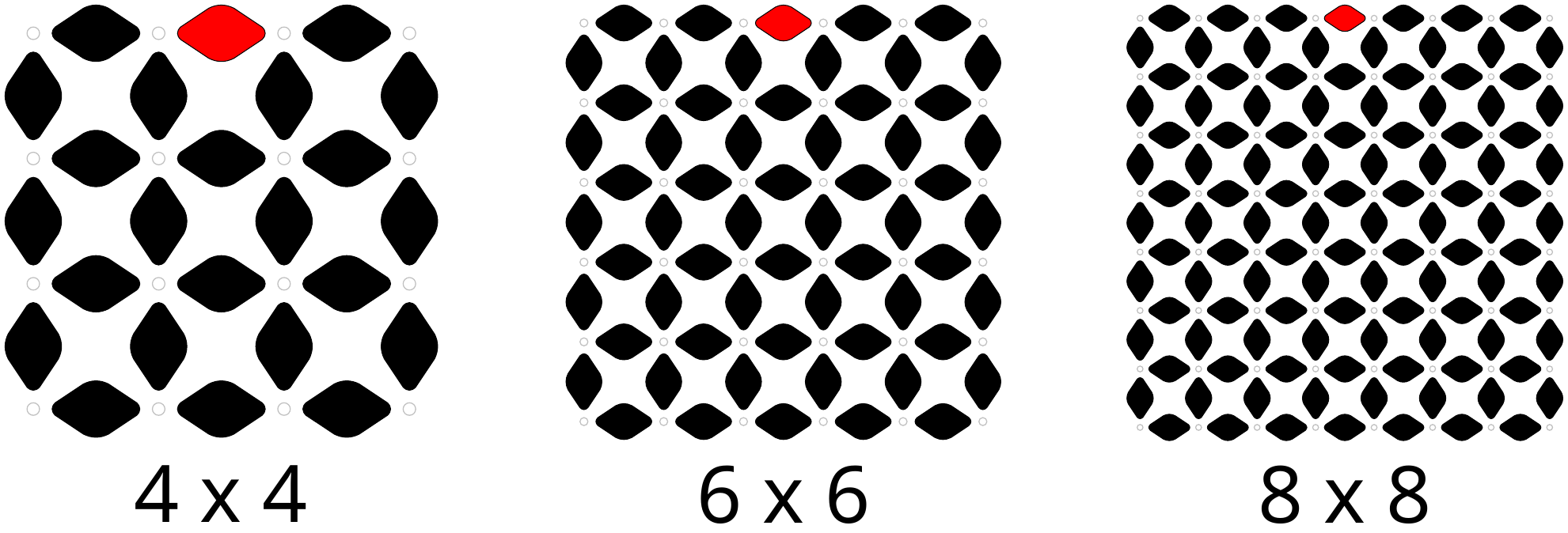}
		\caption{Qubit layouts used to simulate scaling of MPS simulation cost.  These grids have $L^2$ matter qubits and $2L(L-1)$ gauge qubits, with $L=4$, 6, and 8.  The resulting total qubit counts are 40, 96, and 176, respectively.  Initial states are analogous to those used in the experiment, with one initially perturbed gauge qubit, as indicated in red.}
		\label{fig:mps_scaling_grids}
	\end{figure}

	Although the MPS simulations can accurately simulate the circuits on the size and timescales of the experiment, due to the exponential scaling of the cost of MPS simulations with linear size, we expect that further scaling up the system size would lead to a lower fidelity for MPS than for the experiment.  To verify the scaling of MPS computational cost, we simulate our model on lattices with $4\times4$, $6\times6$, and $8\times8$ matter qubits, shown in Fig.~\ref{fig:mps_scaling_grids}.  As in the experiment, in the initial state we perturb the energy in one location at the edge of the system; the location of the initial excitation is also shown in the figure.
	
	At each size, we first simulate the full LGT model including all $L^2$ matter and $2L(L-1)$ gauge qubits, giving total system sizes of 40, 96, and 176 qubits for $L=4$, 6, and 8, respectively.  We simulate both the single-sector and superposition initial states by initializing all matter qubits in the states $|+x\rangle$ and $|+z\rangle$, respectively.  We simulate digital (Trotter) evolution.  Note that for convenience we use first-order Trotter evolution, rather than second-order as in the experiment; this does not affect the cost scaling of the simulations.  
	
\begin{figure*}
		\centering
		\includegraphics[width = \textwidth]{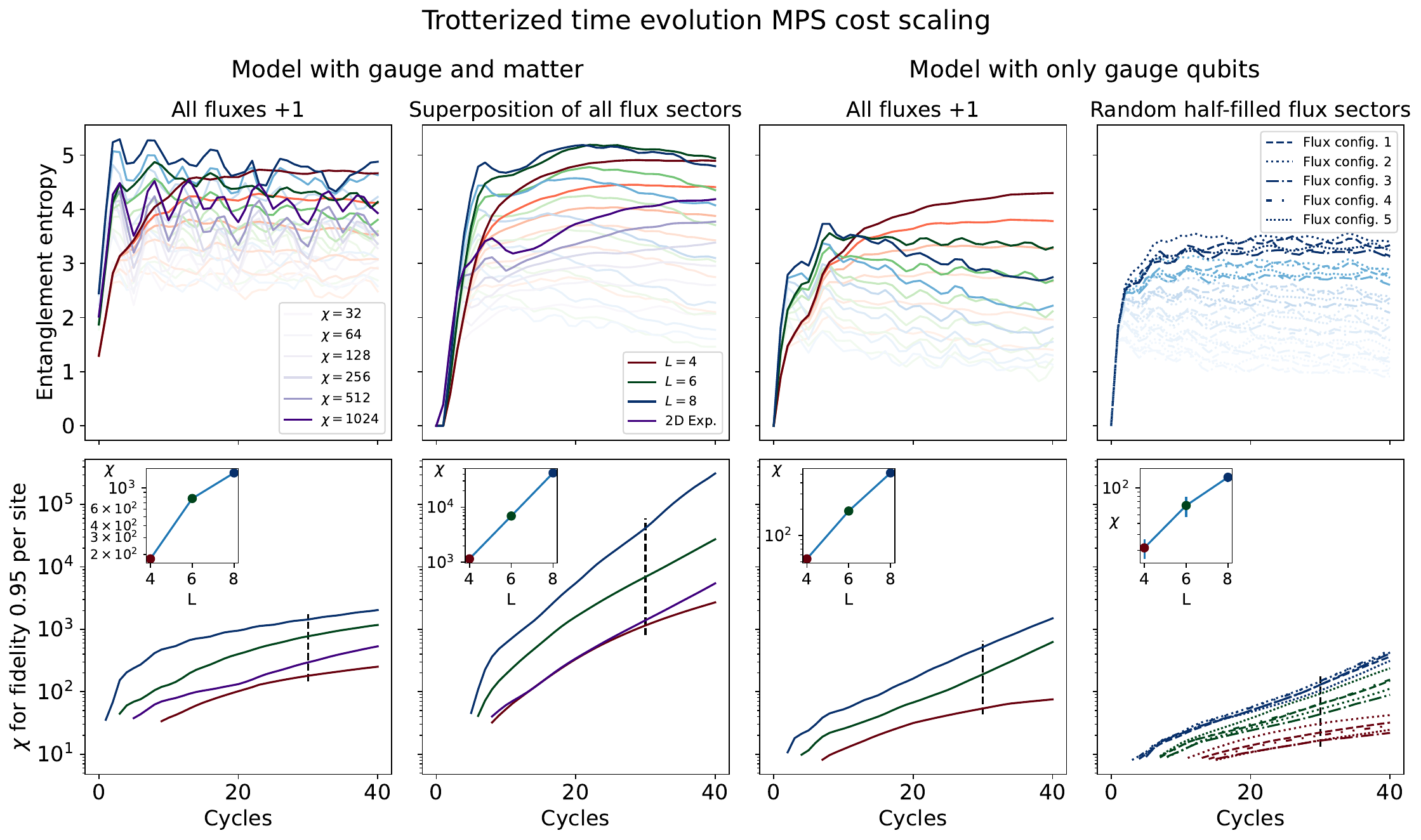}
		\caption{Cost scaling for MPS simulations.  We simulate dynamics on square grids of increasing size as shown in Fig.~\ref{fig:mps_scaling_grids}.  We also include the 81-qubit grid from the experiment for context.  In all panels, line colors correspond to system size: red corresponds to $L=4$, green to $L=6$, blue to $L=8$, and purple to the experimental grid.  Darker lines correspond to larger MPS bond dimension $\chi$, as shown in the key in the upper left panel. \textbf{Model types:} The \emph{left half} shows data for simulations using the full LGT model with both gauge and matter qubits, while the \emph{right half} shows simulations in the dual basis with only gauge qubits.  In both cases, the left column shows simulations of the time evolution of the single-sector initial state, while the right column shows simulations relevant to the superposition initial state.  For the full model, we directly simulate the evolution of the superposition state; for the dual basis model, we simulate five representative flux configurations that are typical in the superposition.  \textbf{Computational cost metrics:} The \emph{top row} shows entanglement entropy for the different simulation setups, system sizes, and bond dimensions.  The \emph{bottom row} shows estimates of the required bond dimension to achieve a target fidelity of 0.95 per gauge qubit.  Insets show the cost at cycle 30 as a function of system size, demonstrating exponential scaling in linear size as expected for MPS simulations in two dimensions.}
		\label{fig:mps_scaling_Trotter}
	\end{figure*}

	Useful metrics for understanding the cost of these simulations are shown in the left half of Fig.~\ref{fig:mps_scaling_Trotter}.  In the upper panels, we show the maximum entropy on any cut through the MPS; the cost of a high-fidelity MPS simulation grows approximately exponentially in the entropy.  We observed the expected linear growth of entropy at short times, and at longer times we see that for each system size the entanglement increases by a constant step when the MPS bond dimension is doubled, indicating that the bond dimension is not yet large enough to capture the true entanglement structure accurately.  However, at the smallest size, $L=4$, the shape of the curve of entanglement vs Trotter cycle appears to be converging, indicating that bond dimensions on the order of low thousands are sufficient for an accurate simulation.  
	
	In the lower panels, we estimate the MPS bond dimension needed to achieve a target wavefunction fidelity of 0.95 per gauge qubit.  This calculation proceeds in several steps.  First, we run the time evolution for each of a variety of bond dimensions, $\chi$, from $2^5$ through $2^{10}$, using the procedure described in Section 5.C.1 above.  At each Trotter cycle, this gives an estimate of the fidelity for each bond dimension.  Because we are interested in local properties of physical systems, it is more meaningful to discuss a fidelity per site, rather than global wavefunction fidelity.  As a simple approach to computing a fidelity per site, we simply raise the global fidelity to the power $1/[\text{Num. gauge qubits}]$.
	
	We then fit a function giving $1/\chi$ as a function of fidelity per site, also using the fact that fidelity is guaranteed to go to 1 as $\chi\rightarrow\infty$; to be precise, we fit a function of the form $1/\chi = A(1-f)^B$, with fitting parameters $A$ and $B$, to the points with $\chi \geq 256$, and we use linear interpolation of $(\chi,f)$ pairs for smaller $\chi$.  We then use the fit function to estimate the required $\chi$ to achieve the target fidelity.
	
	Evidently, the required bond dimension for fixed fidelity grows exponentially in time for the superposition initial state, even though in those dynamics the gauge polarization shows localization of the initial perturbation.  The scaling for the single-sector case, where all fluxes are $+1$, is less clear, but it appears that the required bond dimension still grows exponentially in time, just much more slowly.  In particular, this single-sector evolution can be accurately simulated even on the 176-qubit grid out to 30 cycles with a computationally reasonable bond dimension of 1024.  
	
	We also empirically confirm the exponential scaling of cost with linear system size.  The inset in each lower panel shows the required bond dimension for the target fidelity at cycle 30 at each system size.  Particularly in the superposition case, the exponential scaling is clearly shown via the straight line on the log-linear plot.  In the single-sector case, the exponential scaling is less clear.  %the scaling appears sub-exponential because the maximum possible required bond dimension is limited by the accessible Hilbert space size, which is close to being reached for the smallest system.  In particular, the dynamics take place in a particular flux sector that has an effective Hilbert space size of $2^{24}$, giving a max required bond dimension of just $2^{12}$.
	
	Finally, in order to give a clear picture of the difficulty of simulating our actual experiments, we include data from the (second-order Trotter) simulations of the experiment we performed on the two-dimensional 81-qubit grid.  The cost falls between that of the $4\times 4$ and $6\times 6$ grids as expected based on qubit count.  Note in particular that in the plot of required $\chi$ for the superposition sector, at cycle 30 we estimate a bond dimension of around 1024 is needed for a fidelity of 0.95 per gauge site; this can also be seen in Fig.~\ref{fig::mps_simulations} above.
	
	At first glance, the comparison between the simulation costs for the two initial states is surprising.  One might expect that localized dynamics generate less entanglement and hence require lower bond dimension, but here we see the opposite.  The simulation results are in agreement with the experimental results on small 1d systems that we report in Fig.~4 of the main text.  As stated there, the entanglement in the superposition case includes a volume law contribution from the background superposition of fluxes.  Equivalently, from the perspective of the MPS representation of the state, an optimal MPS for a given bond dimension naturally captures the most important parts of Hilbert space.  Hence, in the single-sector case the MPS simulations implicitly make use of the block-diagonal structure of the Hamiltonian.  No such implicit simplification is possible in the superposition case.  
	
	Now we turn to the right half of Fig.~\ref{fig:mps_scaling_Trotter}, where we show similar results for simulations in the dual basis, using only gauge qubits.  In the single-sector simulation, we set all fluxes $g_j$ to $+1$.  The entanglement is smaller than in the corresponding simulation of the full LGT model because we have substantially reduced the total number of qubits in the simulation.  However, for the smallest system we see that entanglement reaches a similar value at long times, possibly reflecting the fact that the effective MPS dynamics for this initial state even in the full model take place within a small portion of the total Hilbert space.  The estimated required bond dimension $\chi$ grows exponentially in time at least for the larger system sizes, but remains modest even at the 176-qubit scale and 40 cycles.
	
\begin{figure*}
		\centering
		\includegraphics[width = \textwidth]{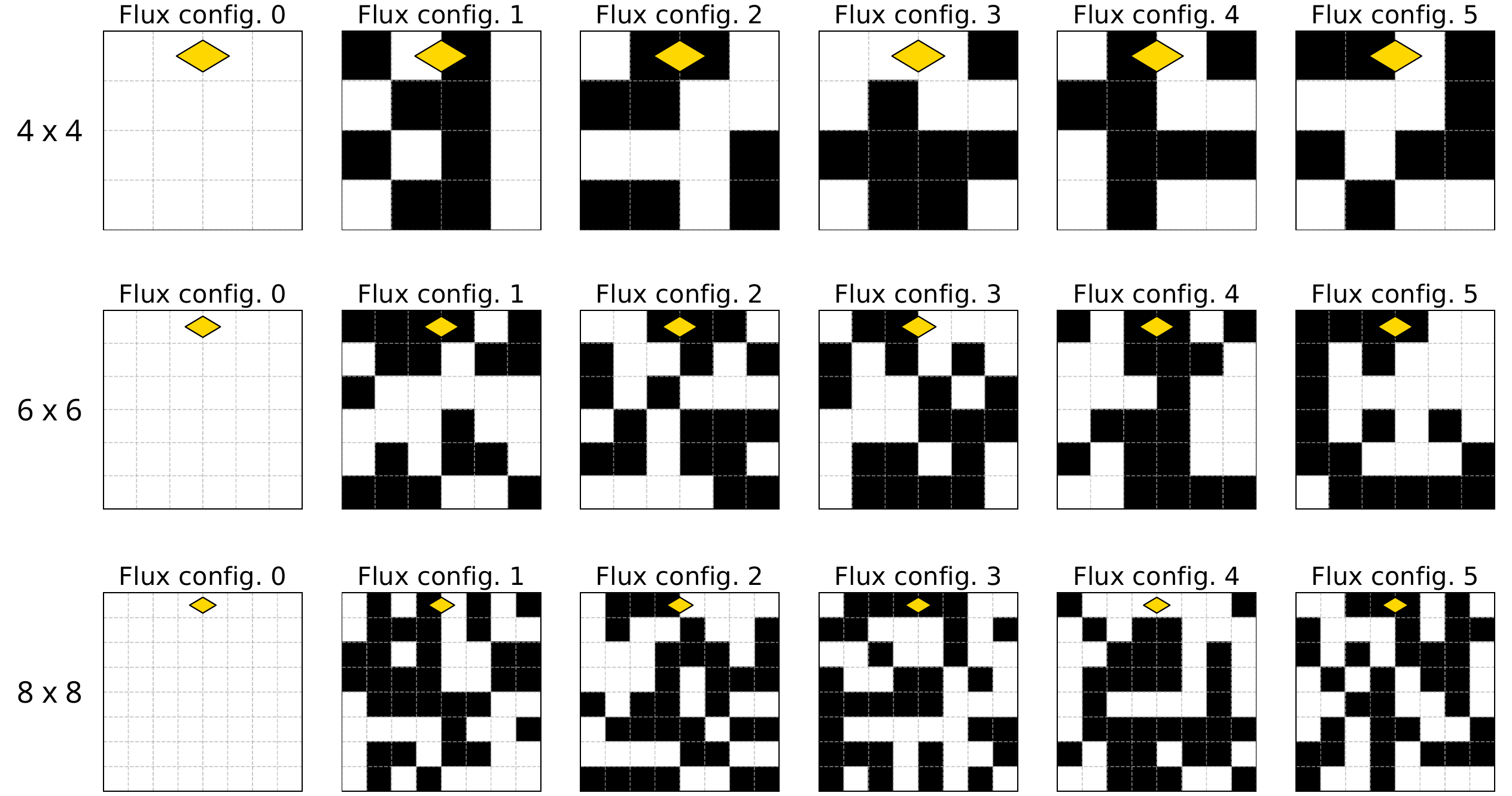}
		\caption{Flux configurations used to estimate the cost of simulating the time evolution of the superposition initial state by averaging over configurations.  The yellow diamond shows the location of the perturbed qubit.  For each grid size, the leftmost grid shows the all fluxes $+1$ sector, corresponding to the delocalized evolution of the single-sector initial state and comprising an exponentially small piece of the superposition dynamics.  The remaining grids show five random flux configurations with half the fluxes being $+1$ and half $-1$, which are representatives of typical flux configurations in the superposition.}
		\label{fig:mps_scaling_flux_configs}
	\end{figure*}

	Finally, we estimate the cost of studying the superposition initial state by sampling different flux configurations.  The superposition as a whole is dominated by an exponential number of flux sectors close to half of the fluxes being $+1$ and half $-1$.  We therefore study, for each system size, the simulation cost for five random flux configurations with exactly half of the fluxes being $-1$.  The exact fluxes used are shown in Fig.~\ref{fig:mps_scaling_flux_configs}.  Different line styles in Fig.~\ref{fig:mps_scaling_Trotter} correspond to the different flux configurations.  As seen in the upper right panel, the variation in entanglement between flux configurations at each bond dimension is less than the effect of doubling bond dimension.  Likewise, as seen in the lower right panel, the variation is less than the effect of increasing system size.  This suggests that all these flux configurations are indeed typical/representative of the superposition.
	
	In accordance with our intuition that evolution of localized systems should be easier to study than evolution of ergodic systems, we see that for the same target fidelity, the required bond dimension by cycle 40 on the largest system is a full order of magnitude smaller than for each of these five disordered flux configurations than for the one with all fluxes equal.  (This confirms the intuition above that the high cost of simulating the evolution of the superposition initial state when using the full LGT model is due to needing to keep contributions from exponentially many flux sectors, as each individual one is cheap to simulate.)  However, the cost still grows exponentially in both time and system size.
	
	Finally, what can we conclude about the possibility of going beyond classical in experiments on disorder-free localization?  The current experiment is not beyond classical; indeed we have simulated it with high fidelity.  However, the scaling shown in Fig.~\ref{fig:mps_scaling_Trotter} clearly demonstrates that on large systems, the MPS simulations have exponential cost in number of Trotter cycles, which is a result of exponentially decaying fidelity in time for any bond dimension less than the maximum required for an exact simulation.  Thus when error per cycle in the experiment can be brought below that of MPS simulations, the experiment will have an exponential advantage at long times.  Also note that simulations with two-dimensional tensor networks would in principle remove the exponential cost of increasing system size; however the exponential cost in time is expected to remain, thus still leaving room for future quantum advantage.
	
	\newpage
	
	\section{Grover's algorithm and DFL}
	\label{sec:grover}
	A key aspect of the DFL construction is that expectation values of operators
	are the disorder-averaged values. Explicitly, we construct a unitary
	\(\ulgt\) such that
	\begin{equation}
		\label{eq:generic-dis-expval}
		\expval{\ulgt^{\dagger} \hat O \ulgt}{0} = \sum_{D} \operatorname{Pr}(D) \expval{\hat O}_D
	\end{equation}
	for some desired disorder distribution \(\operatorname{Pr}\) and single-qubit Pauli observable $\hat{O}$. In general, $\hat O$ can be any hermitian, unitary operator with $\pm 1$ as its only eigenvalues. In our
	particular construction one can view the different disorder sectors as
	evolving independently, so that we end up with
	\begin{equation*}
		\ulgt \ket{0} = \sum_D \sqrt{\operatorname{Pr}(D)} \ket{\psi_D} \otimes \ket{D}
	\end{equation*}
	where the first tensor factor contains the ``dynamical'' degrees of freedom,
	and the second simply labels the disorder. In a series of experiments that
	terminates by measuring observables on each tensor factor independently,
	there is no difference between sampling \(\ulgt \ket{0}\), and sampling \(D\)
	for each experiment independently.
	
	Nevertheless, there is a physically meaningful difference between density
	matrices
	\begin{equation*}
		\rho_{\mathrm{LGT}} = \ulgt \ketbra{0}{0} \ulgt^{\dagger},
		\; \rho_s = \sum_D \operatorname{Pr}(D) \ketbra{\psi_D}{\psi_D} \otimes \ketbra{D}{D}.
	\end{equation*}
	An algorithm which makes a practical use of this distinction is the Grover
	search algorithm \cite{grover_1996}. As an example, we explain how this algorithm can be used to
	estimate \cref{eq:generic-dis-expval} to accuracy \(\sim \epsilon\) in \(\sim 1
	/ \epsilon\) applications of \(\ulgt\) and \(O(1)\) experiments. By contrast,
	naive sampling of \(\rho_s\) would require \(O(1 / \epsilon^2)\) experiments
	where time evolution \(\hat U_D\) is applied once per experiment (i.e. disorder
	realization); in other words, \(O(1 / \epsilon^2)\) applications of \(\hat U_D\).
	
	At a high level, the composition of reflections \(\hat \Gamma\) that appears in an
	instantiation of the Grover algorithm has eigenvalues that reveal the
	expectation value \cref{eq:generic-dis-expval}. We then apply the phase
	estimation algorithm \cite{kitaev_1995_quantum_measurements_abelianstabilizer} to extract these eigenvalues. Explicitly, for
	(efficiently implementable) unitary
	\begin{equation*}
		\hat \Gamma = \ulgt (1 - 2 \ketbra{0}{0} ) \ulgt^{\dagger} \hat O
	\end{equation*}
	the two eigenvectors that have overlap with \(\ulgt \ket{0}\) have eigenvalues
	\(e^{\pm i \lambda}\) satisfying
	\begin{equation*}
		\cos \lambda = - \expval{\ulgt^{\dagger} \hat O \ulgt}{0}.
	\end{equation*}
	The estimation of one such eigenvalue using the standard phase estimation
	algorithm to accuracy \(\epsilon\) consists of a circuit which applies
	\(\ulgt\) \(\sim 2 / \epsilon + 1\) times.
	
	\newpage
	\section{Second R\'enyi entropy}
	\label{sec:renyi_entropy}
	\subsection{Measurement}
	\label{ssec:renyi_measurement}
\begin{figure*}[!b]
		\centering
		\includegraphics{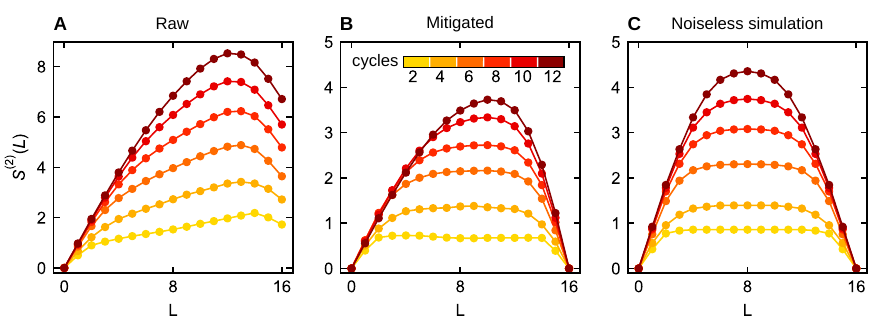}
		\caption{(\textbf{A}) Experimentally measured second R\'enyi entropy as a function of subsystem size ($L$) measured at cycles $2-12$ for a 16-qubit uniform superposition initial state using randomized measurements in single-qubit Clifford basis. Note that the experimentally measured entropy of the ideally pure system grows with the cycle due to decoherence. (\textbf{B}) The entropy of the full system was subtracted from the raw values after rescaling by $L/N$ to mitigate error, which was compared against the simulated values shown in (\textbf{C}). The agreement between mitigated vs. simulated values is stronger at early cycles where the entropy has not yet saturated to a volume law.}
		\label{fig::entropy_sm}
	\end{figure*}

	We utilize a randomized measurement protocol \cite{randomized_measurements_beenakker, entanglement_spectroscopy_troyer,Brydges_2019} to experimentally measure the second R\'enyi entropy without resorting to a full state tomography. We apply random Clifford gates to each qubit prior to measurement. We sample over $N_u$ such choices, measuring each with $N_s$ shots.
	The purity of an $L$-qubit subsystem with reduced density matrix $\hat \rho$ is defined as $P =\Tr[\hat \rho^2]$, and can be estimated as
	\begin{equation}
		P = \frac{2^L}{N_u} \sum_{n=1}^{N_u} \sum_{s,s'}^{2^L} P(s)P(s')(-2)^{H(s,s')},
	\end{equation}
	where $H(s,s')$ is the Hamming distance between bitstrings $s$ and $s'$ in the computational basis, and $P(s)$ is the probability of measuring $s$. There is a bias in this estimation arising from finite sampling, which can be removed via Jacknife resampling \cite{jacknife_resampling} to obtain the following unbiased expression
	\begin{equation}
		P_{\text{unbiased}} = \frac{N_s}{N_s-1} P -\frac{2^L}{N_s-1},
	\end{equation}
	which is then used to compute the second R\'enyi entropy
	\begin{equation}
		S^{(2)} = -\log_2(P_{\text{unbiased}}).
	\end{equation}
	Since for a pure state, the entropy of a subsystem A, $S^{(2)}(A)$, is the same as the entropy of its complement $S^{(2)}(\bar{A})$, the entanglement of subsystems up to half of the system is sufficient to understand the entanglement scaling. Nevertheless, in our experiments, the system undergoes decoherence, resulting in some residual volume-law contribution which can be subtracted, as in Ref.~\cite{mipt_hoke_2023}, to match the exact (simulated) values both qualitatively and quantitatively. This error mitigation technique requires measuring the entropy of the full system. The results for a uniform superposition initial state are shown in Fig.~\ref{fig::entropy_sm}. We note that, especially at larger cycles, the mitigated value consistently underestimates the simulation values. This is expected because the maximum entropy of a $L$-qubit subsystem is $L$, and if the residual contribution to this entropy is large, then subtracting it gives a consistently lower estimate. In fact, after maximum decoherence, this error-mitigation technique falsely gives $0$ entanglement entropy even for highly entangled states. The limitations of this technique are described in more detail in the SM of Ref.~\cite{mipt_hoke_2023}. Despite the limitations of this technique, we obtain a good agreement between experimental and simulated results up to $12$ cycles using first-order Trotter dynamics.
	
	The number of shots required grows exponentially with the subsystem size, which is quite large, especially for $L=16$. For the uniform superposition initial state, which grows to a volume law, we used $N_u=50, N_s=8 \times 10^6$, whereas, for the single-sector initial states, which seem to exhibit an area-law for both uniform and disordered cases, we used $N_u=100$ and $N_s=4 \times 10^6$. The total number of shots $N_u \times N_s$ used for each state was thus fixed at  $4 \times 10^8$. The entropy reported for the disordered gauge-invariant initial state was averaged over disordered configurations and this averaging was done in the same step as averaging over the unitary bases.

	\subsection{Numerical results}
	\label{ssec:renyi_numerical}
	\begin{figure*}[t!]
		\centering
		\includegraphics[scale=0.95]{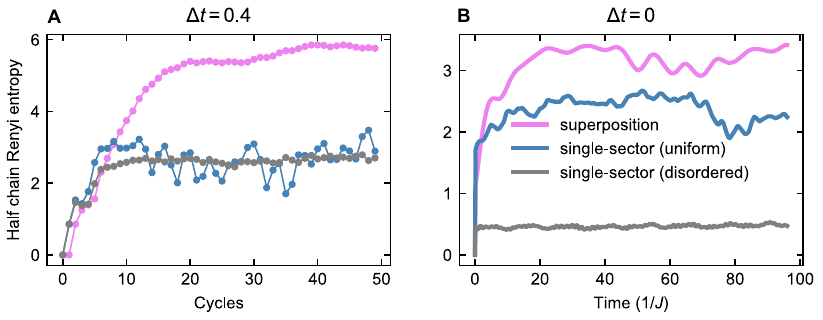}
		\caption{Numerical results for the half chain R\'enyi entropy of a 16-qubit ring, obtained using exact statevector simulation of the (\textbf{A}) Floquet evolution, and (\textbf{B}) Hamiltonian evolution obtained by using Lanczos time-evolution method. }
		\label{sfig:extropy_exact}
	\end{figure*}
	
	We extend the R\'enyi entropy results shown in Fig.~4 of the main text to later times using two methods (Fig.~\ref{sfig:extropy_exact}). 
	\begin{enumerate}[label=(\Alph*)]
		\item Statevector simulation with first-order Trotterized dynamics. We used a larger Trotter step size ($\Delta t=0.4$) to highlight the volume-law entanglement of the superposition initial state. This revealed comparable entropies for uniform and disordered single-sector states at later times, despite the disordered state's slower initial growth.
		
		\item Lanczos method simulation of exact Hamiltonian dynamics: In the limit of exact dynamics ($\Delta t = 0$), the disordered (localized due to MBL) state exhibits lower entropy compared to the delocalized uniform state.
	\end{enumerate}
	
	Interestingly, unlike single-sector states, the superposition state shows a qualitatively different behavior for both cases. This arises because the superposition initial state can have correlations between the background charges and the dynamical degrees since it can exist in an effectively higher dimensional Hilbert space compared to a fixed superselection sector for the single-sector states. Therefore,  DFL doesn't prohibit localized states from having a rapid entanglement growth to a volume-law in contrast to MBL, where the entanglement has been found to grow logarithmically in time \cite{abanin2017recent}. As discussed in the main text, we reiterate that the entropy of the superposition state shown above is not simply equal to the entropy averaged over the disorder realizations.
	
\end{document}